\documentclass[11pt,english,showkeys,showpacs,titlepage]{revtex4-2}
\usepackage[T1]{fontenc}
\usepackage[latin9]{inputenc}
\usepackage{float}
\usepackage{textcomp}
\usepackage{mathtools}
\usepackage{multirow}
\usepackage{amsmath}
\usepackage{amssymb}
\usepackage{graphicx}
\usepackage{wasysym}
\usepackage{esint}
\usepackage[all]{xy}
\usepackage[unicode=true,pdfusetitle,
 bookmarks=true,bookmarksnumbered=false,bookmarksopen=false,
 breaklinks=false,pdfborder={0 0 1},backref=false,colorlinks=false]
 {hyperref}
\usepackage{xcolor}
\usepackage[english]{babel}
\makeatletter

\providecommand{\tabularnewline}{\\}

\usepackage{babel}

\newcommand{\xyR}[1]{%
\makeatletter
\xydef@\xymatrixrowsep@{#1}
\makeatother
} % end of \xyR

\newcommand{\xyC}[1]{%
\makeatletter
\xydef@\xymatrixcolsep@{#1}
\makeatother
} % end of \xyR
\makeatother

\begin{document}
\title{Perturbative Color Correlations in Double Parton Scattering.}
\author{B. Blok}
\email{blok@physics.technion.ac.il}

\author{J. Mehl}
\email{yonatanm@campus.technion.ac.il}

\address{Department of Physics, Technion -- Israel Institute of Technology,
Haifa, Israel}
\begin{abstract}
We study the  contribution of color correlations to Double Parton Scattering (DPS). We show that there
is a specific class of Feynman diagrams related to so called 1 \textrightarrow{}
2 processes when the contribution of these color correlations is not
Sudakov suppressed with the transverse scales. The effective absence of
Sudakov suppression gives hope that although they are small relative
to color singlet correlations, they eventually can be observed.
\end{abstract}
\maketitle
\tableofcontents{}

\section{Introduction\label{sec:Introduction}}

The theory of double parton scattering (DPS) in QCD was the subject
of intensive development in recent years. The first work on DPS was
done in the early 80s \citep{treleani,mekhfi}, and the first detailed
experimental observations of DPS were done in Tevatron. Recently new
detailed experimental studies of DPS were carried out at LHC while
a new theoretical formalism based on pQCD was developed \citep{GS1,Blok2011,diehl1,GS2,Blok2012,Diehl2016,Blok2014,diehl3,Manohar2012}.
In these works, the fundamental role of parton correlations in DPS
scattering was realized and estimated and new physical objects to
study these correlations - two particle generalized parton distribution
(\color{black}  $\phantom{}_{2}GPD$ \color{black}) were introduced. However, most of this work was devoted to the
study of color singlet correlations in DPS processes.

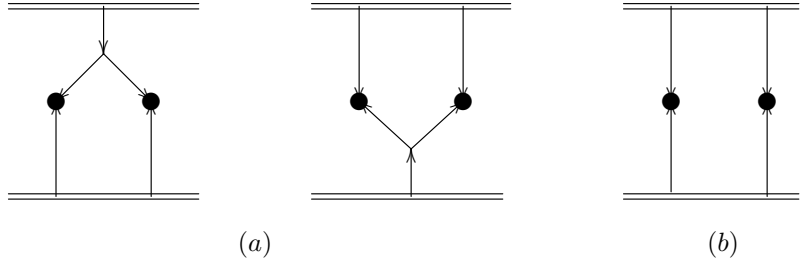
\begin{figure}
\[
\xymatrix{*=0{}\ar@{=}[rrrr] &  & *=0{}\ar@{->}[d] & *=0{} & *=0{} &  & *=0{}\ar@{=}[rrrr] & *=0{}\ar@{->}[dd] &  & *=0{}\ar@{->}[dd] & *=0{} &  & *=0{}\ar@{=}[rrrr] & *=0{}\ar@{->}[dd] &  & *=0{}\ar@{->}[dd] & *=0{}\\
\xyR{1pc}\xyC{1pc} &  & *=0{}\ar@{->}[dl]\ar@{->}[dr]\\
 & *=0{\newmoon} &  & *=0{\newmoon} &  & {\ \ } &  & *=0{\newmoon} & {\ } & *=0{\newmoon} &  & {\ \ } &  & *=0{\newmoon} & {\ } & *=0{\newmoon}\\
 &  &  &  &  &  &  &  & *=0{}\ar@{->}[ul]\ar@{->}[ur]\\
*=0{}\ar@{=}[rrrr] & *=0{}\ar@{->}[uu] &  & *=0{}\ar@{->}[uu] & *=0{} &  & *=0{}\ar@{=}[rrrr] &  & *=0{}\ar@{->}[u] &  &  &  & *=0{}\ar@{=}[rrrr] & \ar@{->}[uu] &  & *=0{}\ar@{->}[uu] & *=0{}\\
 &  &  &  &  & *=0{(a)} &  &  &  &  &  &  &  &  & *=0{(b)}
}
\]

\caption{The different diagrams contributing to double parton scattering (DPS)
$(a)$ the two possible $1+2$ processes and $(b)$ a $2+2$ process.
As explained in the text there is no \textquotedblleft$1+1$\textquotedblright{}
contribution. the $=$ line represents the hadrons \label{fig:1+2-2+2}}
\end{figure}

Recently, a lot of interest was attended to color non-singlet correlations
in proton-proton collisions. The possibility of such correlations
was already discussed in the 80s \cite{artru1,Mekhfi1988}. However,
it was shown that such correlations are strongly Sudakov suppressed
due to a need to change color quantum numbers \color{black} between the amplitude and the complex conjugate \color{black} \citep{Mekhfi1988,artru1}. The color correlations
were shown to be suppressed as

\begin{equation}
\exp\left(-\alpha_{s}\log^{2}\left(\frac{Q^{2}}{\Lambda_{QCD}^{2}}\right)\right)
\end{equation}

where $Q$ is the transverse momenta. As a result, such correlations
are negligible, at least in conventional hard processes, and rapidly
decrease with hard scale. Such correlations were first considered
in \citep{Mekhfi1988,artru1} for conventional hard processes, and
for the so-called $2\rightarrow2$ processes in the DPS (see figure
\ref{fig:1+2-2+2} $(b)$).

More recently it was realized that the color correlations can occur
also in the so-called $1\rightarrow2$ processes and they were studied
in \citep{Diehl2016,Manohar2012,mulders,buffing2021,diehl7}. In recent
work \citep{Diehl2021} it was noted that the two particle \color{black}  $\phantom{}_{2}GPD$ \color{black} that
described color non-singlet correlations can be negative.

Still, there remains a problem to find the contribution of color correlations
in DPS processes. Indeed, the contribution of color correlations is
Sudakov suppressed, so the appearance of color ladders in the scattering
amplitudes is negligible for transverse momenta where one can expect
to observe the DPS processes. On the other hand, the analysis of singlet
correlations in DPS processes shows that a significant part of the
contribution to \color{black}  $\phantom{}_{2}GPD$ \color{black} comes from the processes where the ladder splits
into two short ladders, corresponding to the fundamental solutions
of DGLAP equations for $x\sim1$, leading to two hard processes. In
these ladders, the transverse momenta evolve not from $Q_{0}^{2}\sim0.5\ GeV^{2}$
to $Q^{2}$, but from $k$ to $Q^{2}$ Where $Q_{0}\ll k\ll Q$ of
some indeterminate perturbative scale where the split occurred.

For such processes, one can consider the $1\rightarrow2$ processes
depicted in figure \ref{fig:color ladders} ($a$). Indeed the two
ladders coming from below in figure \ref{fig:1+2-2+2} $(a)$, and
the ladder that splits are not suppressed. Only two ladders that go
to hard processes after the split are colored and will be suppressed,
but Sudakov suppression may be much smaller

\begin{equation}
\sim\exp\left(-\alpha_{s}\log^{2}\left(\frac{Q^{2}}{k^{2}}\right)\right).
\end{equation}

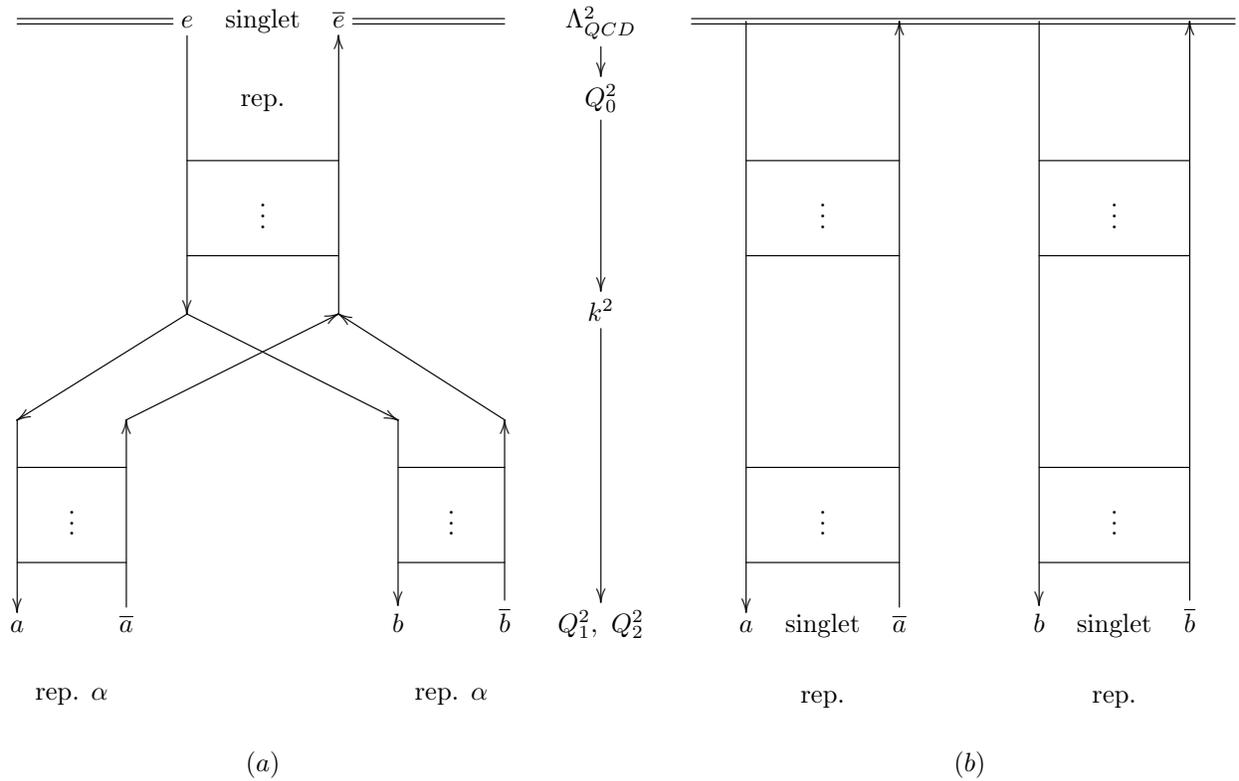
\begin{figure}
\[
\xymatrix{*=0{}\ar@{=}[rrr] &  &  & {e}\ar@{->}[ddddd] & *=0{\text{singlet}} & {\overline{e}}\ar@{<-}[ddddd]\ar@{=}[rrr] &  &  & *=0{} & {\Lambda_{QCD}^{2}}\ar@{->}[d] & *=0{}\ar@{=}[rrrrrrrr] & *=0{}\ar@{->}[ddddddddddd] &  & *=0{}\ar@{<-}[ddddddddddd] &  & *=0{}\ar@{->}[ddddddddddd] &  & *=0{}\ar@{<-}[ddddddddddd] & *=0{}\\
 &  &  &  & {\text{rep.}} &  &  &  &  & {Q_{0}^{2}}\ar@{->}[dddd]\\
 &  &  & *=0{}\ar@{-}[rr] &  & *=0{} &  &  &  &  &  & *=0{}\ar@{-}[rr] &  & *=0{} &  & *=0{}\ar@{-}[rr] &  & *=0{}\\
 &  &  &  & *=0{\vdots} &  &  &  &  &  &  &  & *=0{\vdots} &  &  &  & *=0{\vdots}\\
 &  &  & *=0{}\ar@{-}[rr] &  & *=0{} &  &  &  &  &  & *=0{}\ar@{-}[rr] &  & *=0{} &  & *=0{}\ar@{-}[rr] &  & *=0{}\\
 &  &  & *=0{}\ar@{->}[ddlll]\ar@{->}[ddrrr] &  & *=0{}\ar@{<-}[ddlll]\ar@{<-}[ddrrr] &  &  &  & {k^{2}}\ar@{->}[dddddd]\\
\\
*=0{}\ar@{->}[dddd] &  & *=0{}\ar@{<-}[dddd] &  &  &  & *=0{}\ar@{->}[dddd] &  & *=0{}\ar@{<-}[dddd]\\
*=0{}\ar@{-}[rr] &  & *=0{} &  &  &  & *=0{}\ar@{-}[rr] &  & *=0{} &  &  & *=0{}\ar@{-}[rr] &  & *=0{} &  & *=0{}\ar@{-}[rr] &  & *=0{}\\
 & *=0{\vdots} &  &  &  &  &  & *=0{\vdots} &  &  &  &  & *=0{\vdots} &  &  &  & *=0{\vdots}\\
*=0{}\ar@{-}[rr] &  & *=0{} &  &  &  & *=0{}\ar@{-}[rr] &  & *=0{} &  &  & *=0{}\ar@{-}[rr] &  & *=0{} &  & *=0{}\ar@{-}[rr] &  & *=0{}\\
{a} &  & {\overline{a}} &  &  &  & {b} &  & {\overline{b}} & {Q_{1}^{2},\ Q_{2}^{2}} &  & {a} & *=0{\text{singlet}} & {\overline{a}} &  & {b} & *=0{\text{singlet}} & {\overline{b}}\\
 & *=0{\text{rep. }\alpha} &  &  &  &  &  & *=0{\text{rep. }\alpha} &  &  &  &  & {\text{rep.}} &  &  &  & {\text{rep.}}\\
 &  &  &  & \left(a\right) &  &  &  &  &  &  &  &  &  & \left(b\right)
}
\]

\caption{$\left(a\right)$ The $1\rightarrow2$ process diagram and its complex
conjugate with the DGLAP ladders presented explicitly. Both parton
pairs $a,\overline{a}$ and $b,\overline{b}$ are in some nonsinglet
representation $\alpha$ while $e,\overline{e}$ are in a singlet
state. $\left(b\right)$ the $2\rightarrow2$ diagram and its complex
conjugate, now $a,\overline{a}$ and $b,\overline{b}$ are in a singlet
representation because other representations are Sudakov suppressed.
The scales of the ladder evolution are shown in middle. \label{fig:color ladders}}
\end{figure}

In this paper we shall calculate \color{black}  $\phantom{}_{2}GPD$ \color{black} corresponding to such processes, and
find that such \color{black}  $\phantom{}_{2}GPD$ \color{black} may be indeed large - up to 5-10\% relative to
singlet \color{black}  $\phantom{}_{2}GPD$\color{black}, extensively studied before \citep{Blok2014,gaunt}. Moreover,
this contribution does not decrease with $Q^{2}$ and slowly increases
relative to mean field contribution to \color{black}  $\phantom{}_{2}GPD$ \color{black} like for color singlet
$1\rightarrow2$ processes, thus being present at Tevatron and LHC.
We shall see that these contributions can be both positive and negative,
depending on the representation of color $SU\left(3\right)$.
\par We shall see that the characteristic scale where the singlet ladder in $1\rightarrow2$  is increasing with the hard scale Q 
of the process. This is contrary to singlet split scale, which does not depend on the energy and is of order several GeV.
Due to such energy dependence of the split scale in colour channels, the corresponding $1\rightarrow2$ part 
of $_2$GPD is effectively not suppressed.
\par As a result, we shall have the so called $1v1 $ \cite{Diehl:2017kgu} processes with two  singlet  ladders 
from above and below (i.e. from each of contributing nucleons) split into two colour ladders, being unsuppressed 
with the energy scale increase. Indeed, such contributions were recently considered in \cite{Diehl:2023jje}. Nevertheless
further work is  needed in order to observe the contributions of these processes and distinguish them
from both conventional singlet DPS and conventional $1\rightarrow 1$ hard processes.
%On the other hand, the additional source of suppression of observable
%color correlations is the need for a non-singlet distributions in both hadrons, as is shown in Fig. \ref{fig: interference-1-2}.
%It's argued \cite{Blok2012} that in LO one can't have a $1\rightarrow2$ processes in both hadrons (i.e, no "$1+1$" processes).
%In that case even if we consider one non-singlet $1\rightarrow2$ process the other distribution will necessarily still be Sudakov %suppressed, albeit the total suppression of the final state will be smaller then the 2+2 case.%^Other sources \cite{Diehl:2017kgu} %claim that one can include 1+1 type processes in the cross section but to subtract from them a certain term to avoid over-%counting which again will cause certain suppression.

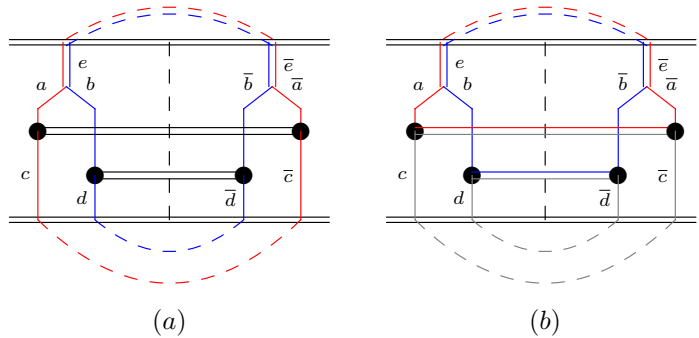
\begin{figure}
\[
\xymatrix{*=0{}\ar@{=}[rrrrrrrrrr] &  & *=0{}\ar@{-}@[blue]@<0.3ex>[dd]^{e}\ar@{-}@[red]@<-0.3ex>[dd]\ar@/^{1pc}/@[red]@<0.3ex>@{--}[rrrrrr]\ar@/^{1pc}/@[blue]@<-0.3ex>@{--}[rrrrrr] &  &  & *=0{}\ar@{--}[dddddddd] &  &  & *=0{}\ar@{-}@[red]@<0.3ex>[dd]^{\overline{e}}\ar@{-}@[blue]@<-0.3ex>[dd] &  & *=0{} &  & *=0{}\ar@{=}[rrrrrrrrrr] &  & *=0{}\ar@{-}@[blue]@<0.3ex>[dd]^{e}\ar@{-}@[red]@<-0.3ex>[dd]\ar@/^{1pc}/@[red]@<0.3ex>@{--}[rrrrrr]\ar@/^{1pc}/@[blue]@<-0.3ex>@{--}[rrrrrr] &  &  & *=0{}\ar@{--}[dddddddd] &  &  & *=0{}\ar@{-}@[red]@<0.3ex>[dd]^{\overline{e}}\ar@{-}@[blue]@<-0.3ex>[dd] &  & *=0{}\\
\xyR{0.2pc}\xyC{0.4pc}\\
 &  & *=0{}\ar@[red]@{-}[dl]_{a}\ar@[blue]@{-}[dr]^{b} &  &  &  &  &  & *=0{}\ar@[blue]@{-}[dl]_{\overline{b}}\ar@[red]@{-}[dr]^{\overline{a}} &  &  &  &  &  & *=0{}\ar@[red]@{-}[dl]_{a}\ar@[blue]@{-}[dr]^{b} &  &  &  &  &  & *=0{}\ar@[blue]@{-}[dl]_{\overline{b}}\ar@[red]@{-}[dr]^{\overline{a}}\\
 & *=0{}\ar@[red]@{-}[d] &  & *=0{}\ar@[blue]@{-}[ddd] &  &  &  & *=0{}{}\ar@[blue]@{-}[ddd] &  & *=0{}{}\ar@[red]@{-}[d] &  &  &  & *=0{}\ar@[red]@{-}[d] &  & *=0{}{}\ar@[blue]@{-}[ddd] &  &  &  & *=0{}{}\ar@[blue]@{-}[ddd] &  & *=0{}{}\ar@[red]@{-}[d]\\
 & *=0{\newmoon}\ar@{-}@[black]@<0.3ex>[rrrrrrrr]\ar@{-}@[black]@<-0.3ex>[rrrrrrrr] &  &  &  &  &  &  &  & *=0{\newmoon} &  &  &  & *=0{\newmoon}\ar@{-}@[red]@<0.3ex>[rrrrrrrr]\ar@{-}@[gray]@<-0.3ex>[rrrrrrrr] &  &  &  &  &  &  &  & *=0{\newmoon}\\
 &  &  &  &  & {\ \ \ } &  &  &  &  &  &  &  &  &  &  &  & {\ \ \ }\\
 &  &  & *=0{\newmoon}\ar@{-}@[black]@<0.3ex>[rrrr]\ar@{-}@[black]@<-0.3ex>[rrrr] &  &  &  & *=0{\newmoon} &  &  &  &  &  &  &  & *=0{\newmoon}\ar@{-}@[blue]@<0.3ex>[rrrr]\ar@{-}@[gray]@<-0.3ex>[rrrr] &  &  &  & *=0{\newmoon}\\
\\
*=0{}\ar@{=}[rrrrrrrrrr] & *=0{}\ar@[red]@{-}[uuuu]^{c}\ar@[red]@/_{2pc}/@{--}[rrrrrrrr] &  & *=0{}\ar@[blue]@{-}[uu]^{d}\ar@[blue]@/_{1pc}/@{--}[rrrr] &  & *=0{} &  & *=0{}\ar@[blue]@{-}[uu]^{\overline{d}} &  & *=0{}\ar@[red]@{-}[uuuu]^{\overline{c}} & *=0{} &  & *=0{}\ar@{=}[rrrrrrrrrr] & *=0{}\ar@[gray]@{-}[uuuu]^{c}\ar@/_{2pc}/@[gray]@{--}[rrrrrrrr] &  & *=0{}\ar@[gray]@{-}[uu]^{d}\ar@/_{1pc}/@[gray]@{--}[rrrr] &  & *=0{} &  & *=0{}\ar@[gray]@{-}[uu]^{\overline{d}} &  & *=0{}\ar@[gray]@{-}[uuuu]^{\overline{c}} & *=0{}\\
 &  &  &  &  & {\phantom{a}} &  &  &  &  &  &  &  &  &  &  &  & {\phantom{a}}\\
 &  &  &  &  & {\phantom{a}} &  &  &  &  &  &  &  &  &  &  &  & {\phantom{a}}\\
 &  &  &  &  & {\left(a\right)} &  &  &  &  &  &  &  &  &  &  &  & {\left(b\right)}\\
}
\]

\caption{Different amplitudes (diagram contracted with it's complex conjugate)
that contribute to the $1+2$ process. Color flow is schematically shown using the red
and blue lines for color neutral final states $\left(a\right)$ and final states that carry color numbers $\left(b\right)$. The $=$ lines represent the final states. Particles in a singlet state are contracted
explicitly by a dashed line (we assume, as will be discussed in the text, that particles not coming from $1\rightarrow2$ processes, i.e the bottom part of the graphs, are in singlet state due to strong Sudakov suppression of the non-singlet states). 
It can be seen that only the singlet structure (i.e $a$ is with the same color as $\overline{a}$
and so are $b$ and $\overline{b}$) can contribute, unless one assume a "1+1" type amplitude or a non-singlet non-perturbative distribution.
 \label{fig: interference-1-2}}
\end{figure}

%TO END
\color{black}

In our calculations,  we use ``DDT-like'' formalism \citep{Dokshitzer1980}.

In our numerical calculations of \color{black}  $\phantom{}_{2}GPD$\color{black}, we concentrate on gluonic \color{black} $\phantom{}_{2}GPD$  \color{black}
since they give a dominant contribution to the considered processes
due to the large color factor, and neglect the contribution of quark
and mixed quark-gluon \color{black}  $\phantom{}_{2}GPD$ \color{black}.

The paper is organized in the following way. In Section \ref{sec:Basic-pQCD-Formalization}
we describe our formalism to calculate the color correlated \color{black}  $\phantom{}_{2}GPD$ \color{black} in
LLA. We study the nonsinglet analogue
of DGLAP equation and the $x\rightarrow1$ asymptotic of its fundamental
solutions. We show how to calculate  nonsinglet \color{black}  $\phantom{}_{2}GPD$ \color{black} for $1\rightarrow2$ processes
using the fundamental  singlet and nonsinglet solutions of the DGLAP
equation. We explain the divergences
in the integrals for \color{black}  $\phantom{}_{2}GPD$ \color{black} and show that these integrals can be calculated
using an analytical continuation procedure.
\par In Section \ref{sec:Numerics} we use the results of Sections \ref{sec:Basic-pQCD-Formalization},
to calculate the non singlet distribution.
Our results are summarized in the conclusion.

In Appendix \ref{sec:Color-Kernels} we review the calculation of
DGLAP kernels for nonsinglet channels. Although the results for kernels
are known we believe it is useful to give the details of derivations.
In Appendix \ref{sec:Formal DGLAP} we rederive nonsinglet DGLAP equation
and show that the contributions of real and virtual emissions have
different color factors, thus leading to Sudakov suppression \citep{Mekhfi1988,artru1}.
In Appendix \ref{sec:-at-the-x-1} we study the relevant asymptotic
for $x\rightarrow1$ for colored channels. In Appendix \ref{sec:Regularizing-Divergent-Integrals}
we discuss the divergent integrals in \color{black}  $\phantom{}_{2}GPD$ \color{black} and explained their regularization
using theory of generalized functions. 
 In Appendix \ref{sec:Rules-for-Color}
we review the properties of color projectors relevant to the calculations done in this paper.
In Appendix \ref{sec:formalisms} we comment on the relation between other formalisms used in the literature and the one used in this paper.

 \section{pQCD Formalism\label{sec:Basic-pQCD-Formalization}}

In this section, we'll develop the formalism for computing the $1\rightarrow2$
distribution when the two partons are color correlated. First, we
review the derivation of the DGLAP equation \citep{Dokshitzer1977,GL,Altarelli1977}
for non-singlet color states and discuss the solutions to this equation,
in particular the fundamental solutions (i.e. with initial conditions
of the form $\sim\delta\left(x-1\right)$ in the $x\rightarrow1$
limit). Then we express the two particle \color{black}  $\phantom{}_{2}GPD$ \color{black} for arbitrary color states,
connected with the $1\rightarrow2$ processes, through the solutions
of the non-singlet DGLAP. This analysis will include divergent integrals,
which we explain how to regularize.

\subsection{Color Non-Singlet DGLAP equation\label{subsec:Color-Non-Singlet-Solutions}}

\color{black}

Consider first the conventional singlet DGLAP equation \citep{Dokshitzer1977,GL,Altarelli1977}. The evolution of a parton A from one energy scale $k_{0}^{2}$ to another parton $B$ at scale $Q^{2}$ (the hard scale of the process) with a fraction of longitudinal momentum $x$ (the Bjorken variable, which is the fraction of longitudinal momentum of the parton compared to the parent hadron) is described by the structure function $D_{A}^{B}\left(x,k_{0}^{2},Q^{2}\right)$ (in the following we'll suppress unnecessary inputs). We take the hard scale $Q^{2}$ to be the transverse scale $ p_{\perp}^{2}$ (which is the transverse momentum of one of the outgoing particles squared). The exact choice of $Q^{2}$ gives only next to leading order effects and therefore is not within the scope of this work. 
In a physical gauge $D_{A}^{B}$
receives contributions from both real (ladder) and virtual (self energy)
diagrams which result in the DGLAP equation:

\color{black}

\begin{equation}
\frac{\partial D_{A}^{B}\left(x,\color{black}Q^{2}\color{black}\right)}{\partial ln\left(\color{black}Q^{2}\color{black}\right)}=\frac{\alpha_{s}\left(\color{black}Q^{2}\color{black}\right)}{4\pi}\sum_{C}\intop_{0}^{1}\frac{dz}{z}\left[\underset{real\ contributions}{\underbrace{\Phi_{C}^{B}\left(z\right)D_{A}^{C}\left(\frac{x}{z},\color{black}Q^{2}\color{black}\right)}}-\underset{virtual\ contributions}{\underbrace{\Phi_{B}^{C}\left(z\right)z^{2}D_{A}^{B}\left(x,\color{black}Q^{2}\color{black}\right)}}\right].\label{eq:Singlet DGLAP}
\end{equation}

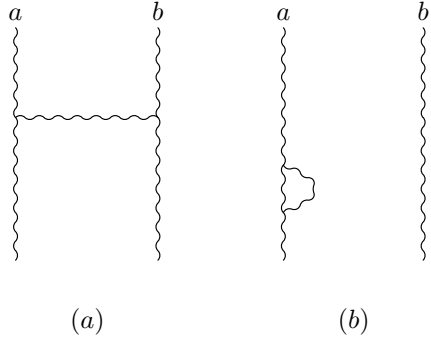
\begin{figure}
 \color{black}
\[
\xymatrix{{a}\ar@{~}[dd] &  & {b}\ar@{~}[dd] &  & {a}\ar@{~}[dd] &  & {b}\ar@{~}[dd]\\
\xyR{1pc}\xyC{1pc}\\
*=0{}\ar@{~}[rr]\ar@{~}[ddd] &  & *=0{}\ar@{~}[ddd] & {\ \ } & *=0{}\ar@{~}[ddd] &  & *=0{}\ar@{~}[ddd]\\
 &  &  &  & *=0{}\\
 &  &  &  & *=0{}\ar@/_{1pc}/@{~}[u]\\
*=0{} &  & *=0{} &  & *=0{} &  & *=0{}\\
 & {(a)} &  &  &  & {(b)}
}
\]
 \color{black}
\caption{Example of different contributions to the DGLAP equation. $(a)$ the
real (ladder) diagram for $\Phi_{G}^{G}$, the color factor for this
diagram might depend on the color state of the gluons $a,\ b$. $(b)$
example of a virtual (self-energy) contribution, these contributions
do not depend on the color state of $a,\ b$.\label{fig:ladder example}}
\end{figure}

The initial conditions for (\ref{eq:Singlet DGLAP}) are $D_{A}^{B}\left(x,k_{0}^{2},k_{0}^{2}\right)=\delta_{A}^{B}\delta\left(1-x\right)$
which represents the fact we are looking for the fundamental solutions (or Green functions)
for these equations. $A,\ B,\ C$ are the different types of partons
(quarks, anti-quarks or gluons) and the sum $\sum_{C}$ runs over
gluons and all $n_{f}$ flavors of quarks and anti-quarks  ($n_{f}$ is the number of active flavors). $\Phi_{A}^{C}$
are the DGLAP kernels without the ``$+$ subscription'' \citep{Dokshitzer1980}
and are given by $\Phi_{A}^{C}\left(z\right)=C_{A}^{C}\cdot V_{A}^{C}\left(z\right)$.
Here $V_{A}^{C}$ is given in table \ref{tab:Vz} and $C_{A}^{C}=\phantom{}^{1}\overline{C}_{A}^{C}$
are the color factors given in table \ref{tab:Color-factors-for}
for the singlet ($\alpha=1$) column. $\alpha_{s}\left(k^{2}\right)$
is the strong coupling constant and is given to leading order by:

\begin{table}
\centering{}%
\begin{tabular}{c|c}
$V_{F}^{F}$  & $2\cdot\frac{1+z^{2}}{1-z}$\tabularnewline
\hline 
$V_{F}^{G}$  & $2\cdot\frac{1+\left(1-z\right)^{2}}{z}$\tabularnewline
\hline 
$V_{G}^{F}$  & $2\cdot\left[z^{2}+\left(1-z\right)^{2}\right]$\tabularnewline
\hline 
$V_{G}^{G}$  & $4\cdot\left[z\left(1-z\right)+\frac{1-z}{z}+\frac{z}{1-z}\right]$\tabularnewline
\end{tabular}\caption{Longitudinal Momentum dependence of the DGLAP kernels in leading order
\label{tab:Vz}}
\end{table}

\begin{equation}
\alpha_{s}\left(k^{2}\right)=\frac{12\pi}{\beta_{0}\cdot ln\left(\frac{k^{2}}{\Lambda_{QCD}^{2}}\right)},
\end{equation}

with $\Lambda_{QCD}=0.22\ GeV$ and $\beta_{0}=11\cdot N-2\cdot n_{f}$.
$N$ is the number of
colors, we take both $N$ and $n_{f}$ to be 3 in the numeric calculations. All the
diagrams in this paper will be computed to leading order in $\alpha_{s}$
and the evolution equations are within the LLA (Leading Logarithmic
Approximation).

We now consider the case where the evolving parton and its complex
conjugate are not in a singlet state. To be more specific assume they
form together some non-singlet irreducible color representation $\alpha$
(here and for the rest of the paper we denote color representations
with Greek letters). Then the virtual and real contributions receive
different color factors as is shown in figure \ref{fig:ladder example}
(as it was first stressed in \citep{Mekhfi1988,artru1}). We therefore,
get the non-singlet DGLAP equation for irreducible $SU\left(3\right)$
representation $\alpha$ (note the representation index is to the
left of the quantities in this equation, i.e $\phantom{}^{\alpha}X$):

\begin{equation}
\frac{\partial\phantom{}^{\alpha}\overline{D}_{A}^{B}\left(x,\color{black}Q^{2}\color{black}\right)}{\partial ln\left(\color{black}Q^{2}\color{black}\right)}=\frac{\alpha_{s}\left(\color{black}Q^{2}\color{black}\right)}{4\pi}\sum_{C}\intop_{0}^{1}\frac{dz}{z}\left[\underset{\text{new kernel}}{\underbrace{\phantom{}^{\alpha}\overline{\Phi}_{C}^{B}\left(z\right)}}\phantom{}^{\alpha}\overline{D}_{A}^{C}\left(\frac{x}{z},\color{black}Q^{2}\color{black}\right)-z^{2}\underset{\text{old kernel}}{\underbrace{\Phi_{B}^{C}\left(z\right)}}\phantom{}^{\alpha}\overline{D}_{A}^{B}\left(x,\color{black}Q^{2}\color{black}\right)\right].\label{eq:Modified-DGLAP}
\end{equation}

Here $\Phi_{A}^{C}$ are the conventional singlet DGLAP kernels (as
described above). On the other hand $\phantom{}^{\alpha}\overline{\Phi}_{A}^{C}$
are the new kernels for the representation $\alpha$, coming from
the real radiation (consequently they appear in the first term, on
the right hand side). These kernels have the same momentum ($z$)
dependence as the singlet ones but have a different color factor:

\begin{equation}
\phantom{}^{\alpha}\overline{\Phi}_{A}^{C}\left(z\right)=\phantom{}^{\alpha}\overline{C}_{A}^{C}\cdot V_{A}^{C}\left(z\right).\label{eq:Color-Factorization}
\end{equation}

\begin{table}
\centering{}%
\begin{tabular}{c|c|c|c|c|c|c}
$\alpha=$  & $1$  & $8_{a}$  & $8_{s}$  & $10$  & $27$  & $0$\tabularnewline
\hline 
$\phantom{}^{\alpha}\overline{C}_{F}^{F}$  & $\frac{N^{2}-1}{2N}$  & $-\frac{1}{2N}$  & $-\frac{1}{2N}$  & $0$  & $0$  & $0$\tabularnewline
\hline 
$\phantom{}^{\alpha}\overline{C}_{F}^{G}$  & $\frac{N^{2}-1}{2N}$  & $\sqrt{\frac{N^{2}-1}{8}}$  & $\frac{1}{2N}\sqrt{\frac{\left(N^{2}-4\right)\left(N^{2}-1\right)}{2}}$  & $0$  & $0$  & $0$\tabularnewline
\hline 
$\phantom{}^{\alpha}\overline{C}_{G}^{F}$  & $\frac{1}{2}$  & $\frac{1}{2}\sqrt{\frac{N^{2}}{2\left(N^{2}-1\right)}}$  & $\frac{1}{2}\sqrt{\frac{N^{2}-4}{2\left(N^{2}-1\right)}}$  & $0$  & $0$  & $0$\tabularnewline
\hline 
$\phantom{}^{\alpha}\overline{C}_{G}^{G}$  & $N$  & $\frac{N}{2}$  & $\frac{N}{2}$  & $0$  & $-1$  & $1$\tabularnewline
\end{tabular}\caption{Color factors for different representations. In this paper, we consider
only irreducible representations, except for the two gluon decuplet
representations where we combine them together and denote $10:=10+\overline{10}$.
\label{tab:Color-factors-for}}
\end{table}

These factors were computed in \citep{Diehl2016} (we also write an
expanded explanation in Appendix \ref{sec:Color-Kernels}) and are
shown in table \ref{tab:Color-factors-for}. 
%TO COPY
\color{black}

Note that while writing the kernels in table \ref{tab:Vz} we use the same regularisation 
as in \cite{Dokshitzer1980} for terms singular when $z\rightarrow 1$:

\begin{equation}
1-z\rightarrow 1-z+\Delta\,\,\, ,\Delta=k_0^2/\sigma_k\sim k_0^2/Q^2 \label{eq:reg}
\end{equation}

As it was shown in \cite{Dokshitzer1980} (\ref{eq:Modified-DGLAP}) includes both double logarithmic and single logarithmic 
terms in hard transverse scale. For the singlet case double logarithmic terms do not contribute.
 Eq. (\ref{eq:Modified-DGLAP}) is well defined after regularisation (\ref{eq:reg}). We can now formally solve this equation using the ansats:

\begin{equation}
\phantom{}^{\alpha}\overline{D}_{A}^{B}\left(x,k_{0}^{2},Q^{2}\right)=\phantom{}^{\alpha}\overline{S}_{A}\left(k_{0}^{2},Q^{2}\right)\phantom{}^{\alpha}\widetilde{D}_{A}^{B}\left(x,k_{0}^{2},Q^{2}\right).\label{eq:Sudakov split}
\end{equation}

If we take for $\phantom{}^{\alpha}\overline{S}_{A}\left(k_{0}^{2},Q^{2}\right)$ the expression

\begin{equation}
\phantom{}^{\alpha}\overline{S}_{A}\left(k_{0}^{2},Q^{2}\right)=e^{-\sum_{C}\left(C_{A}^{C}-\phantom{}^{\alpha}\overline{C}_{A}^{C}\right)\intop_{k_{0}^{2}}^{Q^{2}}\frac{dk^{2}}{k^{2}}\frac{\alpha_{s}\left(k^{2}\right)}{4\pi}\intop_{0}^{1}dzzV_{A}^{C}\left(z\right)}\label{eq:Sudakov}
\end{equation}

we obtain that the function $\tilde D$, which we call the "tree level factor", satisfies the Equation
\begin{equation}
\frac{\partial\phantom{}^{\alpha}\widetilde{D}_{A}^{B}\left(x,Q^{2}\right)}{\partial ln\left(Q^{2}\right)}=\frac{\alpha_{s}\left(Q^{2}\right)}{4\pi}\sum_{C}\intop_{0}^{1}\frac{dz}{z}\left[\underset{\text{new kernel}}{\underbrace{\phantom{}^{\alpha}\overline{\Phi}_{C}^{B}\left(z\right)}}\phantom{}^{\alpha}\widetilde{D}_{A}^{C}\left(\frac{x}{z},Q^{2}\right)-z^{2}\underset{\text{new kernel}}{\underbrace{\phantom{}^{\alpha}\overline{\Phi}_{B}^{C}\left(z\right)}}\phantom{}^{\alpha}\widetilde{D}_{A}^{B}\left(x,Q^{2}\right)\right].\label{eq:Reduced DGLAP}
\end{equation}
which has the form of the conventional DGLAP equation and can be solved by the standard method \cite{Dokshitzer1980}.
\par Consider first the function $\phantom{}^{\alpha}\overline{S}_{A}$.
%This factor is very similar to the conventional Sudakov form factor appearing in the LLA
%renormalization of the parton propagator \citep{Dokshitzer1980},
%which is given by 
%\begin{equation}
%S_{A}\left(k_{0}^{2},Q^{2}\right)=e^{-\sum_{C}C_{A}^{C}\intop_{k_{0}^{2}}^{Q^{2}}\frac{dk^{2}}{k^{2}}\frac{\alpha_{s}\left(k^{2}\right)}{4\pi}\intop_{\color{black}0}^{1}dzzV_{A}^{C}\left(z\right)}.\label{eq:Regular Sudakov}
%\end{equation}
%This Sudakov factor is a doubly logarithmic function usually interpreted
%as the probability that a particle with virtuality $k_{0}^{2}$ experiencing
%a hard process at a scale $Q^{2}$ will not emit any soft gluon. The
%Sudakov factor defined in (\ref{eq:Sudakov}) does not lend itself
%to such a clear physical meaning, but might be interpreted as the
%probability that a system of two partons will not ``forget'' its
%initial color state by the emissions of soft gluons \citep{Mekhfi1988,artru1}.
The direct calculation shows that for DDT regularisation its logarithm is a sum of two pieces: the universal piece
leading to a form factor 

\begin{equation}
\phantom{}^{\alpha}\overline{S}_{A}\left(k_{0}^{2},Q^{2}\right)\approx e^{-\left(C_{A}^{A}-\phantom{}^{\alpha}\overline{C}_{A}^{A}\right)\cdot\frac{12}{\beta_{0}}\left[log\frac{Q^{2}}{\Lambda_{QCD}^{2}}log\left[\frac{log\frac{Q^{2}}{\Lambda_{QCD}^{2}}}{log\frac{k_{0}^{2}}{\Lambda_{QCD}^{2}}}\right]-log\frac{Q^{2}}{k_{0}^{2}}\right]},
\label{eq:Basseto}
\end{equation}
that coincides with a Sudakov form factor obtained in \cite{Bassetto1983}. This piece comes from neglecting the nonsingular terms
in integration of the kernels $V(z)$ in z from 0 to 1, i.e. leaving only $1/(1-z+\Delta)$. There is additional
nonuniversal piece that is single logarithmic and is obtained by integration of nonsingular part of the kernel V over z.
It is interesting to note that numerically the form factor (\ref{eq:Basseto})
is very close to simple double logarithmic expression used in \cite{Mekhfi1988}
\begin{equation}
\phantom{}^{\alpha}\overline{S}_{A}\left(k_{0}^{2},Q^{2}\right)=e^{-\frac{\alpha_{s}}{2\pi}\left(C_{A}^{A}-\phantom{}^{\alpha}\overline{C}_{A}^{A}\right)ln^{2}\left(\frac{Q^{2}}{k_{0}^{2}}\right)}.
\label{eq:Double Logarithem}
\end{equation}
The corrections due to nonsingular parts of the kernel were calculated explicitly for all channels that we considered and
numerically they contributed less than 10 percent to S.  For more detailed discussion of single logarithms and comparison
to the formalism \cite{Manohar2012,Diehl2016,buffing2021} see  Appendix \ref{sec:formalisms}.
\par We can now move to the solution of the DGLAP equation for $\tilde D$.
Equation (\ref{eq:Reduced DGLAP}) has the same form as the conventional singlet
DGLAP equation (\ref{eq:Singlet DGLAP}) but with different color
factors. To solve this equation we simply repeat the steps for solving
the regular DGLAP equations \citep{Dokshitzer1980} and by the analytical
continuation of the arguments replace the singlet color factors $C_{A}^{B}$
with the new, representation dependent, $\phantom{}^{\alpha}\overline{C}_{A}^{B}$.
%TO END
\color{black}
To be more concrete: 
\begin{itemize}
\item Transform to Mellin space using 
\begin{equation}
\phantom{}^{\alpha}\widetilde{D}_{A}^{B}\left(j\right)=\int\phantom{}^{\alpha}\widetilde{D}_{A}^{B}\left(x,k_{0}^{2}\right)x^{j-1}dx
\end{equation}
and change variables from $k_{0}$ and $Q$ to $\xi=\frac{3}{\beta_{0}}ln\left[\frac{\alpha_{s}\left(k_{0}^{2}\right)}{\alpha_{s}\left(Q^{2}\right)}\right]$
($\beta_{0}$ is defined above). 
\item Equation (\ref{eq:Reduced DGLAP}) then transform to a linear system
of first order differential equations in $\xi$ ($\left(1+2n_{f}\right)^{2}$
equations, for each combination of $A,B$, each can be a gluon or
any of $n_{f}$ fermions or anti-fermions). The system has a Hamiltonian
$H$ (a $\left(1+2n_{f}\right)\times\left(1+2n_{f}\right)$ matrix)
that only depends on $j$ and can be worked out analytically. The
solution to this system is simply $exp\left(\xi H\right)$. 
\item Solve $\phantom{}^{\alpha}\widetilde{D}_{A}^{B}\left(j,\xi\right)$
analytically by diagonalizing $H$, taking the exponent at this base,
and then transforming it back. We get that $\phantom{}^{\alpha}\widetilde{D}_{A}^{B}\left(j,\xi\right)$
will be a linear combination of exponents of the eigenvalues of $\xi H$
(there are only 3 independent eigenvalues). 
\item Return to $x$ space using the (numeric) inverse Mellin transform:
\begin{equation}
\phantom{}^{\alpha}\widetilde{D}_{C}^{B}\left(x,\xi\right)=\intop_{-\infty}^{\infty}\frac{dj}{2\pi i}x^{-j}\phantom{}^{\alpha}\widetilde{D}_{C}^{B}\left(j,\xi\right).
\end{equation}
This integral must be taken to the right of every singularity of $\phantom{}^{\alpha}\widetilde{D}_{C}^{B}\left(j,\xi\right)$.
For numerical purposes we take \citep{Vogt:2004ns,Dokshitzer1977}
\begin{equation}
j\left(t\right)=Max\left[1.5,\frac{16}{3}\frac{C_{F}^{F}}{1-x}\right]+\begin{cases}
\left(i-1\right)t & t>0\\
\left(1+i\right)t & t<0
\end{cases}.
\end{equation}
\end{itemize}
Note that although $\phantom{}^{\alpha}\widetilde{D}_{A}^{B}$ can
be written only as a function of $x$ and $\xi$, which encode the
dependence on both $k_{0}$ and $Q$, this is not true for $^{\alpha}\overline{D}_{A}^{B}\left(x,k_{0}^{2},Q^{2}\right)$
due to the Sudakov factor.

To conclude: the solution $\phantom{}^{\alpha}\overline{D}_{A}^{B}$
to the non-singlet DGLAP equation (\ref{eq:Modified-DGLAP}) is a
Sudakov suppression factor $\phantom{}^{\alpha}\overline{S}_{A}$
given in (\ref{eq:Sudakov}) times $\phantom{}^{\alpha}\widetilde{D}_{C}^{B}$,
the solution of (\ref{eq:Reduced DGLAP}), which can be numerically
evaluated for each value of $x,\ k_{0}^{2}$ and $Q^{2}$ in a given
representation. It's important to note that for non-singlet state
$\phantom{}^{\alpha}\overline{D}_{A}^{B}$ no longer represents a
physical distribution and therefore might even be negative \citep{Diehl2021}.

For the region $x\sim1$, if $\phantom{}^{\alpha}\overline{C}_{A}^{A}>0$
we can analytically solve (\ref{eq:Reduced DGLAP}) using the saddle
point method \citep{Dokshitzer1980} (see also Appendix \ref{sec:-at-the-x-1}
for detailed derivation). The saddle point for the inverse Mellin
transform integral is $j_{0}=\frac{4\xi\phantom{}^{\alpha}\overline{C}_{A}^{A}}{1-x}\gg1$
which is on the right of all the singularities of $\widetilde{D}\left(j\right)$.
We then get an analytical expression for $\widetilde{D}_{F}^{F}\left(x,\xi\right)$
and $\widetilde{D}_{G}^{G}\left(x,\xi\right)$ at this region. For
the case $\phantom{}^{\alpha}\overline{C}_{A}^{A}\leq0$ this argument
does not hold, the saddle point $j_{0}$ is then negative and therefore
cannot be taken as the primary value for the inverse integral (which
must be taken to the right of every singularity). However, we can
take the analytical continuation of the analytical expression we got
at the region $\phantom{}^{\alpha}\overline{C}_{A}^{A}>0$ (it's also
satisfying to know that for the region $x\sim0.9-0.999$, where $\widetilde{D}$
can be numerically evaluated to good accuracy, this analytical continuation
proves to be a very good approximation):

\begin{subequations}
\label{eq:x-1-limit}

\begin{equation}
\phantom{}^{\alpha}\widetilde{D}_{F}^{F}\left(x,\xi\right)\underset{x\sim1}{=}\frac{e^{-\xi\left[\left(4\gamma_{E}-\frac{17}{3}\right)\phantom{}^{\alpha}\overline{C}_{F}^{F}+\frac{8}{3}\phantom{}^{\alpha}\overline{C}_{F}^{G}\right]}}{\Gamma\left(4\xi\phantom{}^{\alpha}\overline{C}_{F}^{F}\right)}\cdot\frac{1}{\left(1-x\right)^{1-4\xi\phantom{}^{\alpha}\overline{C}_{F}^{F}}}
\end{equation}

\begin{equation}
\phantom{}^{\alpha}\widetilde{D}_{G}^{G}\left(x,\xi\right)\underset{x\sim1}{=}\frac{e^{\xi\left[\left(\frac{11}{3}-4\gamma_{E}\right)\phantom{}^{\alpha}\overline{C}_{G}^{G}-\frac{4}{3}n_{f}\phantom{}^{\alpha}\overline{C}_{G}^{F}\right]}}{\Gamma\left(4\xi\phantom{}^{\alpha}\overline{C}_{G}^{G}\right)}\cdot\frac{1}{\left(1-x\right)^{1-4\xi\phantom{}^{\alpha}\overline{C}_{G}^{G}}}
\end{equation}
\end{subequations}

Note that the dependence on $x$ in the limit $x\rightarrow1$ is
given by $1-x$ to some power. We therefore see that taking the integral
of these functions over $x$ will diverge if $\phantom{}^{\alpha}\overline{C}_{A}^{A}\leq0$.
This will prove to be a problem in the next section, for which we
introduce a regularization at section \ref{subsec:Regularization}.
The mixed gluon fermion fundamental solutions like $\phantom{}^{\alpha}\widetilde{D}_{F}^{G}$
are suppressed relative to diagonal ones by factors $\left(1-x\right)$.
They are nonsingular and their contribution to two particle \color{black}  $\phantom{}_{2}GPD$ \color{black} for
$1\rightarrow2$ processes is negligible, so we shall not need the
explicit expressions for them.

\subsection{Generalized Parton Distribution \label{subsec:Generalized-Parton-Distribution}}

Recall that the cross section of DPS is expressed through two particle
Generalized Parton distributions. Consider first the singlet case
\citep{Blok2012,Blok2011}. The generalized two parton distribution
(\color{black}  $\phantom{}_{2}GPD$ \color{black}) in a hadron is a sum of these two distributions \citep{Blok2012,Blok2011}

\begin{equation}
D(x_{1},x_{2},Q_{1},Q_{2},\vec{\Delta})=\phantom{}_{\left[1\right]}\overline{D}_{h}\left(x_{1},x_{2},Q_{1}^{2},Q_{2}^{2}\right)+\phantom{}_{\left[2\right]}\overline{D}_{h}\left(x_{1},x_{2},Q_{1}^{2},Q_{2}^{2},\vec{\Delta}\right).\label{eq:GPD}
\end{equation}
Here $x_{1},x_{2}$ the Bjorken variables for the partons, $Q_{1}$,
$Q_{2}$ the \color{black} transverse scales \color{black}, and $\vec{\Delta}$
is conjugated to the distance between the hard processes. The first
term describes the $2\rightarrow2$ processes when the two partons
come directly from the nonperturbative wave function of the nucleon
and then evolve from the hard process, while the second term corresponds
to the parton from the nonperturbative wave functions that evolved
to some perturbative scale $k$ where it splits into 2 perturbative
partons that evolve.

The total DPS cross section is schematically 
\begin{eqnarray}
\sigma_{DPS} & = & \sigma_{1}\sigma_{2}\times\frac{\int d^{2}\Delta}{(2\pi)^{2}}{}_{\left[2\right]}D(x_{1},x_{2},Q_{1},Q_{2},\Delta)_{\left[2\right]}D(x_{3},x_{4},Q_{1},Q_{2},\Delta)\nonumber \\[10pt]
 & + & _{\left[2\right]}D(x_{1},x_{2},Q_{1},Q_{2},\Delta)_{1}D(x_{3},x_{4},Q_{1},Q_{2},\Delta)+_{\left[1\right]}D(x_{1},x_{2},Q_{1},Q_{2},\Delta)_{\left[2\right]}D(x_{3},x_{4},Q_{1},Q_{2},\Delta).\nonumber \\[10pt]
\label{eq: singlet DPS}
\end{eqnarray}
Note the absence of terms $_{\left[1\right]}D_{\left[1\right]}D$
that do not contribute to LLA approximation \citep{GS2,Blok2012}
(note however the discussion in \citep{diehl3} on the subject).

The first term in Eq. \ref{eq:GPD} can be calculated in the mean
field approximation \citep{Blok2011}: 
\begin{equation}
\phantom{}_{\left[2\right]}D_{h}^{AB}\left(x_{1},x_{2},Q_{1}^{2},Q_{2}^{2},\vec{\Delta}\right)=G_{h}^{A}\left(x_{1},Q_{1}^{2}\right)G_{h}^{B}\left(x_{2},Q_{2}^{2}\right)\left[F_{2g}\left(\vec{\Delta}\right)\right]^{2}.\label{eq:parton model}
\end{equation}

Here $G(x,Q^{2})$ are the parton distribution functions in the nucleon
(PDFs) and $F_{2g}\left(\vec{\Delta}\right)$ is the so called ``two
gluon form factor'' \citep{Frankfurt2002}. With good accuracy it
will be enough to use the dipole parametrization of the two gluon
form factor. In this parametrization the two gluon form
factor has the form:

\begin{equation}
F_{2g}\left(\vec{\Delta}\right)=\left(1+\frac{\Delta^{2}}{m_{g}^{2}}\right)^{-2},\label{eq:formfactor}
\end{equation}
\color{black} where $m_{g}$ is a parameter that can be extracted from hadron photoproduction
at the HERA and FNAL experiments, and is approximately $m_{g}\approx1.1\ GeV$\citep{Frankfurt2002,Blok2011,Blok2012}. \color{black} We neglect the weak dependence of $m_{g}$ on $Q^{2}$
and $x$ (the dependence on $x$ of individual $m_{g}$ of different
partons cancel out, and the remaining dependence on $Q^{2}$ is negligible
\citep{Frankfurt2002})

The DPS cross section for $2\rightarrow2$ processes in the mean field
approximation is usually written in the form

\begin{subequations}
\begin{equation}
\sigma_{DPS}=\frac{\sigma_{1}\sigma_{2}}{\sigma_{eff}},\label{eq: singlet factorization}
\end{equation}

\begin{equation}
\sigma_{1}=G^{G}\left(x_{1},Q_{1}^{2}\right)G^{G}\left(x_{3},Q_{1}^{2}\right)\hat{\sigma_{1}},
\end{equation}

\begin{equation}
\sigma_{2}=G^{G}\left(x_{2},Q_{2}^{2}\right)G^{G}\left(x_{4},q_{4}^{2}\right)\hat{\sigma_{2}},
\end{equation}
\end{subequations}
 where in the mean field approximation 
\begin{equation}
\frac{1}{\sigma_{eff}}=\int\frac{d^{2}\vec{\Delta}}{\left(2\pi\right)^{2}}\left[F_{gg}\left(\vec{\Delta}\right)\right]^{4}=\frac{m_{g}^{2}}{28\pi}.\label{eq:F integral-4}
\end{equation}

Consider now the calculation of $1\rightarrow2$ part of the cross
section, corresponding to the second and third terms in Eq. \ref{eq: singlet DPS}.
The integral over $\vec{\Delta}$ can be easily taken since the two gluon
form factors decrease much more rapidly with $\vec{\Delta}$ than perturbative
\color{black}  $\phantom{}_{2}GPD$\color{black}, that can be taken at the point $\vec{\Delta}=0$. The $\vec{\Delta}$ integral
is easily taken giving the fact that 
\begin{equation}
\int\frac{d^{2}\vec{\Delta}}{\left(2\pi\right)^{2}}\left[F_{gg}\left(\vec{\Delta}\right)\right]^{2}=\frac{m_{g}^{2}}{12\pi}.\label{eq:F integral-2}
\end{equation}
So the $1\rightarrow2$ part of the cross section is 
\begin{equation}
\sigma_{DPS}^{1\rightarrow2}=\hat{\sigma}_{1}\hat{\sigma}_{2}\frac{m_{g}^{2}}{12\pi}(G(x_{3},Q_{1}^{2})G(x_{4},Q_{2}^{2})_{\left[1\right]}D(x_{1},x_{2},Q_{1}^{2},Q_{2}^{2},0)+G(x_{1},Q_{1}^{2})G(x_{2},Q_{2}^{2})_{\left[1\right]}D(x_{3},x_{4},Q_{1}^{2},Q_{2}^{2},0)\label{eq: DPS 1+2}
\end{equation}
Note that we get ``geometric'' enhancement of $1\rightarrow2$ relative
$2\rightarrow2$ by a factor $2\times\frac{7}{3}$,
where factor 2 comes from two terms in (\ref{eq: DPS 1+2}).

These results can be immediately extended to color correlations. We
define the free indexed $\phantom{}_{\left[2\right]}D_{h;AB}^{a\overline{a}b\overline{b}}$, $\phantom{}_{\left[1\right]}D_{h;AB}^{a\overline{a}b\overline{b}}$
(see figure \ref{fig:General}). These \color{black}  $\phantom{}_{2}GPD$\color{black}'s are normalized as: 
\begin{equation}
\phantom{}_{\left[2\right]}D_{h;AB}^{a\overline{a}b\overline{b}}=\phantom{}_{\left[2\right]}D_{h}^{AB}P_{a\overline{a};b\overline{b}}^{1}\label{eq:D2 completness}
\end{equation}
for singlet case, where $P^{1}$ is a color projector into the singlet
representation.

Note that due to Sudakov suppression \citep{artru1,Mekhfi1988} the
colored $_{\left[2\right]}D$ are small and we shall take into account
only the singlet \color{black}  $\phantom{}_{2}GPD$ \color{black} in the $2\rightarrow2$ part of the DPS process.
For the $_{\left[1\right]}D$ case non-singlet distributions might
have weaker suppression (as explained in the introduction). We shall
show and compute them explicitly in the next section.

\subsection{$1\rightarrow2$ for Color Non-Singlet Channels\label{subsec:-for-Color}}

Using the DDT \citep{Dokshitzer1977} approximation for TMD (transverse
momentum distribution) one can get a formulation for the contribution
to the cross section from $1\rightarrow2$ process, i.e $\phantom{}_{\left[1\right]}D_{h}$,
when all ladders contributing to the evolution equations are the singlet
ones \citep{Blok2014,Blok2012}.

This analysis can be generalized to non-singlet evolution $\phantom{}_{\left[1\right]}D_{h;AB}^{a\overline{a}b\overline{b}}$
as described above. Define the $t$-channel \citep{Diehl2021,Mekhfi1988}
distribution function as shown in figure \ref{fig:complex conjugate}:

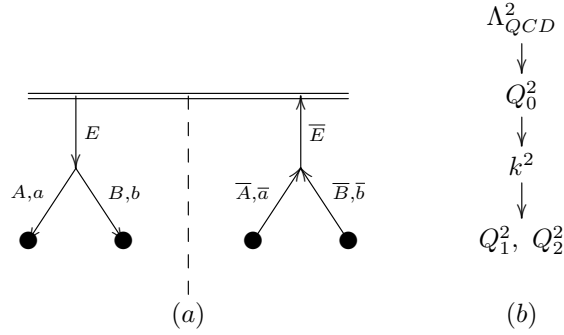
\begin{figure}
\[
\xymatrix{ &  &  &  &  &  &  &  & {\Lambda_{QCD}^{2}}\ar@{->}[d]\\
*=0{}\ar@{=}[rrrrrr] & *=0{}\ar@{->}[d]^{E} &  & *=0{}\ar@{--}[ddd] &  & *=0{}\ar@{<-}[d]^{\overline{E}} & *=0{} & {\ \ \ \ } & {Q_{0}^{2}}\ar@{->}[d]\\
 & *=0{}\ar@{->}[dl]_{A,a}\ar@{->}[dr]^{B,b} &  &  &  & *=0{}\ar@{<-}[dl]_{\overline{A},\overline{a}}\ar@{<-}[dr]^{\overline{B},\overline{b}} &  &  & {k^{2}}\ar@{->}[d]\\
*=0{\newmoon} &  & *=0{\newmoon} & *=0{} & *=0{\newmoon} &  & *=0{\newmoon} &  & {Q_{1}^{2},\ Q_{2}^{2}}\\
 &  &  & {\left(a\right)} &  &  &  &  & {\left(b\right)}
}
\]

\caption{$(a)$ General diagram for the $1\rightarrow2$ process and its complex
conjugate, the line represents either a quark or a gluon and the $\newmoon$
represents the hard process. The indices $E,\overline{E},A,\overline{A},B,\overline{B}$
represent the types of the partons while $a,\overline{a},b,\overline{b}$
represent the color indices of this parton (these indices could be
either quark or gluon indices). $\left(b\right)$ the different characteristic
energy scales of each part of the process. \label{fig:General}}
\end{figure}

\begin{equation}
\phantom{}_{\left[1\right]}^{\alpha}\overline{D}_{h}^{AB}=\frac{1}{K^{\alpha}}P_{a\overline{a};b\overline{b}}^{\alpha}\phantom{}_{\left[1\right]}D_{h;BC}^{a\overline{a}b\overline{b}}.\label{eq:defenition D alpha}
\end{equation}

Here $P^{\alpha}$ is the projector of the color representation $\alpha$,
and we normalized the projectors by $K^{\alpha}=P_{c\overline{c};c\overline{c}}^{\alpha}$
- the dimension of the representation $\alpha$ (see Appendix \ref{sec:Rules-for-Color}
for explicit values of $K^{\alpha}$). The completeness relation of
the color projectors asserts that:

\begin{equation}
\phantom{}_{\left[1\right]}D_{h;AB}^{a\overline{a}b\overline{b}}=\sum_{\alpha}\phantom{}_{\left[1\right]}^{\alpha}\overline{D}_{h}^{AB}P_{a\overline{a};b\overline{b}}^{\alpha},\label{eq:completeness}
\end{equation}

as was the case for $\phantom{}_{\left[1\right]}D_{h;AB}^{a\overline{a}b\overline{b}}$
in (\ref{eq:D2 completness}). We shall take a closer look at the
normalization of $\phantom{}_{\left[1\right]}^{\alpha}\overline{D}_{h}^{BC}$
when computing them, the self consistency of this normalization will
be checked in Section \ref{subsec:Evaluation-of-}. This relation
means we can look at the contribution of each representation alone
and then sum them together to get the total contribution from the
$1\rightarrow2$ process.

The distribution for this process when $\alpha=1$ (i.e, $a,\overline{a}$
and $b,\overline{b}$ are in a singlet state) was obtained in \citep{Blok2012,Blok2014}
as:

\begin{eqnarray}
\phantom{}_{\left[1\right]}^{1}\overline{D}_{h}^{AB}\left(x_{1},x_{2},Q_{1},Q_{2}\right) & = & \underset{E,A^{\prime},B^{\prime}}{\sum}\intop_{Q_{0}^{2}}^{min\left(Q_{1}^{2},Q_{2}^{2}\right)}\frac{dk^{2}}{k^{2}}\frac{\alpha_{s}\left(k^{2}\right)}{2\pi}\int\frac{dy}{y}G_{h}^{E}\left(y;k^{2}\right)\nonumber \\[10pt]
 & \times & \int\frac{dz}{z\left(1-z\right)}\Phi_{E}^{A^{\prime}}\left(z\right)D_{A^{\prime}}^{A}\left(\frac{x_{1}}{zy};Q_{1}^{2},k^{2}\right)D_{B^{\prime}}^{B}\left(\frac{x_{2}}{\left(1-z\right)y};Q_{2}^{2},k^{2}\right).\nonumber \\[10pt]
\label{eq: PH}
\end{eqnarray}

Here (See figure \ref{fig:General} for notations): 
\begin{itemize}
\item The sum over $E,A^{\prime},B^{\prime}$ runs over gluons and fermions \color{black} but only if the splitting $E\rightarrow A^{\prime},B^{\prime}$ is allowed by LO QCD. \color{black}
\item $k^{2}$ is the scale at which the splitting occurs, $Q_{0}$ is some
minimal scale which we take to be $Q_{0}=0.7\ GeV$ and $Q_{1},Q_{2}$
the transverse \color{black} scales \color{black}  and $x_{1},x_{2}$ the Bjorken variables. 
\item $G_{h}^{E}$ are the parton distribution functions (PDF's) of the
hadron $h$, we take their values as given in \citep{Gluck1995,GRV1}.
\item $D$ are the solutions of the singlet DGLAP equation (\ref{eq:Singlet DGLAP}),
and $\Phi_{E}^{B^{\prime}}$ are the singlet splitting functions.
\end{itemize}
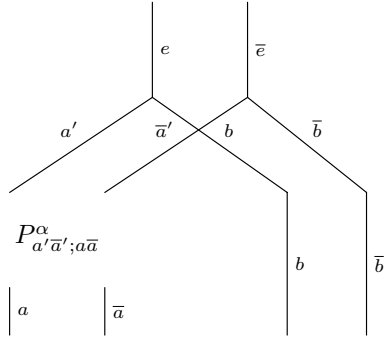
\begin{figure}
\[
\xymatrix{ &  &  & *=0{}\ar@{-}[dd]^{e} &  & *=0{}\ar@{-}[dd]^{\overline{e}}\\
\\
 &  &  & *=0{}\ar@{-}[ddlll]_{a^{\prime}}\ar@{-}[ddrrr]^{b} &  & *=0{}\ar@{-}[ddlll]_{\overline{a}^{\prime}}\ar@{-}[ddrrr]^{\overline{b}}\\
\\
*=0{} &  & *=0{} &  &  &  & *=0{}\ar@{-}[ddd]^{b} &  & *=0{}\ar@{-}[ddd]^{\overline{b}}\\
 & *=0{P_{a^{\prime}\overline{a}^{\prime};a\overline{a}}^{\alpha}}\\
*=0{}\ar@{-}[d]^{a} &  & *=0{}\ar@{-}[d]^{\overline{a}}\\
*=0{} &  & *=0{} &  &  &  & *=0{} &  & *=0{}
}
\]

\caption{General diagram for the $1\rightarrow2$ process and its complex conjugate,
with a projector of the representation $\alpha$ between them. \label{fig:complex conjugate}}
\end{figure}

Now assume we put the color projector of some other non-singlet representation
$\alpha\in\left\{ 8_{a},10,1,8_{s},27,0\right\} $ as explained in
(\ref{eq:defenition D alpha}). This projector means we need to change
the evolution functions $D_{A^{\prime}}^{A}$ to the (representation
dependent) non-singlet evolution $\phantom{}^{\alpha}\overline{D}_{A^{\prime}}^{A}$
that solves the non singlet DGLAP equation (\ref{eq:Modified-DGLAP}).
The color ``flows'' through this diagram so the evolution function
for the partons $B,\overline{B}$ must be in the same representation
$\alpha$ as the one for $A,\overline{A}$ so we also change $D_{B^{\prime}}^{B}\rightarrow^{\alpha}\overline{D}_{B^{\prime}}^{B}$.
This argument follows from the processes we have chosen to minimize
Sudakov suppression: the initial ladder that splits into two nonsinglet
ones is itself a singlet. Unlike $B$ and $A$ the evolution of $E$
is still governed by singlet ladders, as $E$ and $\overline{E}$
in figure \ref{fig:General} are directly connected and not through
a projector, so $G_{h}^{E}$ stays the same as in the singlet case.

The last change we need to account for is the splitting vertex $\Phi_{E}^{A^{\prime}}\left(z\right)=C_{E}^{A^{\prime}}\cdot V_{E}^{A^{\prime}}\left(z\right)$.
The momentum part $V$ stays the same as it's not color dependent.
The color factor however needs to be changed as it's no longer equivalent
to a singlet splitting. Looking at the process from the point of view
of particle $A$ the splitting is exactly like a ladder $\overline{\Phi}_{A^{\prime}}^{B^{\prime}}$
but with one difference. In a regular ladder, we average over the
initial particle color state and sum over both the final and the ladder
particle color state. But in this diagram, we should still sum over
the color state of the particles $A,B$ and average over that of $E$.
To conclude the color factor must compensate for that, and we get
a total color factor of:

\begin{equation}
\frac{n_{A^{\prime}}}{n_{E}}\phantom{}^{\alpha}\overline{C}_{A^{\prime}}^{B^{\prime}}.
\end{equation}

Here we define $n_{F}=N,\ n_{G}=N^{2}-1$ is the number of color states
for the parton. $\phantom{}^{\alpha}\overline{C}_{A^{\prime}}^{B^{\prime}}$
are defined in table \ref{tab:Color-factors-for}. The choice to look
from the point of particle $A$ and not from $B$ is of course arbitrary,
but our choice of color factors makes sure the result is the same
either way:

\begin{equation}
\frac{n_{A^{\prime}}}{n_{E}}\phantom{}^{\alpha}\overline{C}_{A^{\prime}}^{B^{\prime}}=\frac{n_{B^{\prime}}}{n_{E}}\phantom{}^{\alpha}\overline{C}_{B^{\prime}}^{A^{\prime}}.
\end{equation}

To conclude we write the non-singlet $1\rightarrow2$ part of two
particle \color{black}  $\phantom{}_{2}GPD$ \color{black} as:

\begin{eqnarray}
\phantom{}_{\left[1\right]}^{\alpha}\overline{D}_{h}^{AB}\left(x_{1},x_{2},Q_{1},Q_{2}\right) & = & \underset{E,A^{\prime},B^{\prime}}{\sum}\intop_{Q_{0}^{2}}^{min\left(Q_{1}^{2},Q_{2}^{2}\right)}\frac{dk^{2}}{k^{2}}\frac{\alpha_{s}\left(k^{2}\right)}{2\pi}\int\frac{dy}{y}G_{h}^{E}\left(y;k^{2}\right)\nonumber \\[10pt]
 & \times & \int\frac{dz}{z\left(1-z\right)}V_{E}^{A^{\prime}}\left(z\right)\frac{n_{A^{\prime}}}{n_{E}}\phantom{}^{\alpha}\overline{C}_{A^{\prime}}^{B^{\prime}}\nonumber \\
 & \times & \phantom{}^{\alpha}\overline{D}_{A^{\prime}}^{A}\left(\frac{x_{1}}{zy};Q_{1}^{2},k^{2}\right)\phantom{}^{\alpha}\overline{D}_{B^{\prime}}^{B}\left(\frac{x_{2}}{\left(1-z\right)y};Q_{2}^{2},k^{2}\right).\nonumber \\[10pt]
\label{eq: PH-1}
\end{eqnarray}

Using (\ref{eq:Sudakov split}) and writing $z_{1}=\frac{x_{1}}{zy}$,
$z_{2}=\frac{x_{2}}{\left(1-z\right)y}$ this equation can be rewritten
as:

\begin{eqnarray}
\phantom{}_{\left[1\right]}^{\alpha}\overline{D}_{h}^{AB}\left(x_{1},x_{2},Q_{1},Q_{2}\right) & = & \underset{E,A^{\prime},B^{\prime}}{\sum}\intop_{Q_{0}^{2}}^{min\left(Q_{1}^{2},Q_{2}^{2}\right)}\frac{dk^{2}}{k^{2}}\frac{\alpha_{s}\left(k^{2}\right)}{2\pi}\overline{S}_{A^{\prime}}\left(k^{2},Q_{1}^{2}\right)\overline{S}_{B^{\prime}}\left(k^{2},Q_{2}^{2}\right)\nonumber \\[10pt]
 & \times & \intop_{x_{1}}^{1}dz_{1}\intop_{x_{2}}^{1}dz_{2}\frac{x_{1}x_{2}}{z_{1}^{2}z_{2}^{2}\left(\frac{x_{1}}{z_{1}}+\frac{x_{2}}{z_{2}}\right)}G_{h}^{E}\left(\frac{x_{1}}{z_{1}}+\frac{x_{2}}{z_{2}};k^{2}\right)\nonumber \\[10pt]
 & \times & V_{E}^{A^{\prime}}\left(\frac{x_{1}}{z_{1}\left(\frac{x_{1}}{z_{1}}+\frac{x_{2}}{z_{2}}\right)}\right)\frac{n_{A^{\prime}}}{n_{E}}\overline{C}_{A^{\prime}}^{B^{\prime}}\widetilde{D}_{A^{\prime}}^{A}\left(z_{1};Q_{1}^{2},k^{2}\right)\widetilde{D}_{B^{\prime}}^{B}\left(z_{2};Q_{2}^{2},k^{2}\right).\nonumber \\
\label{eq:D 1-2}
\end{eqnarray}

Here it's understood that $G_{h}^{E}\left(\frac{x_{1}}{z_{1}}+\frac{x_{2}}{z_{2}};k^{2}\right)=0$
for $\frac{x_{1}}{z_{1}}+\frac{x_{2}}{z_{2}}>1$. We have suppressed
the representation index on the r.h.s although $\overline{S},\overline{C}$
and $\widetilde{D}$ are representation dependent. There is one problem
with this equation, as can be seen from (\ref{eq:x-1-limit}) and
table \ref{tab:Color-factors-for}: for certain representations, $\widetilde{D}\propto\frac{1}{\left(1-z\right)^{\lambda}}$
with $\lambda>1$. Therefore the integrals over $z_{1\backslash2}$
might diverge at the limit $z_{1\backslash2}\rightarrow1$. We delay
the solution to this problem for section \ref{subsec:Regularization}
and for now assume these integrals are finite.

\subsubsection*{Self Consistency of the $\protect\phantom{}_{\left[1\right]}^{\alpha}\overline{D}_{h}^{AB}$
Normalization \label{subsec:Evaluation-of-}}

Equation (\ref{eq:D 1-2}) cannot be solved analytically, we delay
its numerical solution to section \ref{sec:Numerics}. We can solve
it, however, in the approximation of no evolution. In this approximation,
we neglect any $k^{2}$ dependence of physical quantities. We call
this approximation ``$0^{th}$ order''. This approximation can also
help us check the self consistency of the normalizations defined at
(\ref{eq:defenition D alpha}). At this approximation

\begin{equation}
\phantom{}^{\alpha}\overline{D}_{A}^{B}\left(x\right)\rightarrow\delta_{A}^{B}\delta\left(1-x\right),
\end{equation}

Note the r.h.s is independent of $\alpha$ since the representation
affects only the evolution equation. we also set $G_{h}^{E}\left(y;k^{2}\right)\rightarrow G_{h}^{E}\left(y\right)$
to be independent of $k^{2}$. We can now write (\ref{eq:D 1-2})
as (ignoring the $k^{2}$ integral):

\begin{equation}
\phantom{}_{\left[1\right]}^{\alpha}\overline{D}_{h}^{AB}=G_{h}^{E}\left(x_{1}+x_{2}\right)\frac{V_{E}^{A}\left(\frac{x_{1}}{x_{1}+x_{2}}\right)}{x_{1}+x_{2}}\frac{n_{A}}{n_{E}}\phantom{}^{\alpha}\overline{C}_{A}^{B}.\label{eq:D 0th}
\end{equation}

Now there is no summation over $B^{\prime}$ , $A^{\prime}$ and $E$
is completely determined by charge conservation.

On the other hand, we can compute $\phantom{}_{\left[1\right]}D_{h;BC}^{a\overline{a}b\overline{b}}$,
which is defined by figure \ref{fig:General} to the same approximation.
We first divide it to a momentum part which depends on the Bjorken
variables $x_{1}$ and $x_{2}$ and a color part which depends on
the color indices $a,\overline{a},b,\overline{b}$:

\begin{equation}
\phantom{}_{\left[1\right]}D_{h;AB}^{a\overline{a}b\overline{b}}\left(x_{1},x_{2}\right)=U_{h;AB}\left(x_{1},x_{2}\right)\cdot T^{a\overline{a}b\overline{b}}.
\end{equation}

The $U$ factor describes a parton $E$ originating from a hadron
with Bjorken variable $x_{1}+x_{2}$ which then splits to $A$ and
$B$ with Bjorken variables $x_{1}$ and $x_{2}$ respectively. It
can therefore be written as:

\begin{equation}
U_{h;BC}\left(x_{1},x_{2}\right)=G_{h}^{E}\left(x_{1}+x_{2}\right)\frac{V_{E}^{A}\left(\frac{x_{1}}{x_{1}+x_{2}}\right)}{x_{1}+x_{2}}.
\end{equation}

Here $G_{h}^{E}\left(x_{1}+x_{2}\right)$ is the distribution of $E$
in the hadron. $V_{E}^{A}\left(\frac{x_{1}}{x_{1}+x_{2}}\right)$
is the splitting kernel from $E$ to $A$ ($B$) with fraction of
longitudinal momentum $\frac{x_{1}}{x_{1}+x_{2}}$ ($\frac{x_{2}}{x_{1}+x_{2}}$)
compared to $E$. The division by $\frac{1}{x_{1}+x_{2}}$ comes from
the fact that $E$ has actually a Bjorken variable of $x_{1}+x_{2}$
and not $1$ as assumed when computing $V_{A}^{B}$ in table \ref{tab:Vz}.
The color factor $T^{a\overline{a}b\overline{b}}$ depend on the type
of particles (gluons or fermions). For example assume $A=B=G$ ($A$
and $B$ are gluons) which then means $E=G$ also (by charge conservation)
so the diagram in figure \ref{fig:gluons diagram} has the color factor
of:

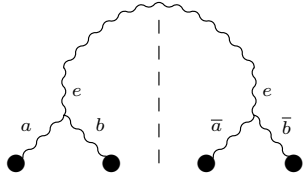
\begin{figure}
\[
\xymatrix{ &  &  & *=0{}\ar@{--}[ddd]\\
 & *=0{}\ar@{~}[d]^{e}\ar@/^{2pc}/@{~}[rrrr] &  &  &  & *=0{}\ar@{~}[d]^{e}\\
 & *=0{}\ar@{~}[dl]_{a}\ar@{~}[dr]^{b} &  &  &  & *=0{}\ar@{~}[dl]_{\overline{a}}\ar@{~}[dr]^{\overline{b}}\\
*=0{\newmoon} &  & *=0{\newmoon} & *=0{} & *=0{\newmoon} &  & *=0{\newmoon}
}
\]

\caption{Diagram for the $1\rightarrow2$ process and its complex conjugate
where all particles are gluons, $\newmoon$ represents the hard process.
The upper gluon connecting $e$ is only symbolic of the color connection
and does not represent an actual gluon. \label{fig:gluons diagram}}
\end{figure}

\begin{equation}
T^{a\overline{a}b\overline{b}}=f^{eab}f^{e\overline{a}\overline{b}}=\sum_{\alpha}c_{\alpha}P_{a\overline{a}b\overline{b}}^{\alpha}.
\end{equation}

Here $c_{\alpha}$ are given by the contraction identity (Appendix
\ref{sec:Rules-for-Color}) $c_{\alpha}=\phantom{}^{\alpha}\overline{C}_{G}^{G}=\left\{ \frac{N}{2},0,N,\frac{N}{2},-1,1\right\} $
so we can write the whole distribution as:

\begin{equation}
\phantom{}_{\left[1\right]}D_{h;GG}^{a\overline{a}b\overline{b}}\left(x_{1},x_{2}\right)=G_{h}^{G}\left(x_{1}+x_{2}\right)\frac{V_{G}^{G}\left(\frac{x_{1}}{x_{1}+x_{2}}\right)}{x_{1}+x_{2}}\sum_{\alpha}\phantom{}^{\alpha}\overline{C}_{G}^{G}P_{a\overline{a}b\overline{b}}^{\alpha}=\sum_{\alpha}\phantom{}_{\left[1\right]}^{\alpha}\overline{D}_{h}^{GG}P_{a\overline{a}b\overline{b}}^{\alpha}.
\end{equation}

Where in the last step we used (\ref{eq:D 0th}), those proving (\ref{eq:completeness})
for the 0th order case and seeing the normalizations are all correct.
Equation (\ref{eq:D 0th}) has another interesting property: we see
that for certain representations $\phantom{}_{\left[1\right]}^{\alpha}\overline{D}_{h}^{AB}$
should be negative. This is not really a surprise as non singlet or
spin dependent distributions can, in general, be negative \citep{Diehl2021}
and only the observable cross section, after considering all color
channels, should be positive.

Let us recall that we need $1\rightarrow2$ \color{black}  $\phantom{}_{2}GPD$ \color{black} only at $\vec{\Delta}=0$,
since in our total cross section this \color{black}  $\phantom{}_{2}GPD$ \color{black} is convoluted with nonperturbative
two gluon form factors from the second nucleon, that decrease in \ensuremath{\Delta}
much more rapidly than the perturbative $1\rightarrow2$ piece \citep{Blok2014,Blok2012}.

\subsection{Regularization of $z_{i}\rightarrow1$ Singularity\label{subsec:Regularization}}

%TO COPY
\color{black}

As we have seen before the functions $\bar D^A_B(z,Q^2,k_0^2)$ are the building blocks for 
the $1\rightarrow 2$ part of the $\phantom{}_{2}GPD$, corresponding to upper parts of the Feynman diagrams of figs. \ref{fig:1+2-2+2}-\ref{fig: interference-1-2}. Note that z dependence is fully contained in $\tilde  D^A_B(z,Q^2,k_0^2)$ since $\phantom{}^{\alpha}\overline{S}_{A}$ do not depend in z. Although we can 
calculate $\tilde D^A_B(z,Q^2,k_0^2)$ numerically their asymptotics at $z\rightarrow 1$ was found analyutically,
see eq. (\ref{eq:x-1-limit}). It has a form $1/(1-z)^{1-g}$, where $g$ is proportional to running coupling constant,
and is $g>0$ for singlet case, so the integral (\ref{eq:D 1-2}) converges in the $z_1\rightarrow 1$ and $z_2\rightarrow 1$ limit.
However for the nonsinglet case $g\sim \xi \bar C^A_A$ can be negative, since the colour factoers $\bar C^A_A$
are negative. We assume that after integration over z ($z_1,z_2$ ) the resulting function is an analytic function of 
$g$ (i.e. of running coupling constant). Then the theory of generalised functions \cite{Kanwal1983,Gelfand1966}] guarantees the unique 
analytical continuation into $g<0$ domain. Note that it is possible to calculate $_{2}GPD$ also by solving numerically the evolution
equation. In this way one can indeed avoid dealing with singularities in nonsinglet fundamental solutions.
Proof of equivalence of both approaches is given in  Appendix \ref{sec:formalisms}.
\par The analytical continuation method was already used in connection with diverging integrals in z for observables related to Double Drell Yan
in \cite{Manohar2012}, see also \cite{Becher:2014oda} for detailed discussion.
\par Let us address the issue of divergent untegrals in more detail, since the analytical continuation in $g$
was explicitly considered in \cite{Kanwal1983,Manohar2012,Becher:2014oda} only for one-dimensional integrals.

%TO END
\color{black}

In the 1-dimensional case, such integrals can be made finite in the
following way. Consider a function $f\left(z\right)$ that is bounded
in the segment $\left[0,1\right]$ and smooth in some neighborhood
of $0$. We then define the integral

\begin{equation}
\int_{0}^{1}dx\frac{f\left(x\right)}{x^{\lambda}}:=\int_{0}^{1}\frac{f\left(x\right)-f\left(0\right)}{x^{\lambda}}dx+\frac{f\left(0\right)}{1-\lambda}
\end{equation}

for every $\lambda<2$ except $\lambda=1,0$. For $\lambda<1$ this
relation clearly holds as both the l.h.s and the r.h.s integrals converge.
For $1<\lambda<2$ the l.h.s formally diverges but the r.h.s converges
and is the analytical continuation of the l.h.s as a function of the
complex variable $\lambda$. We'll now generalize this method to 2
dimensional functions. Let $F\left(z_{1},z_{2}\right)$ be a function
that is well defined and bounded in the region $\left(z_{1},z_{2}\right)\in\left[x_{1},1\right]\times\left[x_{2},1\right]$
for $0<x_{1},x_{2}<1$. And also has a finite derivative at the lines
$z_{1}=1$ and $z_{2}=1$. Note that we do not require $F$ to be
continuous except at these lines. Then look at the integral

\begin{equation}
I=\intop_{x_{1}}^{1}dz_{1}\intop_{x_{2}}^{1}dz_{2}\frac{F\left(z_{1},z_{2}\right)}{\left(1-z_{1}\right)^{1-g_{1}}\left(1-z_{2}\right)^{1-g_{2}}}\label{eq:regularization}
\end{equation}

for $g_{1},g_{2}>0$ this integral is well defined and can be written
as a sum of four integrals:

\begin{subequations}
\begin{equation}
I:=I_{A}+I_{B}+I_{C}+I_{D},
\end{equation}

\begin{align}
I_{A}= & \intop_{x_{1}}^{1}dz_{1}\intop_{x_{2}}^{1}dz_{2}\frac{F\left(z_{1},z_{2}\right)-F\left(1,z_{2}\right)-F\left(z_{1},1\right)+F\left(1,1\right)}{\left(1-z_{1}\right)^{1-g_{1}}\left(1-z_{2}\right)^{1-g_{2}}},\\
I_{B}= & \intop_{x_{1}}^{1}dz_{1}\intop_{x_{2}}^{1}dz_{2}\frac{F\left(z_{1},1\right)-F\left(1,1\right)}{\left(1-z_{1}\right)^{1-g_{1}}\left(1-z_{2}\right)^{1-g_{2}}}=\frac{\left(1-x_{2}\right)^{g_{2}}}{g_{2}}\intop_{x_{1}}^{1}dz_{1}\frac{F\left(z_{1},1\right)-F\left(1,1\right)}{\left(1-z_{1}\right)^{1-g_{1}}},\\
I_{C}= & \intop_{x_{1}}^{1}dz_{1}\intop_{x_{2}}^{1}dz_{2}\frac{F\left(1,z_{2}\right)-F\left(1,1\right)}{\left(1-z_{1}\right)^{1-g_{1}}\left(1-z_{2}\right)^{1-g_{2}}}=\frac{\left(1-x_{1}\right)^{g_{1}}}{g_{1}}\intop_{x_{2}}^{1}dz_{2}\frac{F\left(1,z_{2}\right)-F\left(1,1\right)}{\left(1-z_{2}\right)^{1-g_{2}}},\\
I_{D}= & \intop_{x_{1}}^{1}dz_{1}\intop_{x_{2}}^{1}dz_{2}\frac{F\left(1,1\right)}{\left(1-z_{1}\right)^{1-g_{1}}\left(1-z_{2}\right)^{1-g_{2}}}=\frac{\left(1-x_{2}\right)^{g_{2}}}{g_{2}}\frac{\left(1-x_{1}\right)^{g_{1}}}{g_{1}}F\left(1,1\right).
\end{align}
\end{subequations}

We now analytically continue $I_{A},\ I_{B},\ I_{C}$ and $I_{D}$
as functions of $g_{1},g_{2}$ to the region $-1<g_{1},g_{2}<0$.
The integrals on the r.h.s still converge because of the condition
that $\partial_{z_{i}}F|_{z_{i}=1}$ exist. Therefore $I$ is well
defined and finite in this region too, even when the r.h.s of (\ref{eq:regularization})
formally diverges. In our case, we can define:

\begin{subequations}
\begin{eqnarray}
F\left(z_{1},z_{2}\right) & = & \frac{x_{1}x_{2}}{z_{1}^{2}z_{2}^{2}\left(\frac{x_{1}}{z_{1}}+\frac{x_{2}}{z_{2}}\right)}G_{h}^{E}\left(\frac{x_{1}}{z_{1}}+\frac{x_{2}}{z_{2}};k^{2}\right)V_{E}^{A^{\prime}}\left(\frac{x_{1}}{z_{1}\left(\frac{x_{1}}{z_{1}}+\frac{x_{2}}{z_{2}}\right)}\right)\nonumber \\[10pt]
 & \times & \frac{n_{A^{\prime}}}{n_{E}}\phantom{}^{\alpha}\overline{C}_{A^{\prime}}^{B^{\prime}}\frac{\widetilde{D}_{A^{\prime}}^{A}\left(z_{1};Q_{1}^{2},k^{2}\right)\widetilde{D}_{B^{\prime}}^{B}\left(z_{2};Q_{2}^{2},k^{2}\right)}{\left(1-z_{1}\right)^{g_{1}-1}\left(1-z_{2}\right)^{g_{2}-1}},\nonumber \\[10pt]
\label{eq:F(z1,z2)}
\end{eqnarray}

\begin{equation}
g_{1}=4\xi\left(Q_{1},k\right)\overline{C}_{A^{\prime}}^{A},
\end{equation}

\begin{equation}
g_{2}=4\xi\left(Q_{2},k\right)\overline{C}_{B^{\prime}}^{B}.
\end{equation}
\end{subequations}

These quantities satisfy the rules above due to (\ref{eq:x-1-limit})
as long as

\begin{subequations}
\begin{equation}
4\xi\overline{C}_{G}^{G},\ 4\xi\overline{C}_{F}^{F}<-1,
\end{equation}

\begin{equation}
4\xi\overline{C}_{F}^{G}-1<4\xi\overline{C}_{F}^{F},
\end{equation}

\begin{equation}
4\xi\overline{C}_{G}^{F}-1<4\xi\overline{C}_{G}^{G}.
\end{equation}
\end{subequations}

For the values given in table \ref{tab:Color-factors-for}, these
conditions are fulfilled for $\xi<0.2$ (in our choice of $Q_{0}$
this condition is equivalent to $k\approx200\ GeV$). This procedure
gives a regularization for the integrals in (\ref{eq:D 1-2}), which
we'll implement in the numerical calculation. For higher values of
$k$ one can use a second order version of this regularization (for
which the $1d$ case is given in \citep{Gelfand1966,Kanwal1983})
but we'll not need that in this paper.

\section{Numerics\label{sec:Numerics}}

In this section, we'll use the results of the previous chapters
to give numeric evaluation of the non-singlet channels distributions
relative to the singlet channel distributions. We'll use (\ref{eq:D 1-2})
to compute $_{\left[1\right]}^{\alpha}\overline{D}_{h}^{AB}$. However, it'll be more instructive to actually investigate
\begin{equation}
R^{\alpha}\coloneqq\frac{7}{3}\frac{\phantom{}_{\left[1\right]}^{\alpha}\overline{D}^{GG}}{G^{G}G^{G}},\label{eq:Ralpha}
\end{equation}
which is analogous to $R$ used in \cite{Blok2014} and directly enters the cross section.

In order to make the computation
we need first to express all the components of (\ref{eq:D 1-2}) in
a way that can be numerically computable. \color{black} $\phantom{}^{\alpha}\overline{S}_{A}$ can be computed numerically using (\ref{eq:Sudakov}) and the regularization (\ref{eq:reg}).
 \color{black}
$\widetilde{D}$ that were defined by (\ref{eq:Sudakov split}) have
no close expression to use but we explained in section \ref{subsec:Color-Non-Singlet-Solutions}
how they can be numerically evaluated. For $G^{A}$ the single parton
distribution we use the expressions given in \citep{Gluck1995} and
we remember to take the $z_{1}$ and $z_{2}$ integrals in (\ref{eq:D 1-2})
according to the prescription given in section \ref{subsec:Regularization}.

Eq. (\ref{eq:D 1-2}) has an interesting property: since the lower
momentum in the suppression factor $\overline{S}$ is $k^{2}$ and
not $Q_{0}^{2}$, and $k^{2}$ ranges all the way up to $Q^{2}=min\left(Q_{1}^{2},Q_{2}^{2}\right)$,
the phase space gets larger for larger $Q^{2}$. So we expect $\left|_{\left[1\right]}^{\alpha}\overline{D}_{h}^{CB}\right|$
to moderately grow with\textbf{ $Q$,} this statement is in direct
contrast to the regular color non-singlet $2\rightarrow2$ process
(as explained in section \ref{subsec:Generalized-Parton-Distribution})
which should strongly decrease with the increasing $Q$ \citep{Mekhfi1988,artru1}.
Moreover, when $Q_{1}\approx Q_{2}\approx Q$, there is always a region
at $k\sim Q$ for which the suppression factor $\overline{S}\left(k^{2},Q_{1}^{2}\right)\overline{S}\left(k^{2},Q_{2}^{2}\right)$
is not significant, even when $Q$ is very large.
\color{black}

We consider the kinematics where the scattering angle  in the c.m. 
in each hard process is $\pi/2$. We work with leading logarithmic accuracy,
so we take $p_T\sim Q$, where $p_T$  is the transverse momenta of  the %colliding?
final states hard 
partons. Then we have in our kinematics
\begin{equation}
x=\sqrt{\frac{4p_T^2}{s}}\sim\sqrt{\frac{4Q^2}{s}},
\end{equation}
see \cite{Ellis1996} for the details of kinematics in 2 to 2 hard collisions.
\color{black}Where $s$ is the center of mass energy of the hadron collision. We'll
check the contribution of non singlet color channels at several different
kinematics: 
\begin{itemize}
\item At LHC where $s=1.96\times10^{8}\ \left[GeV^{2}\right]$ 
\item At the Tevatron where $s=4\cdot10^{6}\ \left[GeV^{2}\right]$ 
\item And at even lower hadron energy where $s=1\cdot10^{5}\ \left[GeV^{2}\right]$.
We look at this energy scale because we expect the contributions to
be relatively high in this region, but there might not be enough DPS
processes to measure it in experiments. 
\end{itemize}
We expect the non-singlet contributions to be largest especially when
the two different hard processes are at similar scales, i.e $Q_{1}^{2}=Q_{2}^{2}:=Q^{2}$
and $x_{1}=x_{2}=x_{3}=x_{4}:=x$. At these scales, the values of
the $R^{\alpha}$ (which are $_{\left[1\right]}^{\alpha}\overline{D}_{h}^{GG}$
normalized by $\frac{7}{3}\cdot\frac{1}{\left[G_{p}^{G}\right]^{2}}$
as explained in (\ref{eq:Ralpha})) are shown in figure \ref{fig: Dalpha graph}
as a function of $Q$. 
\begin{center}
\begin{figure}
\begin{centering}
\includegraphics[scale=0.75]{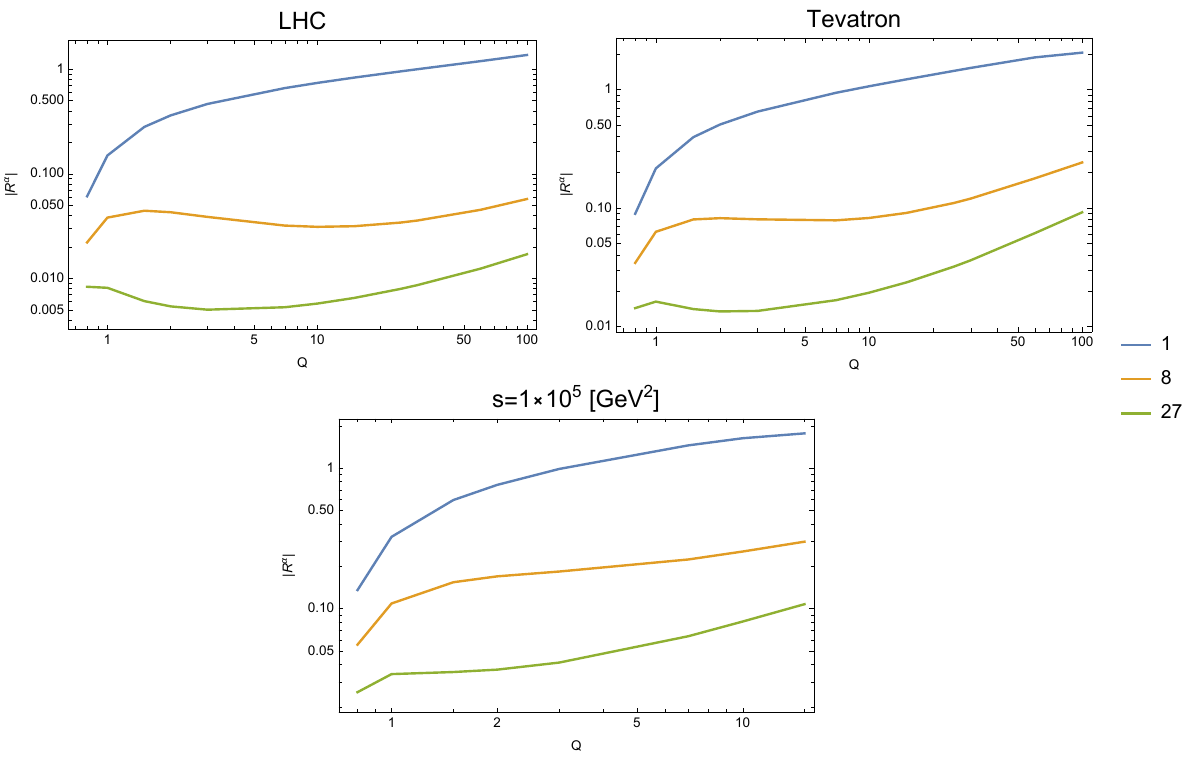}
\par\end{centering}
\caption{Absolute values of $R^{\alpha}$ ($R^{27}$
is negative, the others are strictly positive) at LHC, Tevatron and
$s=1\cdot10^{5}\ \left[GeV^{2}\right]$ kinematics for $Q_{1}^{2}=Q_{2}^{2}=Q^{2}$,
$x_{1}=x_{2}=x$ and $N=3$. The values of $R^{8_{a}}$ and $R^{8_{s}}$
are very similar and therefore only $R^{8_{a}}$ is shown on the graph.
The plot is logarithmic in both axes. \label{fig: Dalpha graph}}
\end{figure}
\par\end{center}

We see that the singlet distribution is much larger than the octet
and $27$ representation. But we also see that the non-singlet distribution
values stay the same and even slowly increase at high energy (as predicted).
This behavior is unlike what is expected from the non-singlet DPDs
that should be highly suppressed at high energies \citep{Mekhfi1988,artru1}.
The source of the seen minimum value of the non-singlet functions
is due to the normalization, $G_{p}^{G}\left(x,Q^{2}\right)$ as a
function of $Q$ (\color{black} where we take $x=\sqrt{\frac{4\cdot Q^{2}}{s}}$ \color{black}).

It's also interesting to look at the $k$ behavior of the integrand
of (\ref{eq:D 1-2}) to explicitly see which splitting scales contribute
the most to $_{\left[1\right]}^{\alpha}\overline{D}_{h}^{GG}$. The
value of this integrand (after performing the $z_{1}$ and $z_{2}$
integrals) is shown on the left of figure \ref{fig: DalphaK graph} for the LHC
kinematics and $Q=50\ [GeV]$ (similar behavior is also seen for different
kinematics). It can be seen from this figure that
the singlet distribution gets most of the contribution from the low
$k$ region. In contrast, the non singlet distributions are suppressed
at low $k$ and get notable contributions only starting at $k>5\ [GeV]$. The $k$ for which the maximal value of the integrand is obtained is shown in figure \ref{fig: DalphaK graph} on the right for different values of $Q$ at the LHC kinematics. As is expected the dominant splitting scale increases with $Q$.
\begin{center}
\begin{figure}
\begin{centering}
\includegraphics[scale=0.75]{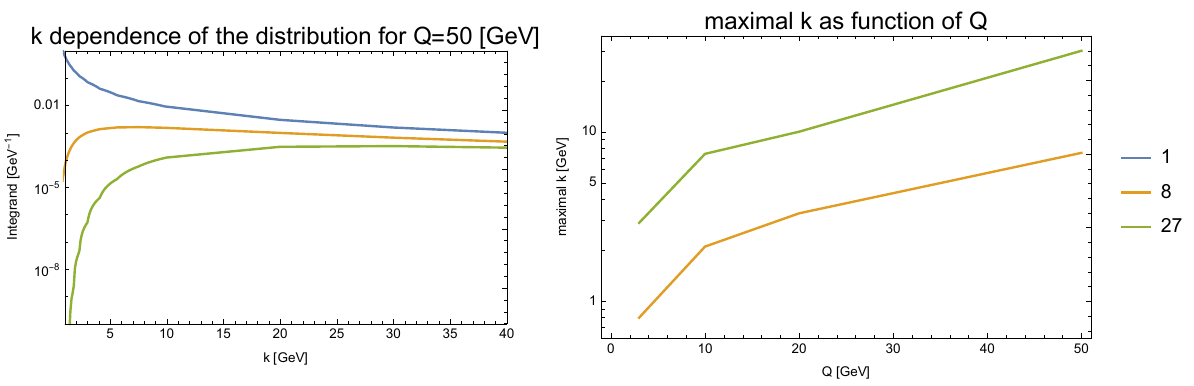}
\par\end{centering}
\caption{On the left: the integrand of (\ref{eq:D 1-2}), after performing the $z_{1}$
and $z_{2}$ integrals, in the LHC kinematics and $Q=50\ [GeV]$ for
different representations and values of $k$. On the right: The value of $k$ for which the maximal value of this integrand is obtained as a function of $Q$. \label{fig: DalphaK graph}}
\end{figure}
\par\end{center}

\section{Conclusion\label{sec:Conclusion}}
%TO COPY
\color{black}

In this paper, we studied the the color correlations in the nucleon. We have considered
for simplicity the case of symmetric kinematics, when $x_{1}=x_{2}=x_{3}=x_{4}=x$
and $Q_{1}=Q_{2}=Q$.
\color{black}

We have seen that it is possible to avoid the conventional Sudakov
suppression of color correlations, by considering specific $1\rightarrow2$
processes depicted in figure \ref{fig:color ladders}. The reason is the increase of  the split scale
with the increase of the hard scale, in difference to the singlet case.

 Although
these contributions  do not decrease with the increase of the transverse
scale, they are still small due to relative smallness of the color factors relative to the singlet case. 

We saw that in the Tevatron and LHC
kinematics they amount to about 1-10\% of the singlet $1\rightarrow2$ distribution. 
%which by itself only account for about 50\% of the total cross section.

The color correlation contribution increases if we go to smaller energies,
however, we must remember that for smaller energies there might be
not enough statistics to observe DPS.

Nevertheless, the fact that perturbative color correlations do not
decrease with transverse scale makes it potentially possible to observe
them and thus have further insight into the precise QCD description
of the nucleon wave functions.
\par One possibility to observe the color correlations contributions due to 
$1\rightarrow2$ processes discussed here is the so called $1v1$ processes,
with two $1\rightarrow 2$ processes from the side of each colliding nucleons.
These processes were considered recently in  \cite{Diehl:2017kgu}, although 
further work may be needed to distinguish this particular contribution in the process 
cross section.

\begin{acknowledgments}
We thank Yu Dokshitzer and M. Strikman for useful discussions and
M. Strikman for reading the article. This research was supported by
ISF grant 2025311 and BSF grant 2020115. 
\end{acknowledgments}

\appendix
%dummy comment inserted by tex2lyx to ensure that this paragraph is not empty

\section{Color Kernels\label{sec:Color-Kernels}}

We now turn to compute explicitly the factors $\phantom{}^{\alpha}\overline{C}_{A}^{B}$
for different representations. These factors were derived partially
in \citep{Manohar2012} for polarized particles and completely in
\citep{buffing2021}. In this section we rederive them in our formalism
for completeness. Let it be remind that for the singlet representation
(as in the original DGLAP case), these factors are shown in table
\ref{tab:singlet color}.

\begin{table}[H]
\begin{centering}
\begin{tabular}{c|c}
R  & 1\tabularnewline
\hline 
$C_{F}^{F}$  & $\frac{N^{2}-1}{2N}$\tabularnewline
\hline 
$C_{F}^{G}$  & $\frac{N^{2}-1}{2N}$\tabularnewline
\hline 
$C_{G}^{F}$  & $\frac{1}{2}$\tabularnewline
\hline 
$C_{G}^{G}$  & $N$\tabularnewline
\end{tabular}
\par\end{centering}
\caption{Color factors for the singlet representation \label{tab:singlet color}}
\end{table}

We start by considering the simplest case of $R=27$, $q\overline{q}$
cannot be in this representation, and therefore $\phantom{}^{27}\overline{C}_{F}^{F}=\phantom{}^{27}\overline{C}_{G}^{F}=\phantom{}^{27}\overline{C}_{F}^{G}=0$.
In order to find the only non-zero kernel $\phantom{}^{27}\overline{C}_{G}^{G}$
attach the $27$ projector to a gluon ladder, as shown in fig. \ref{fig:C27GG}.
The color factor for such a diagram is well known \citep{Dokshitzer2022,Dokshitzer2005}
and equal to $\phantom{}^{27}\overline{C}_{G}^{G}=-1$ . This result
is a special case of the ``interaction force identity'', as will
be explained in Appendix \ref{sec:Rules-for-Color}.

\begin{figure}
\[
\xymatrix{\xyR{1pc}\xyC{1pc} & {P^{27}}\\
*=0{}\ar@{~}[d]\ar@{~}[ur] &  & *=0{}\ar@{~}[d]\ar@{~}[ul]\\
*=0{}\ar@{~}[rr]\ar@{~}[d] &  & *=0{}\ar@{~}[d] & {=} & {\overline{C}_{G}^{G}\left(27\right)\times P_{27}}\\
*=0{} &  & *=0{}
}
\]

\caption{The ladder diagram that determines $\overline{C}_{G}^{G}\left(27\right)$,
$P^{27}$ is the 2 gluon projector to the irreducible representation
$27$.\label{fig:C27GG}}
\end{figure}

By the same line of argument $\phantom{}^{10}\overline{C}_{F}^{F}=\phantom{}^{10}\overline{C}_{G}^{F}=\phantom{}^{10}\overline{C}_{F}^{G}=\phantom{}^{10}\overline{C}_{G}^{G}=0$
where $10$ is the sum of the irreducible decouple representations
\citep{Dokshitzer2005}, we, therefore, conclude that the representation
$10$ does not evolve in ladders and should not contribute to physical
processes. We now turn to consider the octet representation of quarks
and the symmetric octet ($8_{s}$) and anti-symmetric octet ($8_{a}$)
of gluons. Considering these representations raises two problems that
were absent when we considered the $27$ and $10$ representations: 
\begin{itemize}
\item There is now a mixing of quark and gluons, this poses a certain ambiguity
to the $\overline{C}_{G}^{F},\ \overline{C}_{F}^{G}$ kernels. 
\item The second problem is that the single quark octet representation mixes
with the two gluon octet representations. This effect potentially
could make $\overline{C}_{F}^{G}$ a matrix connecting two representations
rather than a simple factor, which would ruin the analysis made in
section \ref{subsec:Color-Non-Singlet-Solutions}. 
\end{itemize}
We'll delay the solution of the first problem to the end of this section,
and until then keep the freedom of normalization of these kernels.
As for the second problem, luckily it can be solved easily. This solution
is done by considering, instead of quark or anti-quark ladders separately,
the sum of quark and anti quark ladders, as shown in fig. \ref{fig:DGLAP-ladders}.
This solution, although very elegant, works only as long as there
is symmetry between quarks and anti-quarks. When such symmetry breaks
(for example when considering the valence parton distribution) it
will no longer work. In this paper, we consider kinematic regions
where valence distribution is small compared to gluon or sea distributions.

\begin{figure}
\[
\xymatrix{ &  &  &  & {}\ar@{~}[d]^{a} &  & {}\ar@{~}[d]^{b} & \xyR{1pc}\xyC{1pc}\\
 &  &  &  & {}\ar@{~}[rr]\ar@{~}[d]^{c} &  & {}\ar@{~}[d]^{d} & {=f^{aec}f^{bde}=V_{ab;cd}} & {(a)}\\
 &  &  &  & {} &  & {}\\
{}\ar@{~}[d]^{a} &  & {}\ar@{~}[d]^{b} &  & {}\ar@{~}[d]^{a} &  & {}\ar@{~}[d]^{b}\\
{}\ar@{<-}[rr]\ar@{->}[d]^{i} &  & {}\ar@{<-}[d]^{j} & {+} & {}\ar@{->}[rr]\ar@{<-}[d]^{i} &  & {}\ar@{->}[d]^{j} & {=t_{ik}^{a}t_{kj}^{b}+t_{jk}^{b}t_{ki}^{a}} & {(b)}\\
{} &  & {} &  & {} &  & {}\\
{}\ar@{->}[d]^{i} &  & {}\ar@{<-}[d]^{j} &  & {}\ar@{<-}[d]^{i} &  & {}\ar@{->}[d]^{j}\\
{}\ar@{->}[rr]\ar@{~>}[d]^{c} &  & {}\ar@{~>}[d]^{d} & {+} & {}\ar@{<-}[rr]\ar@{~>}[d]^{c} &  & {}\ar@{~>}[d]^{d} & {=t_{jk}^{d}t_{ki}^{c}+t_{ik}^{c}t_{kj}^{d}} & {(c)}\\
{} &  & {} &  & {} &  & {}\\
{}\ar@{->}[d]^{i} &  & {}\ar@{<-}[d]^{j} &  & {}\ar@{<-}[d]^{i} &  & {}\ar@{->}[d]^{j}\\
{}\ar@{~}[rr]\ar@{->}[d]^{m} &  & {}\ar@{<-}[d]^{n} & {+} & {}\ar@{~}[rr]\ar@{<-}[d]^{m} &  & {}\ar@{->}[d]^{n} & {=t_{mi}^{e}t_{jn}^{e}+t_{im}^{e}t_{nj}^{e}} & {(d)}\\
{} &  & {} &  & {} &  & {}
}
\]

\caption{The DGLAP ladders associated with $\overline{C}_{G}^{G}\ (a)$, $\overline{C}_{G}^{F}\ (b)$,
$\overline{C}_{F}^{G}\ (c)$ and $\overline{C}_{F}^{F}\ (d)$, where
we have summed quark and anti-quark ladders together. \label{fig:DGLAP-ladders}}
\end{figure}

\subsection{Color Factor for the $8_{a}$ Representation}

We start by concretely looking at the $8_{a}$ representation. Consider
the two structures $G_{a}$ and $Q_{a}$ that represent this representation
for quarks and gluons, as shown in fig \ref{fig:Ga and Qa}. We'll
show that these structures propagate within the ladder and do not
mix with any other structures. The color kernels for $8_{a}$ will
be the contractions of these structures with the ladders given in
figure \ref{fig:DGLAP-ladders}.

\begin{figure}
\[
\xymatrix{\xyR{1pc}\xyC{1pc} & *=0{}\ar@{~}[dd]^{e} &  &  &  & *=0{}\ar@{~}[dd]^{e} &  &  &  & *=0{}\ar@{~}[dd]^{e}\\
\\
{\alpha\cdot} & *=0{}\ar@{~}[dr]^{d} &  &  & {\beta\cdot} & *=0{}\ar@{<-}[dr]^{l} &  & {-} &  & *=0{}\ar@{->}[dr]^{l}\\
{}\ar@{~}[ur]^{c} &  & {} & {\ \ } & {}\ar@{<-}[ur]^{k} &  & *=0{} &  & *=0{}\ar@{->}[ur]^{k} &  & {}\\
 & *=0{G_{a}^{dc}=i\alpha f^{dce}} &  &  &  &  &  & *=0{Q_{a}^{kl}=\beta\left(t_{kl}^{e}-t_{lk}^{e}\right)}\\
 & {(a)} &  &  &  &  &  & {(b)}
}
\]

\caption{The anti-symmetric octet ($8_{a}$) structures $G_{a}\ (a)$ and $Q_{a}\ (b)$.
$\alpha$ and $\beta$ are arbitrary real numbers. \label{fig:Ga and Qa}}
\end{figure}

\subsubsection{$\overline{C}_{G}^{G}$ }

\begin{figure}
\[
\xymatrix{ & *=0{}\ar@{~>}[dd]^{a}\\
\\
 & *=0{}\ar@{~>}[dr]^{b}\\
{}\ar@{<~}[ur]^{c} &  & {}\\
{}\ar@{~>}[d] &  & {}\ar@{~>}[d]\\
{}\ar@{~>}[rr]^{e}\ar@{~>}[d]^{f} &  & {}\ar@{~>}[d]^{g}\\
{} &  & {}
}
\]

\caption{Contraction of $\Phi_{G}^{G}$ with $G_{a}$\label{fig:Contraction-of-FGG-P8}}
\end{figure}

We now contract the ladder of type $\Phi_{G}^{G}$ with the color
structure $G$ as seen in figure \ref{fig:Contraction-of-FGG-P8}:

\begin{equation}
G_{a}^{bc}\cdot V_{bc;gf}=i\alpha f^{bca}\left(c_{\alpha}P_{bc;gf}^{\alpha}\right).
\end{equation}

We use the contraction property of the projectors (the generalized
version as proved in section \ref{sec:Rules-for-Color}) to cancel
every projector but $8_{a}$ and are left with:

\[
=i\alpha f^{bca}c_{8_{a}}P_{bc;gf}^{8_{a}}=i\alpha c_{8_{a}}\left(f^{bca}\frac{f^{bce}f^{gfe}}{N}\right)=c_{8_{a}}\left(\alpha iN\frac{f^{gfa}}{N}\right)=c_{8_{a}}G_{gf}.
\]

Here, as we'll see in Appendix \ref{sec:Rules-for-Color}, $c_{8_{a}}=\frac{N}{2}$.

\subsubsection{$\overline{C}_{F}^{G}$ }

\begin{figure}
\[
\xymatrix{ & *=0{}\ar@{~>}[dd]^{a} &  &  &  & *=0{}\ar@{~>}[dd]^{a}\\
\\
 & *=0{}\ar@{~>}[dr]^{b} &  &  &  & *=0{}\ar@{~>}[dr]^{b}\\
{}\ar@{<~}[ur]^{c} &  & {} & {+} & {}\ar@{<~}[ur]^{c} &  & {}\\
{}\ar@{~>}[d] &  & {}\ar@{~>}[d] &  & {}\ar@{~>}[d] &  & {}\ar@{~>}[d]\\
{}\ar@{<-}[rr]^{k}\ar@{->}[d]^{i} &  & {}\ar@{<-}[d]^{j} &  & {}\ar@{>-}[rr]^{k}\ar@{-<}[d]^{i} &  & {}\ar@{>-}[d]^{j}\\
{} &  & {} &  & {} &  & {}
}
\]

\caption{Contraction of $\Phi_{F}^{G}$ with $G_{a}$\label{fig:Contraction-of-FGF-P8}}
\end{figure}

This contraction is shown in figure \ref{fig:Contraction-of-FGF-P8}
and is written as:

\begin{equation}
i\alpha f^{bca}\left(t_{ik}^{c}t_{kj}^{b}+t_{jk}^{b}t_{ki}^{c}\right)=-\alpha if^{cba}t_{ik}^{c}t_{kj}^{b}+if^{bca}t_{jk}^{b}t_{ki}^{c}=\alpha\frac{N}{2}\left(t_{ij}^{e}-t_{ji}^{e}\right)=\frac{\alpha}{\beta}\frac{N}{2}Q_{ij}.
\end{equation}

So we get that $\overline{C}_{F}^{G}=\frac{\alpha}{\beta}\frac{N}{2}$.

\subsubsection{$\overline{C}_{F}^{F}$}

This contraction is viewed on figure \ref{fig:Contraction-of-FGF-P8-1}
and is written as:

\begin{figure}
\[
\xymatrix{ & *=0{}\ar@{~>}[dd]^{a} &  &  &  & *=0{}\ar@{~>}[dd]^{e}\\
\\
 & *=0{}\ar@{-<}[dr]^{l} &  &  &  & *=0{}\ar@{->}[dr]^{l}\\
{}\ar@{<-}[ur]^{k} &  & {} & {-} & {}\ar@{>-}[ur]^{k} &  & {}\\
{}\ar@{->}[d] &  & {}\ar@{<-}[d] &  & {}\ar@{-<}[d] &  & {}\ar@{->}[d]\\
{}\ar@{~>}[rr]^{e}\ar@{->}[d]^{i} &  & {}\ar@{<-}[d]^{j} &  & {}\ar@{~>}[rr]^{e}\ar@{-<}[d]^{i} &  & {}\ar@{->}[d]^{j}\\
{} &  & {} &  & {} &  & {}
}
\]

\caption{Contraction of $\Phi_{F}^{F}$ with $Q_{a}$\label{fig:Contraction-of-FGF-P8-1}}
\end{figure}

\begin{align}
Q_{a}^{ij}\left(t_{ik}^{e}t_{lj}^{e}+t_{ki}^{e}t_{jl}^{e}\right) & =\beta\left(t_{kl}^{a}t_{ik}^{e}t_{lj}^{e}-t_{lk}^{a}t_{ki}^{e}t_{jl}^{e}\right)=\beta\left(t_{ik}^{e}t_{kl}^{a}t_{lj}^{e}-t_{jl}^{e}t_{lk}^{a}t_{ki}^{e}\right)\nonumber \\
 & =-\beta\frac{1}{2N}\left(t_{ij}^{a}-t_{ji}^{a}\right)=-\frac{1}{2N}Q_{ij}.
\end{align}

So the factor is $\overline{C}_{F}^{F}=-\frac{1}{2N}$.

\subsubsection{$\overline{C}_{G}^{F}$ }

This contraction is viewed on figure \ref{fig:Contraction-of-FGF-P8-1-1}
and is written as

\begin{figure}
\[
\xymatrix{ & *=0{}\ar@{~>}[dd]^{a} &  &  &  & *=0{}\ar@{~>}[dd]^{a}\\
\\
 & *=0{}\ar@{-<}[dr]^{j} &  &  &  & *=0{}\ar@{->}[dr]^{j}\\
{}\ar@{<-}[ur]^{i} &  & {} & {-} & {}\ar@{>-}[ur]^{i} &  & {}\\
{}\ar@{->}[d] &  & {}\ar@{<-}[d] &  & {}\ar@{<-}[d]^{i} &  & {}\ar@{>-}[d]^{j}\\
{}\ar@{->}[rr]^{k}\ar@{~>}[d]^{c} &  & {}\ar@{~>}[d]^{b} &  & {}\ar@{<-}[rr]^{k}\ar@{~>}[d]^{c} &  & {}\ar@{~>}[d]^{b}\\
{} &  & {} &  & {} &  & {}
}
\]

\caption{Contraction of $\Phi_{G}^{F}$ with $Q_{a}$\label{fig:Contraction-of-FGF-P8-1-1}}
\end{figure}

\begin{align}
Q_{a}^{ij}\left(t_{jk}^{b}t_{ki}^{c}+t_{ik}^{c}t_{kj}^{b}\right) & =\beta\left(t_{ij}^{a}t_{jk}^{b}t_{ki}^{c}-t_{ji}^{a}t_{ik}^{c}t_{kj}^{b}\right)\nonumber \\
 & =\beta\left(tr\left[t^{a}t^{b}t^{c}\right]-tr\left[t^{a}t^{c}t^{b}\right]\right)\nonumber \\
 & =\beta\frac{1}{4}\left(if^{abc}+d^{abc}-if^{acb}-d^{acb}\right)\nonumber \\
 & =\beta\frac{i}{4}\left(f^{abc}-f^{acb}\right)\nonumber \\
 & =\beta\frac{i}{2}f^{bca}\nonumber \\
 & =\frac{\beta}{\alpha}\frac{1}{2}G_{bc}.
\end{align}

So the factor is $\overline{C}_{G}^{F}=\frac{\beta}{\alpha}\frac{1}{2}$.

\subsection{Color Factor for the $8_{s}$ Representation}

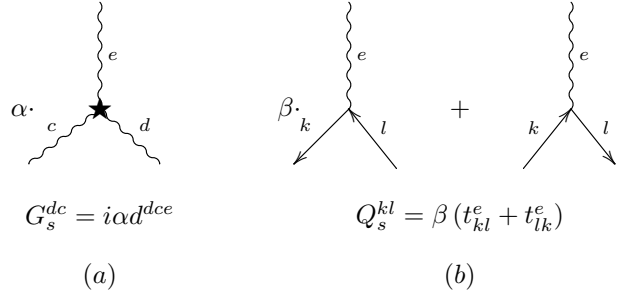
\begin{figure}
\[
\xymatrix{\xyR{1pc}\xyC{1pc} & *=0{}\ar@{~}[dd]^{e} &  &  &  & *=0{}\ar@{~}[dd]^{e} &  &  &  & *=0{}\ar@{~}[dd]^{e}\\
\\
{\alpha\cdot} & *=0{\bigstar}\ar@{~}[dr]^{d} &  &  & {\beta\cdot} & *=0{}\ar@{<-}[dr]^{l} &  & {+} &  & *=0{}\ar@{->}[dr]^{l}\\
{}\ar@{~}[ur]^{c} &  & {} & {\ \ } & {}\ar@{<-}[ur]^{k} &  & *=0{} &  & *=0{}\ar@{->}[ur]^{k} &  & {}\\
 & *=0{G_{s}^{dc}=i\alpha d^{dce}} &  &  &  &  &  & *=0{Q_{s}^{kl}=\beta\left(t_{kl}^{e}+t_{lk}^{e}\right)}\\
 & {(a)} &  &  &  &  &  & {(b)}
}
\]

\caption{The symmetric octet ($8_{s}$) structures $G_{s}\ (a)$ and $Q_{s}\ (b)$.
$\alpha$ and $\beta$ are arbitrary real numbers that are not necessarily
the same as for $8_{a}$. \label{fig:Gs and Qs}}
\end{figure}

We now look at the $8_{s}$ representation. Consider two new structures
$G_{s}$ and $Q_{s}$ that represent this representation for gluon
and quark, as shown in fig \ref{fig:Gs and Qs}. The rest of the analysis
is the same.

\subsubsection{$\overline{C}_{G}^{G}$ }

\begin{figure}
\[
\xymatrix{ & *=0{}\ar@{~>}[dd]^{a}\\
\\
 & *=0{\bigstar}\ar@{~>}[dr]^{b}\\
{}\ar@{<~}[ur]^{c} &  & {}\\
{}\ar@{~>}[d] &  & {}\ar@{~>}[d]\\
{}\ar@{~>}[rr]^{e}\ar@{~>}[d]^{f} &  & {}\ar@{~>}[d]^{g}\\
{} &  & {}
}
\]

\caption{Contraction of $\Phi_{G}^{G}$ with $G_{s}$\label{fig:Contraction-of-FGG-P8-1}}
\end{figure}
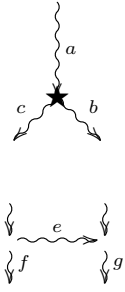

We now contract the ladder of type $\Phi_{G}^{G}$ with the color
structure $G_{s}$ as seen in figure \ref{fig:Contraction-of-FGG-P8-1}:

\begin{equation}
G_{s}^{bc}\cdot V_{bc;gf}=\alpha d^{bca}\left(c_{\alpha}P_{bc;gf}^{\alpha}\right).
\end{equation}

We use the contraction property of the projectors (the generalized
version) to cancel every projector but $8_{s}$ and are left with:

\[
=\alpha d^{bca}c_{8_{s}}P_{bc;gf}^{8_{s}}=\alpha c_{8_{s}}\left(d^{bca}d^{bce}d^{gfe}\frac{N}{N^{2}-4}\right)=c_{8_{s}}\left(\alpha\frac{N^{2}-4}{N}d^{gfa}\frac{N}{N^{2}-4}\right)=c_{8_{s}}G_{gf}^{s}.
\]

Where, as we have seen $c_{8_{s}}=\frac{N}{2}$.

\subsubsection{$\overline{C}_{F}^{G}$ }

\begin{figure}
\[
\xymatrix{ & *=0{}\ar@{~>}[dd]^{a} &  &  &  & *=0{}\ar@{~>}[dd]^{a}\\
\\
 & *=0{}\ar@{~>}[dr]^{b} &  &  &  & *=0{}\ar@{~>}[dr]^{b}\\
{}\ar@{<~}[ur]^{c} &  & {} & {+} & {}\ar@{<~}[ur]^{c} &  & {}\\
{}\ar@{~>}[d] &  & {}\ar@{~>}[d] &  & {}\ar@{~>}[d] &  & {}\ar@{~>}[d]\\
{}\ar@{<-}[rr]^{k}\ar@{->}[d]^{i} &  & {}\ar@{<-}[d]^{j} &  & {}\ar@{>-}[rr]^{k}\ar@{-<}[d]^{i} &  & {}\ar@{>-}[d]^{j}\\
{} &  & {} &  & {} &  & {}
}
\]

\caption{Contraction of $\Phi_{F}^{G}$ with $G_{s}$\label{fig:Contraction-of-FGF-P8-2}}
\end{figure}
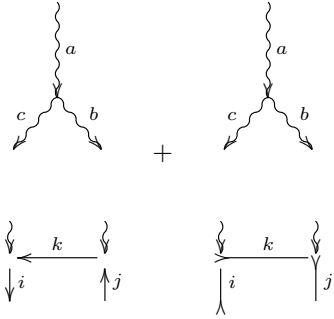

This contraction is shown in figure \ref{fig:Contraction-of-FGF-P8-2}
and is written as:

\begin{equation}
\alpha d^{bca}\left(t_{ik}^{c}t_{kj}^{b}+t_{jk}^{b}t_{ki}^{c}\right)=\alpha\left(d^{cba}t_{ik}^{c}t_{kj}^{b}+d^{bca}t_{jk}^{b}t_{ki}^{c}\right)=\alpha\frac{N^{2}-4}{2N}\left(t_{ij}^{e}+t_{ji}^{e}\right)=\frac{\alpha}{\beta}\frac{N^{2}-4}{2N}Q_{ij}^{s}.
\end{equation}

So we get that $\overline{C}_{F}^{G}=\frac{N^{2}-4}{2N}$.

\subsubsection{$\overline{C}_{F}^{F}$}

This contraction is viewed on figure \ref{fig:Contraction-of-FGF-P8-1-2}
and is written as:

\begin{figure}
\[
\xymatrix{ & *=0{}\ar@{~>}[dd]^{a} &  &  &  & *=0{}\ar@{~>}[dd]^{e}\\
\\
 & *=0{}\ar@{-<}[dr]^{l} &  &  &  & *=0{}\ar@{->}[dr]^{l}\\
{}\ar@{<-}[ur]^{k} &  & {} & {-} & {}\ar@{>-}[ur]^{k} &  & {}\\
{}\ar@{->}[d] &  & {}\ar@{<-}[d] &  & {}\ar@{-<}[d] &  & {}\ar@{->}[d]\\
{}\ar@{~>}[rr]^{e}\ar@{->}[d]^{i} &  & {}\ar@{<-}[d]^{j} &  & {}\ar@{~>}[rr]^{e}\ar@{-<}[d]^{i} &  & {}\ar@{->}[d]^{j}\\
{} &  & {} &  & {} &  & {}
}
\]

\caption{Contraction of $\Phi_{F}^{F}$ with $Q_{s}$\label{fig:Contraction-of-FGF-P8-1-2}}
\end{figure}
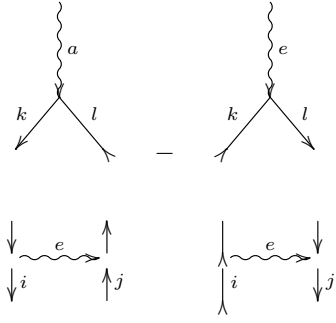

\begin{align}
Q_{s}^{ij}\left(t_{ik}^{e}t_{lj}^{e}+t_{ki}^{e}t_{jl}^{e}\right) & =\beta\left(t_{kl}^{a}t_{ik}^{e}t_{lj}^{e}+t_{lk}^{a}t_{ki}^{e}t_{jl}^{e}\right)\nonumber \\
 & =\beta\left(t_{ik}^{e}t_{kl}^{a}t_{lj}^{e}+t_{jl}^{e}t_{lk}^{a}t_{ki}^{e}\right)\nonumber \\
 & =-\frac{\beta}{2N}\left(t_{ij}^{a}+t_{ji}^{a}\right)=-\frac{1}{2N}Q_{ij}^{s}.
\end{align}

So the factor is $\overline{C}_{F}^{F}=-\frac{1}{2N}$.

\subsubsection{$\overline{C}_{G}^{F}$ }

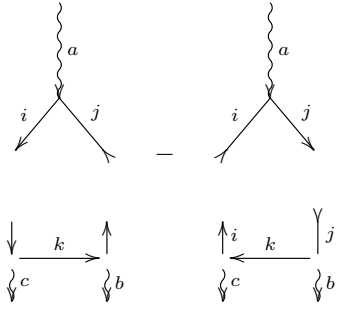
\begin{figure}
\[
\xymatrix{ & *=0{}\ar@{~>}[dd]^{a} &  &  &  & *=0{}\ar@{~>}[dd]^{a}\\
\\
 & *=0{}\ar@{-<}[dr]^{j} &  &  &  & *=0{}\ar@{->}[dr]^{j}\\
{}\ar@{<-}[ur]^{i} &  & {} & {-} & {}\ar@{>-}[ur]^{i} &  & {}\\
{}\ar@{->}[d] &  & {}\ar@{<-}[d] &  & {}\ar@{<-}[d]^{i} &  & {}\ar@{>-}[d]^{j}\\
{}\ar@{->}[rr]^{k}\ar@{~>}[d]^{c} &  & {}\ar@{~>}[d]^{b} &  & {}\ar@{<-}[rr]^{k}\ar@{~>}[d]^{c} &  & {}\ar@{~>}[d]^{b}\\
{} &  & {} &  & {} &  & {}
}
\]

\caption{Contraction of $\Phi_{G}^{F}$ with $Q_{s}$\label{fig:Contraction-of-FGF-P8-1-1-1}}
\end{figure}

This contraction is viewed on figure \ref{fig:Contraction-of-FGF-P8-1-1-1}
and is written as

\begin{align}
Q_{s}^{ij}\left(t_{jk}^{b}t_{ki}^{c}+t_{ik}^{c}t_{kj}^{b}\right) & =\beta\left(t_{ij}^{a}t_{jk}^{b}t_{ki}^{c}+t_{ji}^{a}t_{ik}^{c}t_{kj}^{b}\right)=\nonumber \\
 & =\beta\left(tr\left[t^{a}t^{b}t^{c}\right]+tr\left[t^{a}t^{c}t^{b}\right]\right)\nonumber \\
 & =\beta\frac{1}{4}\left(if^{abc}+d^{abc}+if^{acb}+d^{acb}\right)\nonumber \\
 & =\beta\frac{i}{4}\left(d^{abc}+d^{acb}\right)\nonumber \\
 & =\beta\frac{1}{2}d^{abc}\nonumber \\
 & =\frac{\beta}{\alpha}\frac{1}{2}G_{bc}^{s}
\end{align}

So the factor is $\overline{C}_{G}^{F}=\frac{\beta}{\alpha}\frac{1}{2}G_{bc}^{s}$.

\subsection{Normalization of $\overline{C}_{F}^{G}$ and $\overline{C}_{G}^{F}$}

\begin{figure}[H]
\[
\xymatrix{\xyR{3pc}\xyC{3pc}\\
*=0{}\ar@{->}[d] &  & *=0{}\ar@{-<}[d] & *=0{\downarrow\Phi_{F}^{G}}\\
*=0{}\ar@{->}[rr]\ar@{~}[d] &  & *=0{}\ar@{~}[d]\\
*=0{} &  & *=0{} & *=0{\uparrow\Phi_{G}^{F}}
}
\]

\caption{gluon to quark ladder diagram. \label{fig:gluon-gluon-1}}
\end{figure}
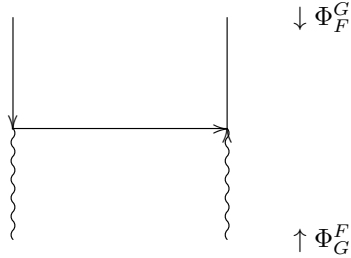

Note that the color factors given above for $\overline{C}_{F}^{G}$
and $\overline{C}_{G}^{F}$ are not well defined and depend on the
ratio of $\frac{\alpha}{\beta}$ which was arbitrary. Therefore only
the multiplication $C_{G}^{F}\cdot C_{F}^{G}$ is well defined. In
order to derive well defined color factors we impose the condition
that a ladder would be well defined independently of whether we go
from quark to gluon or from gluon to quark as shown in figure \ref{fig:gluon-gluon-1}.

In our convention the ladder end in the hard process and therefore
we sum over the color (and flavor) indices of the final (superscript)
parton type and average\textbf{ }over the color (and flavor) indices
of the initial (subscript) parton type. This convention means that
if we take the diagram in figure \ref{fig:gluon-gluon-1} from above
or from below it will be equal only if we compensate for the averaging
of the initial parton (and therefore effectively summing for both
initial and final parton color and flavor indices):

\begin{equation}
\underset{"types"\ of\ quarks}{\underbrace{N}}\overline{\Phi}_{F}^{G}=\underset{"types"\ of\ gluons}{\underbrace{\left(N^{2}-1\right)}}\overline{\Phi}_{G}^{F}.
\end{equation}

The $z$ part of this argument is true\citep{Dokshitzer1980} so we'll
ignore it and get for the color-flavor part:

\begin{equation}
N\overline{C}_{F}^{G}=\left(N^{2}-1\right)\overline{C}_{G}^{F},
\end{equation}

\begin{equation}
\frac{\overline{C}_{F}^{G}}{\overline{C}_{G}^{F}}=\frac{N^{2}-1}{N}.
\end{equation}

Note that this equation is actually independent of the color channel
and therefore should be true for both the singlet and octet channels.
Indeed for the singlet channel we have: 
\begin{equation}
\frac{\overline{C}_{F}^{G}}{\overline{C}_{G}^{F}}=\frac{\frac{N^{2}-1}{2N}}{\frac{1}{2}}=\frac{N^{2}-1}{N}.
\end{equation}

\subsubsection{$8_{a}$}

Now the color factor multiplication is $\overline{C}_{F}^{G}\overline{C}_{G}^{F}=\frac{N}{4}$,
therefore we can solve:

\begin{subequations}
\begin{equation}
\phantom{}^{8_{a}}\overline{C}_{G}^{F}=\frac{1}{2}\sqrt{\frac{N^{2}}{2\left(N^{2}-1\right)}},
\end{equation}

\begin{equation}
\phantom{}^{8_{a}}\overline{C}_{F}^{G}=\sqrt{\frac{N^{2}-1}{8}}.
\end{equation}
\end{subequations}

\subsubsection{$8_{s}$}

Now the color factor multiplication is $\overline{C}_{F}^{G}\overline{C}_{G}^{F}=\frac{N^{2}-4}{4N}$,
therefore we can solve:

\begin{subequations}
\begin{equation}
\phantom{}^{8_{s}}\overline{C}_{G}^{F}=\frac{1}{2}\sqrt{\frac{N^{2}-4}{2\left(N^{2}-1\right)}},
\end{equation}

\begin{equation}
\phantom{}^{8_{s}}\overline{C}_{F}^{G}=\frac{1}{2N}\sqrt{\frac{\left(N^{2}-4\right)\left(N^{2}-1\right)}{2}}.
\end{equation}
\end{subequations}

\section{Formal Derivation of the Non-Singlet DGLAP Equation\label{sec:Formal DGLAP}}

In this section, we give a more detailed and formal derivation of
(\ref{eq:Modified-DGLAP}). We start with the Bethe-Salpeter equation
for $D_{A}^{B}$, the singlet structure function, that comes from
considering ladder type diagrams \citep{Dokshitzer1980}:

\begin{equation}
D_{A}^{B}\left(x,Q^{2}\right)=\delta_{A}^{B}\left(1-x\right)d_{A}\left(k_{0}^{2}\right)d_{B}^{-1}\left(Q^{2}\right)+d_{B}^{-1}\left(Q^{2}\right)\sum_{C}\intop_{zk_{0}^{2}}^{Q^{2}}\frac{dk^{2}}{k^{2}}\intop_{0}^{1}\frac{dz}{z}A_{A}^{C}d_{B}\left(k^{2}\right)\Phi_{C}^{B}\left(z\right)D_{A}^{C}\left(\frac{x}{z},k^{2}\right).\label{eq:Ladder}
\end{equation}

Here: 
\begin{itemize}
\item $d_{A}$ is the renormalization of particle $A$. 
\item $A_{A}^{C}=d_{A}\left(\frac{k^{2}}{z}\right)d_{C}\left(k^{2}\right)d_{G}\Gamma^{2}$
is the renormalization of the ladder, which we take to be $\frac{\alpha\left(k^{2}\right)}{4\pi}$. 
\item $\Phi_{A}^{C}\left(z\right)=C_{A}^{B}\cdot V\left(z\right)$ is the
singlet splitting function as explained in the text. 
\item $A,B,C$ are types of partons, e.g gluon, quark, anti-quark. 
\end{itemize}
$d_{A}$ obeys the relation \citep{Dokshitzer1980}:

\begin{equation}
\frac{\partial}{\partial ln\left(Q^{2}\right)}d_{B}\left(Q^{2}\right)=d_{B}\left(Q^{2}\right)\sum_{C}\intop_{0}^{1}z\frac{\alpha\left(Q^{2}\right)}{4\pi}\Phi_{C}^{B}\left(z\right)dz.\label{eq:Sudakov2}
\end{equation}

This equation represents virtual corrections of the parton to itself
and therefore is true even if the parton is in a non-singlet state
with its complex conjugate (see figure \ref{fig:ladder example}.$(b)$
). therefore for a general channel the only part of equation (\ref{eq:Ladder})
that changes is the one dependent on the real (ladder, see figure
\ref{fig:ladder example}.$(a)$ ) contributions:

\begin{equation}
\overline{D}_{A}^{B}\left(x,Q^{2}\right)=\delta_{A}^{B}\left(1-x\right)d_{A}\left(k_{0}^{2}\right)d_{B}^{-1}\left(Q^{2}\right)+d_{B}^{-1}\left(Q^{2}\right)\sum_{C}\intop_{zk_{0}^{2}}^{Q^{2}}\frac{dk^{2}}{k^{2}}\intop_{0}^{1}\frac{dz}{z}A_{A}^{C}d_{B}\left(k^{2}\right)\overline{\Phi}_{C}^{B}\left(z\right)\overline{D}_{A}^{C}\left(\frac{x}{z},k^{2}\right).\label{eq:OctetLadder}
\end{equation}

Note that $d_{A}$ still obeys (\ref{eq:Sudakov2}) with the singlet
($\Phi_{A}^{C}$) splitting function. Trivially: 
\begin{equation}
d_{B}^{-1}\left(Q^{2}\right)\frac{\partial}{\partial ln\left(Q^{2}\right)}\left[\overline{D}_{A}^{B}\left(x,Q^{2}\right)d_{B}\left(Q^{2}\right)\right]=\overline{D}_{A}^{B}\left(x,Q^{2}\right)d_{B}^{-1}\left(Q^{2}\right)\frac{\partial d_{B}\left(Q^{2}\right)}{\partial ln\left(Q^{2}\right)}+\frac{\partial\overline{D}_{A}^{B}\left(x,Q^{2}\right)}{\partial ln\left(Q^{2}\right)}.
\end{equation}

Using equation (\ref{eq:Sudakov2}) and (\ref{eq:OctetLadder}) we
can write:

\begin{equation}
\sum_{C}\intop_{0}^{1}\frac{dz}{z}\frac{\alpha\left(Q^{2}\right)}{4\pi}\overline{\Phi}_{C}^{B}\left(z\right)\overline{D}_{A}^{C}\left(\frac{x}{z},Q^{2}\right)=\overline{D}_{A}^{B}\left(x,Q^{2}\right)\sum_{C}\intop_{0}^{1}z\frac{\alpha\left(Q^{2}\right)}{4\pi}\Phi_{C}^{B}\left(z\right)dz+\frac{\partial\overline{D}_{A}^{B}\left(x,Q^{2}\right)}{\partial ln\left(Q^{2}\right)},
\end{equation}

\begin{equation}
\frac{\partial\overline{D}_{A}^{B}\left(x,Q^{2}\right)}{\partial ln\left(Q^{2}\right)}=\frac{\alpha\left(Q^{2}\right)}{4\pi}\sum_{C}\intop_{0}^{1}\frac{dz}{z}\left[\overline{\Phi}_{C}^{B}\left(z\right)\overline{D}_{A}^{C}\left(\frac{x}{z},Q^{2}\right)-z^{2}\Phi_{B}^{C}\left(z\right)\overline{D}_{A}^{B}\left(x,Q^{2}\right)\right].\label{eq:Modified-DGLAP-1}
\end{equation}

Using (\ref{eq:Color-Factorization}) eq. (\ref{eq:Modified-DGLAP-1})
can be written as:

\begin{subequations}
\begin{equation}
\frac{\partial D_{A}^{B}\left(x,Q^{2}\right)}{\partial ln\left(Q^{2}\right)}=K_{1}+K_{2},
\end{equation}

\begin{equation}
K_{1}=\frac{\alpha\left(Q^{2}\right)}{4\pi}\sum_{C}\intop_{0}^{1}\frac{dz}{z}\overline{\Phi}_{C}^{B}\left(z\right)\overline{D}_{A}^{C}\left(\frac{x}{z},Q^{2}\right)-z^{2}\overline{\Phi}_{B}^{C}\left(z\right)\overline{D}_{A}^{B}\left(x,Q^{2}\right),
\end{equation}

\begin{equation}
K_{2}=\overline{D}_{A}^{B}\left(x,Q^{2}\right)\frac{\alpha\left(Q^{2}\right)}{4\pi}\sum_{C}\intop_{0}^{1} dzV_{B}^{C}\left(z\right)z\left[C_{B}^{C}-\overline{C}_{B}^{C}\right].
\end{equation}
\end{subequations}

The first part is finite as $z\rightarrow1$ and is just the regular
DGLAP equation with the new kernels (which are given in table \ref{tab:Color-factors-for}).
The second part diverges as $z\rightarrow1$ and therefore we need
to regularize it. We'll keep the regularization given in\citep{Dokshitzer1980}: 
\begin{itemize}
\item Integration of $z$ up to $1-\frac{\lambda^{2}}{k_{0}^{2}}$, for
$\lambda$ an IR cutoff regulator 
\item Adding a non zero $\Delta$ in $V_{C}^{B}\left(z\right)\rightarrow V_{C}^{B}\left(z,\Delta\right)$
for $\Delta=\frac{zk_{0}^{2}}{\sigma_{k_{0}}}$ (where $\sigma_{k_{0}}=\frac{\left(2k_{0}c\right)^{2}}{c^{2}}\approx Q^{2}$,
for $c$ the gauge fixing vector) 
\end{itemize}
Define the Sudakov suppression factor:

\begin{equation}
\overline{S}_{A}\left(k_{0}^{2},Q^{2}\right)=e^{-\sum_{C}\left(C_{A}^{C}-\overline{C}_{A}^{C}\right)\intop_{k_{0}^{2}}^{Q^{2}}\frac{dk^{2}}{k^{2}}\frac{\alpha\left(k^{2}\right)}{4\pi}\intop_{0}^{1-\frac{\lambda^{2}}{k_{0}^{2}}}dzzV_{A}^{C}\left(z,\frac{zk_{0}^{2}}{\sigma_{k_{0}}}\right)}
\end{equation}

Which is a properly regularized version of the version appearing in
the text. Then for $\overline{D}_{A}^{B}\left(x,k_{0}^{2},Q^{2}\right)=\overline{S}_{A}\left(k_{0}^{2},Q^{2}\right)\widetilde{D}_{A}^{B}\left(x,k_{0}^{2}\right)$
eq. (\ref{eq:Modified-DGLAP-1}) is reduced to (\ref{eq:Reduced DGLAP}).

\section{$\widetilde{D}_{A}^{B}$ at the Limit $x\rightarrow1$\label{sec:-at-the-x-1}}

In this section, we write a derivation of (\ref{eq:x-1-limit}) as
the derivation in \citep{Dokshitzer1980} is only partial, hard to
generalize for different color factors, and does not include $\widetilde{D}_{G}^{G}$.
For reference we write the standard solutions of (\ref{eq:Reduced DGLAP})
in Mellin space:

\begin{subequations}
\begin{align}
\widetilde{D}_{F}^{F}\left(j,\xi\right) & =\widetilde{D}^{sea}\left(j,\xi\right)+\widetilde{D}^{val}\left(j,\xi\right)\\
\widetilde{D}^{val}\left(j,\xi\right) & =e^{\overline{\nu}_{0}\xi}\\
\widetilde{D}^{sea}\left(j,\xi\right) & =\frac{1}{2n_{f}}\left[\frac{\overline{\nu}_{0}-\overline{\nu}_{-}}{\overline{\nu}_{+}-\overline{\nu}_{-}}e^{\overline{\nu}_{+}\xi}+\frac{\overline{\nu}_{+}-\overline{\nu}_{0}}{\overline{\nu}_{+}-\overline{\nu}_{-}}e^{\overline{\nu}_{-}\xi}-e^{\overline{\nu}_{0}\xi}\right]\\
\widetilde{D}_{G}^{G}\left(j,\xi\right) & =\frac{\overline{\nu}_{0}-\overline{\nu}_{-}}{\overline{\nu}_{+}-\overline{\nu}_{-}}e^{\overline{\nu}_{-}\xi}+\frac{\overline{\nu}_{+}-\overline{\nu}_{0}}{\overline{\nu}_{+}-\overline{\nu}_{-}}e^{\overline{\nu}_{+}\xi}\\
\widetilde{D}_{F}^{G}\left(j,\xi\right) & =\frac{\overline{\Phi}_{F}^{G}\left(j\right)}{\overline{\nu}_{+}-\overline{\nu}_{-}}\left[e^{\overline{\nu}_{+}\xi}-e^{\overline{\nu}_{-}\xi}\right]\\
\widetilde{D}_{G}^{F}\left(j,\xi\right) & =\frac{\overline{\Phi}_{G}^{F}\left(j\right)}{\overline{\nu}_{+}-\overline{\nu}_{-}}\left[e^{\overline{\nu}_{+}\xi}-e^{\overline{\nu}_{-}\xi}\right]
\end{align}
\end{subequations}

Where $\xi=\frac{3}{\beta_{0}}ln\left[\frac{\alpha\left(k_{0}^{2}\right)}{\alpha\left(Q^{2}\right)}\right]$
and:

\begin{subequations}
\begin{align}
\overline{\nu}_{0} & =\overline{\nu}_{F}=\frac{1}{3}\left(-\left(\frac{6}{j}+\frac{6}{j+1}+12(\psi(j)+\gamma_{E})-17\right)\overline{C}_{F}^{F}-8\overline{C}_{F}^{G}\right)\\
\overline{\nu}_{G} & =-\overline{C}_{G}^{G}\left(-\frac{8\left(j^{2}+j+1\right)}{j\left(j^{2}-1\right)(j+2)}+4\psi(j+1)-\frac{11}{3}+4\gamma_{E}\right)-\frac{4}{3}n_{f}\overline{C}_{G}^{F}\\
\overline{\Phi}_{F}^{G}\left(j\right) & =2\overline{C}_{F}^{G}\frac{j^{2}+j+1}{j\left(j^{2}-1\right)}\\
\overline{\Phi}_{G}^{F}\left(j\right) & =\overline{C}_{G}^{F}\frac{j^{2}+j+2}{j\left(j+1\right)\left(j+2\right)}\\
\overline{\nu}_{\pm} & =\frac{1}{2}\left\{ \overline{\nu}_{F}+\overline{\nu}_{G}\pm\sqrt{\left(\overline{\nu}_{F}-\overline{\nu}_{G}\right)^{2}+8n_{f}\overline{\Phi}_{F}^{G}\left(j\right)\overline{\Phi}_{G}^{F}\left(j\right)}\right\} 
\end{align}
\end{subequations}

Transforming back from Mellin to $x-space$ is by numerically evaluating
the inverse Mellin transform:

\begin{equation}
\widetilde{D}_{A}^{B}\left(x,\xi\right)=\intop\frac{dj}{2\pi i}x^{-j}\widetilde{D}_{A}^{B}\left(j,\xi\right).\label{eq:Mellin}
\end{equation}

For the region $x\sim1$ this integral can be analytically evaluated.
We'll start with $\widetilde{D}^{val}$ , let's write (\ref{eq:Mellin})
explicitly in this case:

\begin{align}
\widetilde{D}^{val}\left(x,\xi\right) & =\intop\frac{dj}{2\pi i}x^{-j}e^{\overline{\nu}_{0}\xi}\nonumber \\
 & =\intop\frac{dj}{2\pi i}e^{\frac{1}{3}\left(-\left(\frac{6}{j}+\frac{6}{j+1}+12(\psi(j)+\gamma_{E})-17\right)\overline{C}_{F}^{F}-8\overline{C}_{F}^{G}\right)\xi-jln\left(x\right)}.
\end{align}

This integral gets the highest contribution form its saddle point,
we want to prove that this saddle point is at $j\gg1$. In order to
do that we'll find the maximum of the argument (aside from the singularity
at $j=0$). We'll assume that this maximum is achieved at very large
$j$ and then we will see that there is indeed a maximum at this region,
which will justify the assumption. At this region $\frac{6}{j}+\frac{6}{j+1}$
are negligible and we can write $\psi\left(j\right)\approx ln\left(j\right)-\frac{1}{2j}\approx ln\left(j\right)$,
therefore:

\begin{align}
0 & =\frac{d}{dj}\left[\frac{1}{3}\left(-\left(12(ln(j)+\gamma_{E})-17\right)\overline{C}_{F}^{F}-8\overline{C}_{F}^{G}\right)\xi-jln\left(x\right)\right]_{j=j_{0}}\nonumber \\
 & =-4\xi\overline{C}_{F}^{F}\frac{1}{j_{0}}-ln\left(x\right).
\end{align}

So the maximum is at:

\begin{equation}
j_{0}=-\overline{C}_{F}^{F}\frac{4\xi}{ln\left(x\right)}\approx\overline{C}_{F}^{F}\frac{4\xi}{1-x}.
\end{equation}

Since $x\sim1$, $j_{0}$ is very large, and so our assumption that
the maximum is at very large $j$ is justified, but only for $\overline{C}_{F}^{F}>0$
(we'll return to this point later on). If we now assume the major
contribution to this integral comes from the proximity of this saddle
point we can write it as:

\begin{equation}
\widetilde{D}^{val}\left(x,\xi\right)\approx e^{-\xi\left[-\left(\frac{17}{3}-4\gamma_{E}\right)\overline{C}_{F}^{F}+\frac{8}{3}\overline{C}_{F}^{G}\right]}\frac{1}{2\pi i}\int e^{-4\xi\overline{C}_{F}^{F}ln\left(j\right)-jln\left(x\right)}dj.
\end{equation}

Changing $t=jln\left(x\right)$ we get:

\begin{equation}
\widetilde{D}^{val}\left(x,\xi\right)\approx-\frac{1}{2\pi i}\frac{e^{-\xi\left[-\left(\frac{17}{3}-4\gamma_{E}\right)\overline{C}_{F}^{F}+\frac{8}{3}\overline{C}_{F}^{G}\right]}}{ln\left(\frac{1}{x}\right)^{-4\xi\frac{N^{2}-1}{2N}+1}}\int\left(-t\right)^{-4\xi\overline{C}_{F}^{F}}e^{-t}dt.
\end{equation}

Using the integral representation of the gamma function $\frac{-2\pi i}{\Gamma\left(z\right)}=\int\left(-t\right)^{-z}e^{-t}dt$
and $ln\left(x\right)\approx x-1$ we write:

\begin{equation}
\widetilde{D}^{val}\left(x,\xi\right)\overset{x\sim1}{\approx}\frac{e^{-\xi\left[\left(4\gamma_{E}-\frac{17}{3}\right)\overline{C}_{F}^{F}+\frac{8}{3}\overline{C}_{F}^{G}\right]}}{\left(1-x\right)^{1-4\xi\overline{C}_{F}^{F}}\Gamma\left(4\xi\overline{C}_{F}^{F}\right)}.\label{eq:Dval-x-1}
\end{equation}

We can now try to evaluate the other structure functions, in order
to do that we'll evaluate first their components at the region of
$j_{0}$. $j_{0}$ is very large so we'll only keep leading terms
in the expressions for $\Phi_{F}^{G}$ and $\Phi_{G}^{F}$:

\begin{subequations}
\begin{equation}
\Phi_{F}^{G}\left(j_{0}\right)=2\overline{C}_{F}^{G}\frac{1}{j_{0}},
\end{equation}

\begin{equation}
\Phi_{G}^{F}\left(j_{0}\right)=2\overline{C}_{G}^{F}\frac{1}{j_{0}}.
\end{equation}
\end{subequations}

Both are very small and therefore we can approximate using $\sqrt{1+\epsilon}\approx1+\frac{1}{2}\epsilon$
:

\begin{multline}
\sqrt{\left[\overline{\nu}_{F}-\overline{\nu}_{G}\right]^{2}+8n_{f}\Phi_{F}^{G}\Phi_{G}^{F}}=\left(\overline{\nu}_{F}-\overline{\nu}_{G}\right)\sqrt{1+8n_{f}\frac{\left(\frac{2\overline{C}_{F}^{G}}{j_{0}}\right)\left(\frac{2\overline{C}_{G}^{F}}{j_{0}}\right)}{\left(\overline{\nu}_{F}-\overline{\nu}_{G}\right)^{2}}}\\
\approx\left(\overline{\nu}_{F}-\overline{\nu}_{G}\right)\left[1+\frac{1}{2}8n_{f}\frac{\left(\frac{2\overline{C}_{F}^{G}}{j_{0}}\right)\left(\frac{2\overline{C}_{G}^{F}}{j_{0}}\right)}{\left(\overline{\nu}_{F}-\overline{\nu}_{G}\right)^{2}}\right]=\left(\overline{\nu}_{F}-\overline{\nu}_{G}\right)+16n_{f}\frac{\overline{C}_{F}^{G}\overline{C}_{G}^{F}}{\left(\overline{\nu}_{F}-\overline{\nu}_{G}\right)j_{0}^{2}}.
\end{multline}

Which gives:

\begin{subequations}
\begin{equation}
\overline{\nu}_{+}\approx\overline{\nu}_{F}+8n_{f}\overline{C}_{F}^{G}\overline{C}_{G}^{F}\frac{1}{\left(\overline{\nu}_{F}-\overline{\nu}_{G}\right)j_{0}^{2}},
\end{equation}

\begin{equation}
\overline{\nu}_{-}\approx\overline{\nu}_{G}-8n_{f}\overline{C}_{F}^{G}\overline{C}_{G}^{F}\frac{1}{\left(\overline{\nu}_{F}-\overline{\nu}_{G}\right)j_{0}^{2}}.
\end{equation}
\end{subequations}

Define:

\begin{equation}
\Delta_{0}=\overline{\nu}_{F}\left(j_{0}\right)-\overline{\nu}_{G}\left(j_{0}\right).
\end{equation}

Using the approximation:

\begin{equation}
\int djf\left(j\right)e^{Mg\left(j\right)}\approx f\left(j_{0}\right)\int dje^{Mg\left(j\right)}
\end{equation}

for the saddle point $j_{0}$ we can write:

\begin{subequations}
\begin{align}
\frac{1}{2\pi i}\int x^{-j}e^{\xi\overline{\nu}_{G}} & =\frac{1}{2\pi i}\int x^{-j}e^{\xi\left(\overline{\nu}_{G}-\overline{\nu}_{F}\right)}e^{\xi\overline{\nu}_{F}}\approx e^{-\xi\Delta_{0}}D^{val}\left(x,\xi\right),\\
\frac{1}{2\pi i}\int x^{-j}e^{\xi\overline{\nu}_{-}} & \approx e^{-\xi\Delta_{0}-\xi8n_{f}\overline{C}_{F}^{G}\overline{C}_{G}^{F}\frac{1}{\Delta_{0}j_{0}^{2}}}D^{val}\left(x,\xi\right),\\
\frac{1}{2\pi i}\int x^{-j}e^{\xi\overline{\nu}_{+}} & \approx e^{\xi8n_{f}\overline{C}_{F}^{G}\overline{C}_{G}^{F}\frac{1}{\Delta_{0}j_{0}^{2}}}D^{val}\left(x,\xi\right).
\end{align}
\end{subequations}

We can also write, at the saddle point $j_{0}$:

\begin{subequations}
\begin{equation}
\frac{\overline{\nu}_{+}-\overline{\nu}_{0}}{\overline{\nu}_{+}-\overline{\nu}_{-}}\approx\frac{1}{\Delta_{0}+\frac{1}{\Delta_{0}}\frac{16n_{f}\overline{C}_{F}^{G}\overline{C}_{G}^{F}}{j_{0}^{2}}}\frac{1}{\Delta_{0}}4n_{f}\frac{N^{2}-1}{2N}\frac{1}{j_{0}^{2}}=\frac{\frac{1}{\Delta_{0}}\frac{8n_{f}\overline{C}_{F}^{G}\overline{C}_{G}^{F}}{j_{0}^{2}}\left(\Delta_{0}-\frac{1}{\Delta_{0}}\frac{16n_{f}\overline{C}_{F}^{G}\overline{C}_{G}^{F}}{j_{0}^{2}}\right)}{\left(\Delta_{0}\right)^{2}-\left(\frac{1}{\Delta_{0}}\frac{16n_{f}\overline{C}_{F}^{G}\overline{C}_{G}^{F}}{j_{0}^{2}}\right)^{2}},
\end{equation}

\begin{equation}
\frac{\overline{\nu}_{0}-\overline{\nu}_{-}}{\overline{\nu}_{+}-\overline{\nu}_{-}}\approx\frac{\Delta_{0}+\frac{1}{\Delta_{0}}\frac{8n_{f}\overline{C}_{F}^{G}\overline{C}_{G}^{F}}{j_{0}^{2}}}{\Delta_{0}+\frac{1}{\Delta_{0}}\frac{16n_{f}\overline{C}_{F}^{G}\overline{C}_{G}^{F}}{j_{0}^{2}}}=\frac{\left(\Delta_{0}+\frac{1}{\Delta_{0}}\frac{8n_{f}\overline{C}_{F}^{G}\overline{C}_{G}^{F}}{j_{0}^{2}}\right)\left(\Delta_{0}-\frac{1}{\Delta_{0}}\frac{16n_{f}\overline{C}_{F}^{G}\overline{C}_{G}^{F}}{j_{0}^{2}}\right)}{\left(\Delta_{0}\right)^{2}-\left(\frac{1}{\Delta_{0}}\frac{16n_{f}\overline{C}_{F}^{G}\overline{C}_{G}^{F}}{j_{0}^{2}}\right)^{2}}.
\end{equation}
\end{subequations}

Neglecting terms of order $\frac{1}{j_{0}^{4}}$ in both numerator
and denominator we arrive at:

\begin{subequations}
\begin{equation}
\frac{\overline{\nu}_{+}-\overline{\nu}_{0}}{\overline{\nu}_{+}-\overline{\nu}_{-}}\approx\frac{8n_{f}\overline{C}_{F}^{G}\overline{C}_{G}^{F}}{j_{0}^{2}\Delta_{0}^{2}},
\end{equation}

\begin{equation}
\frac{\overline{\nu}_{0}-\overline{\nu}_{-}}{\overline{\nu}_{+}-\overline{\nu}_{-}}\approx\frac{\Delta_{0}^{2}-\frac{8n_{f}\overline{C}_{F}^{G}\overline{C}_{G}^{F}}{j_{0}^{2}}}{\Delta_{0}^{2}}=1-8n_{f}\overline{C}_{F}^{G}\overline{C}_{G}^{F}\frac{1}{\Delta_{0}^{2}j_{0}^{2}}.
\end{equation}
\end{subequations}

Using All these together let us write:

\begin{equation}
\widetilde{D}_{F}^{G}\left(x,\xi\right)\approx\frac{\Phi_{F}^{G}\left(j_{0}\right)}{\overline{\nu}_{+}-\overline{\nu}_{-}}\left[e^{\frac{\xi8n_{f}\overline{C}_{F}^{G}\overline{C}_{G}^{F}}{\Delta_{0}j_{0}^{2}}}\widetilde{D}^{val}\left(x,\xi\right)-e^{-\xi\Delta_{0}-\frac{\xi8n_{f}\overline{C}_{F}^{G}\overline{C}_{G}^{F}}{\Delta_{0}j_{0}^{2}}}\widetilde{D}^{val}\left(x,\xi\right)\right].
\end{equation}

Keeping terms only up to $\frac{1}{j_{0}}$ we have:

\begin{equation}
\widetilde{D}_{F}^{G}\left(x,\xi\right)\approx\frac{2\overline{C}_{F}^{G}}{\Delta_{0}j_{0}}\left[1-e^{-\xi\Delta_{0}}\right]D^{val}\left(x,\xi\right).
\end{equation}

For $\widetilde{D}^{sea}\left(x,\xi\right)$ we'll have to keep terms
up to $\frac{1}{j_{0}^{2}}$ and to keep the equations short write
$\overline{Z}_{0}=\frac{8n_{f}\overline{C}_{F}^{G}\overline{C}_{G}^{F}}{\Delta_{0}^{2}}$:

\begin{align}
D^{sea}\left(x,\xi\right) & \approx\frac{1}{2n_{f}}\left[\left(1-\frac{\overline{Z}_{0}}{j_{0}^{2}}\right)e^{\frac{\xi\Delta_{0}\overline{Z}_{0}}{j_{0}^{2}}}+\frac{\overline{Z}_{0}}{j_{0}^{2}}e^{-\xi\Delta_{0}-\frac{\xi\Delta_{0}\overline{Z}_{0}}{j_{0}^{2}}}-1\right]\widetilde{D}^{val}\left(x,\xi\right)\nonumber \\
 & \approx\frac{1}{2n_{f}}\left[e^{\xi8n_{f}\overline{C}_{F}^{G}\overline{C}_{G}^{F}\frac{1}{\Delta_{0}j_{0}^{2}}}-1+\frac{\overline{Z}_{0}}{j_{0}^{2}}\left(-e^{\frac{\xi\Delta_{0}\overline{Z}_{0}}{j_{0}^{2}}}+e^{-\xi\Delta_{0}-\frac{\xi\Delta_{0}\overline{Z}_{0}}{j_{0}^{2}}}\right)\right]\widetilde{D}^{val}\left(x,\xi\right)\nonumber \\
 & \approx\frac{1}{2n_{f}}\left[\left(e^{\frac{\xi\Delta_{0}\overline{Z}_{0}}{j_{0}^{2}}}-1\right)+\frac{\overline{Z}_{0}}{j_{0}^{2}}\left(-e^{\frac{\xi\Delta_{0}\overline{Z}_{0}}{j_{0}^{2}}}+e^{-\xi\Delta_{0}}\left(e^{-\frac{\xi\Delta_{0}\overline{Z}_{0}}{j_{0}^{2}}}\right)\right)\right]\widetilde{D}^{val}\left(x,\xi\right)\nonumber \\
 & \approx\frac{1}{2n_{f}}\left[\frac{\xi\Delta_{0}\overline{Z}_{0}}{j_{0}^{2}}+\frac{\overline{Z}_{0}}{j_{0}^{2}}\left(-1-\frac{\xi\Delta_{0}\overline{Z}_{0}}{j_{0}^{2}}+e^{-\xi\Delta_{0}}\left(1-\frac{\xi\Delta_{0}\overline{Z}_{0}}{j_{0}^{2}}\right)\right)\right]\widetilde{D}^{val}\left(x,\xi\right)\nonumber \\
 & \approx4\frac{\overline{C}_{F}^{G}\overline{C}_{G}^{F}}{\Delta_{0}^{2}j_{0}^{2}}\left[\Delta_{0}\xi-1+e^{-\xi\Delta_{0}}+\frac{\xi\Delta_{0}\overline{Z}_{0}}{j_{0}^{2}}\left(1-e^{-\xi\Delta_{0}}\right)\right]\widetilde{D}^{val}\left(x,\xi\right).
\end{align}

Neglecting again any $\frac{1}{j_{0}^{4}}$ terms we arrive at:

\begin{equation}
\widetilde{D}^{sea}\left(x,\xi\right)\approx4\frac{\overline{C}_{F}^{G}\overline{C}_{G}^{F}}{\Delta_{0}^{2}j_{0}^{2}}\left[\Delta_{0}\xi-1+e^{-\xi\Delta_{0}}\right]\widetilde{D}^{val}\left(x,\xi\right).
\end{equation}

To compute $\widetilde{D}_{G}^{G}$ and $\widetilde{D}_{F}^{G}$ we
would want to take the saddle point at the value suitable to $\overline{\nu}_{G}$
instead of $\overline{\nu}_{F}$. The change is very easy and we have:

\begin{equation}
j_{G}=\overline{C}_{G}^{G}\frac{4\xi}{1-x},
\end{equation}

\begin{equation}
\frac{1}{2\pi i}\int djx^{-j}e^{\xi\overline{\nu}_{G}}\approx e^{\xi\left[\overline{C}_{G}^{G}\left(\frac{11}{3}-4\gamma_{E}\right)-\frac{4}{3}n_{f}\overline{C}_{G}^{F}\right]}\frac{\left(1-x\right)^{4\xi\overline{C}_{G}^{G}-1}}{\Gamma\left(4\xi\overline{C}_{G}^{G}\right)}:=G\left(x,\xi\right).
\end{equation}

By the same reasoning we now approximate:

\begin{subequations}
\begin{gather}
\frac{1}{2\pi i}\int djx^{-j}e^{\xi\overline{\nu}_{F}}\approx e^{\Delta_{0}\xi}G\left(x,\xi\right),\\
\overline{\nu}_{+}\approx\overline{\nu}_{F}+\frac{8n_{f}\overline{C}_{F}^{G}\overline{C}_{G}^{F}}{\Delta_{0}j_{G}^{2}},\\
\overline{\nu}_{-}\approx\overline{\nu}_{G}-\frac{8n_{f}\overline{C}_{F}^{G}\overline{C}_{G}^{F}}{\Delta_{0}j_{G}^{2}}.
\end{gather}
\end{subequations}

And so on. We can therefore compute:

\begin{align}
\widetilde{D}_{G}^{G}\left(x,\xi\right) & \approx\frac{8n_{f}\overline{C}_{F}^{G}\overline{C}_{G}^{F}}{\Delta_{0}^{2}j_{0}^{2}}e^{\Delta_{0}\xi+\frac{\xi8n_{f}\overline{C}_{F}^{G}\overline{C}_{G}^{F}}{\Delta_{0}j_{G}^{2}}}G\left(x,\xi\right)+\left(1-\frac{8n_{f}\overline{C}_{F}^{G}\overline{C}_{G}^{F}}{\Delta_{0}^{2}j_{0}^{2}}\right)e^{-\frac{\xi8n_{f}\overline{C}_{F}^{G}\overline{C}_{G}^{F}}{\Delta_{0}j_{G}^{2}}}G\left(x,\xi\right)\nonumber \\
 & \approx\left[\frac{8n_{f}\overline{C}_{F}^{G}\overline{C}_{G}^{F}}{\Delta_{0}^{2}j_{0}^{2}}\left(1+\Delta_{0}\xi+\frac{\xi8n_{f}\overline{C}_{F}^{G}\overline{C}_{G}^{F}}{\Delta_{0}j_{G}^{2}}\right)+\left(1-\frac{8n_{f}\overline{C}_{F}^{G}\overline{C}_{G}^{F}}{\Delta_{0}^{2}j_{0}^{2}}\right)\left(1-\frac{\xi8n_{f}\overline{C}_{F}^{G}\overline{C}_{G}^{F}}{\Delta_{0}j_{G}^{2}}\right)\right]G\left(x,\xi\right).
\end{align}

The leading term in this solution is clearly:

\begin{equation}
\widetilde{D}_{G}^{G}\left(x,\xi\right)\approx G\left(x,\xi\right).
\end{equation}

Which is analog to $D^{val}$ for quark. We also have:

\begin{align}
\widetilde{D}_{G}^{F} & \approx\frac{2\overline{C}_{G}^{F}}{j_{G}\Delta_{0}}\left[e^{\Delta_{0}\xi+\frac{\xi8n_{f}\overline{C}_{F}^{G}\overline{C}_{G}^{F}}{\Delta_{0}j_{G}^{2}}}D_{G}^{G}\left(x,\xi\right)-e^{-\frac{\xi8n_{f}\overline{C}_{F}^{G}\overline{C}_{G}^{F}}{\Delta_{0}j_{G}^{2}}}D_{G}^{G}\left(x,\xi\right)\right]\nonumber \\
 & \approx\frac{2\overline{C}_{G}^{F}}{j_{G}\Delta_{0}}\left[e^{\Delta_{0}\xi}\left(1+\frac{\xi8n_{f}\overline{C}_{F}^{G}\overline{C}_{G}^{F}}{\Delta_{0}j_{G}^{2}}\right)-\left(1-\frac{\xi8n_{f}\overline{C}_{F}^{G}\overline{C}_{G}^{F}}{\Delta_{0}j_{G}^{2}}\right)\right]D_{G}^{G}\left(x,\xi\right)\nonumber \\
 & \approx\frac{2\overline{C}_{G}^{F}}{j_{G}\Delta_{0}}\left[e^{\Delta_{0}\xi}-1\right]\widetilde{D}_{G}^{G}\left(x,\xi\right).
\end{align}

Here we neglected any $\frac{1}{j_{G}^{2}}$ terms. To summarize:

\begin{subequations}
\begin{align}
\widetilde{D}^{val}\left(x,\xi\right) & \overset{x\sim1}{\approx}\widetilde{D}_{F}^{F}\left(x,\xi\right)\approx\frac{e^{-\xi\left[\left(4\gamma_{E}-\frac{17}{3}\right)\overline{C}_{F}^{F}+\frac{8}{3}\overline{C}_{F}^{G}\right]}}{\left(1-x\right)^{1-4\xi\overline{C}_{F}^{F}}\Gamma\left(4\xi\overline{C}_{F}^{F}\right)},\\
\widetilde{D}^{sea}\left(x,\xi\right) & \approx4\frac{\overline{C}_{F}^{G}\overline{C}_{G}^{F}}{\Delta_{0}^{2}j_{0}^{2}}\left[\Delta_{0}\xi-1+e^{-\xi\Delta_{0}}\right]\widetilde{D}^{val}\left(x,\xi\right),\\
\widetilde{D}_{F}^{G}\left(x,\xi\right) & \approx\frac{2\overline{C}_{F}^{G}}{\Delta_{0}j_{0}}\left[1-e^{-\xi\Delta_{0}}\right]D^{val}\left(x,\xi\right),\\
\widetilde{D}_{G}^{G}\left(x,\xi\right) & \approx e^{\xi\left[\overline{C}_{G}^{G}\left(\frac{11}{3}-4\gamma_{E}\right)-\frac{4}{3}n_{f}\overline{C}_{G}^{F}\right]}\frac{\left(1-x\right)^{4\xi\overline{C}_{G}^{G}-1}}{\Gamma\left(4\xi\overline{C}_{G}^{G}\right)},\\
\widetilde{D}_{G}^{F} & \approx\frac{2\overline{C}_{G}^{F}}{j_{G}\Delta_{0}}\left[e^{\Delta_{0}\xi}-1\right]\widetilde{D}_{G}^{G}\left(x,\xi\right).
\end{align}
\end{subequations}

At $x\rightarrow1$ $j_{0},j_{G}\propto\frac{1}{1-x}$ and $\Delta_{0}\propto ln\left(1-x\right)$
and therefore we get (\ref{eq:Dval-x-1}).

\section{Regularizing Divergent Integrals\label{sec:Regularizing-Divergent-Integrals}}

In section \ref{subsec:Regularization} we have introduced the regularization:

\begin{equation}
\int_{0}^{1}dx\frac{f\left(x\right)}{x^{\lambda}}:=\int_{0}^{1}\frac{f\left(x\right)-f\left(0\right)}{x^{\lambda}}dx+\frac{f\left(0\right)}{1-\lambda}.\label{eq:1d reg}
\end{equation}

It's natural to ask whether regularization is well defined, particularly
if we change the range of integration. We'll prove in the 1 dimensional
case that this regularization is well defined and get the same value
for every range of integration that contains all the values for which
$f\left(x\right)\neq0$. Let $f\left(x\right)=\theta\left(x-x_{0}\right)\widetilde{f}\left(x\right)$
be a function that is smooth and bounded in the region $\left[0,x_{0}\right)$,
we need to understand how to generalize (\ref{eq:1d reg}) to a general
integration interval. The generalization of the first term on the
r.h.s is trivial: just change the integration to the new region. The
generalization of the second term is a bit less trivial, this term
is defined as the analytical continuation of the integral 
\begin{equation}
\int_{0}^{1}dx\frac{f\left(0\right)}{x^{\lambda}}=\frac{f\left(0\right)}{1-\lambda}
\end{equation}
for $\lambda<1$. When we change the integration limit we then get
\begin{equation}
\int_{0}^{y}dx\frac{f\left(0\right)}{x^{\lambda}}=y^{1-\lambda}\frac{f\left(0\right)}{1-\lambda},
\end{equation}
so we can define for $y>x_{0}$:

\begin{equation}
I_{y}=\int_{0}^{y}dx\frac{f\left(x\right)}{x^{\lambda}}:=\int_{0}^{y}\frac{f\left(x\right)-f\left(0\right)}{x^{\lambda}}dx+y^{1-\lambda}\frac{f\left(0\right)}{1-\lambda}.
\end{equation}

Note that we changed the integration limit in the first term and the
coefficient in the second term, as discussed above. Now we need to
prove that $I_{y}=I_{x_{0}}$ for $1<\lambda<2$ (for $\lambda<1$
this statement is trivial). Write:

\begin{equation}
I_{y}=\int_{0}^{x_{0}}\frac{f\left(x\right)-f\left(0\right)}{x^{\lambda}}dx+\int_{x_{0}}^{y}\frac{f\left(x\right)-f\left(0\right)}{x^{\lambda}}dx+y^{1-\lambda}\frac{f\left(0\right)}{1-\lambda},
\end{equation}

using the fact that $f\left(x\right)=0$ for $x_{0}<x<y$ we write
this equation as:

\begin{equation}
I_{y}=\int_{0}^{x_{0}}\frac{f\left(x\right)-f\left(0\right)}{x^{\lambda}}dx-\int_{x_{0}}^{y}\frac{f\left(0\right)}{x^{\lambda}}+y^{1-\lambda}\frac{f\left(0\right)}{1-\lambda}.
\end{equation}

The integral in the middle term converges because $x_{0}>0$ so we
can write it as:

\begin{alignat}{1}
I_{y} & =\int_{0}^{x_{0}}\frac{f\left(x\right)-f\left(0\right)}{x^{\lambda}}dx-y^{1-\lambda}\frac{f\left(0\right)}{1-\lambda}+x_{0}^{1-\lambda}\frac{f\left(0\right)}{1-\lambda}+y^{1-\lambda}\frac{f\left(0\right)}{1-\lambda}\\
 & =\int_{0}^{x_{0}}\frac{f\left(x\right)-f\left(0\right)}{x^{\lambda}}dx+x_{0}^{1-\lambda}\frac{f\left(0\right)}{1-\lambda}=I_{x_{0}}.
\end{alignat}

As was needed. This computation proves the 1 dimensional case of this
regularization is well defined, note however we needed to change both
terms of (\ref{eq:1d reg}) and not only the integration limit of
the first term. In the 2 dimensional case, the proof is harder because
integration regions in $2d$ can be much more complex, we'll therefore
only see an example of this phenomenon. The definition of $F\left(z_{1},z_{2}\right)$
in (\ref{eq:F(z1,z2)}) contain the hadron structure function $G_{h}^{A^{\prime}}\left(\frac{x_{1}}{z_{1}}+\frac{x_{2}}{z_{2}};k^{2}\right)$
which satisfy $G_{h}^{A^{\prime}}\left(\frac{x_{1}}{z_{1}}+\frac{x_{2}}{z_{2}};k^{2}\right)\equiv0$
for $\frac{x_{1}}{z_{1}}+\frac{x_{2}}{z_{2}}>1$. We could therefore
take the integrals in (\ref{eq:regularization}) with this condition
instead of over the entire region $\left[x_{1},1\right]\times\left[x_{2},1\right]$.
This condition would mean to write the four parts of this integral
as:

\begin{equation}
\widetilde{I}=\underset{\frac{x_{1}}{z_{1}}+\frac{x_{2}}{z_{2}}>1}{\intop\intop}dz_{1}dz_{2}\frac{F\left(z_{1},z_{2}\right)}{\left(1-z_{1}\right)^{1-g_{1}}\left(1-z_{2}\right)^{1-g_{2}}},\label{eq:regularization-1}
\end{equation}

\begin{subequations}
\begin{equation}
\widetilde{I}=\widetilde{I}_{A}+\widetilde{I}_{B}+\widetilde{I}_{C}+\widetilde{I}_{D},
\end{equation}

\begin{align}
\widetilde{I}_{A} & =\underset{\frac{x_{1}}{z_{1}}+\frac{x_{2}}{z_{2}}>1}{\intop\intop}dz_{1}dz_{2}\frac{F\left(z_{1},z_{2}\right)-F\left(1,z_{2}\right)-F\left(z_{1},1\right)+F\left(1,1\right)}{\left(1-z_{1}\right)^{1-g_{1}}\left(1-z_{2}\right)^{1-g_{2}}},\\
\widetilde{I}_{B} & =\underset{\frac{x_{1}}{z_{1}}+\frac{x_{2}}{z_{2}}>1}{\intop\intop}dz_{1}dz_{2}\frac{F\left(z_{1},1\right)-F\left(1,1\right)}{\left(1-z_{1}\right)^{1-g_{1}}\left(1-z_{2}\right)^{1-g_{2}}},\nonumber \\
 & =\intop_{x_{1}}^{1}dz_{1}\frac{F\left(z_{1},1\right)-F\left(1,1\right)}{\left(1-z_{1}\right)^{1-g_{1}}}\intop_{\frac{x_{2}}{1-\frac{x_{1}}{z_{1}}}}^{1}dz_{2}\frac{1}{\left(1-z_{2}\right)^{1-g_{2}}}\nonumber \\
 & =\intop_{x_{1}}^{1}dz_{1}\frac{F\left(z_{1},1\right)-F\left(1,1\right)}{\left(1-z_{1}\right)^{1-g_{1}}}C\left(x_{1},x_{2},z_{1},g_{2}\right)\\
\widetilde{I}_{C} & =\underset{\frac{x_{1}}{z_{1}}+\frac{x_{2}}{z_{2}}>1}{\intop\intop}dz_{1}dz_{2}\frac{F\left(1,z_{2}\right)-F\left(1,1\right)}{\left(1-z_{1}\right)^{1-g_{1}}\left(1-z_{2}\right)^{1-g_{2}}},\nonumber \\
 & =\intop_{x_{2}}^{1}dz_{2}\frac{F\left(1,z_{2}\right)-F\left(1,1\right)}{\left(1-z_{2}\right)^{1-g_{2}}}dz_{2}\intop_{\frac{x_{1}}{1-\frac{x_{2}}{z_{2}}}}^{1}dz_{1}\frac{1}{\left(1-z_{1}\right)^{1-g_{1}}}\nonumber \\
 & =\intop_{x_{2}}^{1}dz_{2}\frac{F\left(1,z_{2}\right)-F\left(1,1\right)}{\left(1-z_{2}\right)^{1-g_{2}}}C\left(x_{2},x_{1},z_{2},g_{1}\right)\\
\widetilde{I}_{D} & =\underset{\frac{x_{1}}{z_{1}}+\frac{x_{2}}{z_{2}}>1}{\intop\intop}dz_{1}dz_{2}\frac{F\left(1,1\right)}{\left(1-z_{1}\right)^{1-g_{1}}\left(1-z_{2}\right)^{1-g_{2}}}=D\left(x_{1},x_{2},g_{1},g_{2}\right)F\left(1,1\right).
\end{align}
\end{subequations}

Where $C$ and $D$ are the analytic solutions of these integrals,
which has a complex form as a combination of hypergeometric functions
that can be extended to every value of $g_{1}$ and $g_{2}$. Although
this expression is much more complex than (\ref{eq:regularization})
numerical calculations show it's actually the same.

\section{Rules for Color Projectors\label{sec:Rules-for-Color}}

\subsection{Exact Form of Projectors}

We define the usual color projectors for two gluons into irreducible
representations which commonly appear in the literature (see for example
\citep{Mekhfi1988,Cvitanovic2011,Ioffe:2010zz}). We use however the
notations of \citep{Dokshitzer2022,Dokshitzer2005} both for the names
of the representations $\alpha=\left\{ 8_{a},10,1,8_{s},27,0\right\} $
(the order is important) and for their definitions. The names are
of the ``$SU\left(3\right)$'' dimensions (even though in the paper
we keep the general $SU\left(N\right)$ representations). Also as
in \citep{Dokshitzer2022,Dokshitzer2005} $10$ is actually the (direct)
sum of the two irreducible representations $\overline{10}+10$. In
the following, we sum over repeated indices both representation indices
(marked in Greek letters) and color (marked in small Latin characters).

\begin{subequations}
\label{eq: explicit forms}

\begin{align}
P_{ab;cd}^{1}= & \frac{1}{N^{2}-1}\delta_{ab}\delta_{cd},\\
P_{ab;cd}^{8_{a}}= & -\frac{f^{aeb}f^{cde}}{N},\\
P_{ab;cd}^{8_{s}}= & \frac{N}{N^{2}-4}d_{abe}d_{cde},\\
P_{ab;cd}^{10}= & \frac{1}{2}\left(\mathfrak{1}_{ab;cd}-X_{ab;cd}\right)-P_{ab;cd}^{8_{a}},\\
P_{ab;cd}^{27}= & \frac{1}{4}\left(\mathfrak{1}_{ab;cd}+X_{ab;cd}\right)-\frac{N-2}{2N}P_{ab;cd}^{8_{s}}-\frac{N-1}{2N}P_{ab;cd}^{1}+\left(W_{ab;cd}^{+}+W_{ab;cd}^{-}\right),\\
P_{ab;cd}^{0}= & \frac{1}{4}\left(\mathfrak{1}_{ab;cd}+X_{ab;cd}\right)-\frac{N+2}{2N}P_{ab;cd}^{8_{s}}-\frac{N+1}{2N}P_{ab;cd}^{1}+\left(W_{ab;cd}^{+}+W_{ab;cd}^{-}\right).
\end{align}
\end{subequations}

Where $N$ is the number of colors in $SU\left(N\right)$ and:

\begin{subequations}
\begin{align}
\mathfrak{1}_{ab;cd}= & \delta_{ac}\delta_{bd},\\
X_{ab;cd}= & \delta_{ad}\delta_{bc},\\
W_{ab;cd}^{-}= & Tr\left(t^{b}t^{c}t^{a}t^{d}\right),\label{eq:W+-}\\
W_{ab;cd}^{+}= & Tr\left(t^{b}t^{d}t^{a}t^{c}\right).
\end{align}
\end{subequations}

\subsection{Properties}

We list here some of the properties of these projectors we have used
in the body of the text. Unless otherwise noted they are cited from
\citep{Dokshitzer2005} where we direct to the equations in that paper.

\subsubsection{Projectors}

The above defined are projectors (eq. $A.17$), i.e:

\[
P_{ab;cd}^{\alpha}P_{cd;ef}^{\beta}=\delta_{\alpha\beta}P_{ab;ef}^{\alpha}.
\]

\subsubsection{Symmetries}

Each representation is either symmetric or anti-symmetric with respect to interchanging
incoming (outgoing) gluons. this property means that:

\begin{equation}
P_{ab;cd}^{\alpha}=r_{\alpha}P_{ba;cd}^{\alpha}=r_{\alpha}P_{ab;dc}^{\alpha}.
\end{equation}

Here $r_{\alpha}=\left\{ -1,-1,1,1,1,1\right\} $ depend on the representation.
Since each projector is either symmetric or anti-symmetric if we interchange
both its ``incoming'' (first two) or ``outgoing'' (last two)
indices it'll remain the same (anti-symmetric projectors acquire a
$\left(-1\right)^{2}=1$ factor):

\begin{equation}
P_{ab;cd}^{\alpha}=P_{ba;dc}^{\alpha}.
\end{equation}

Also, all of the projectors are symmetric in incoming and outgoing
indices (proof in section \ref{subsec:Proof-of-Incoming-Outgoing}):

\begin{equation}
P_{ab;cd}^{\alpha}=P_{cd;ab}^{\alpha}.
\end{equation}

\subsubsection{Change in Basis}

We could also work in $t$ or $u$-channel projectors which are simply:

\begin{equation}
P_{ab;cd}=P_{ac;bd}^{\left(t\right)}=P_{ac;db}^{\left(u\right)},
\end{equation}

graphically seen in figure \ref{fig:stu}.We can use the so called
``re-projection'' matrix (eq $2.31$) to find that:

\begin{figure}
\[
\xymatrix{{a}\ar@{~}[ddr] &  &  & {c} &  & {a}\ar@{~}[dr] &  & {b} &  & {a}\ar@{~}[dr] &  & {d}\\
\xyR{1pc}\xyC{1pc} &  &  &  &  &  & *=0{}\ar@{~}[dd]\ar@{~}[ur] &  &  &  & *=0{}\ar@{~}[dd]\ar@{~}[dddr]\\
 & *=0{}\ar@{~}[r] & *=0{}\ar@{~}[uur]\ar@{~}[ddr] &  & {=} &  &  &  & {=}\\
 &  &  &  &  &  & *=0{}\ar@{~}[dr] &  &  &  & *=0{}\ar@{~}[uuur]\\
{b}\ar@{~}[uur] &  &  & {d} &  & {c}\ar@{~}[ur] &  & {d} &  & {c}\ar@{~}[ur] &  & {b}
}
\]

\caption{$s,t,u$ channels \label{fig:stu}}
\end{figure}
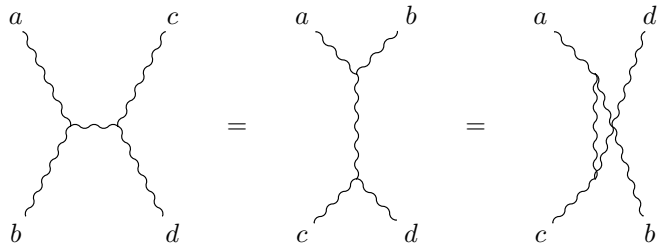

\begin{equation}
P_{ab;cd}^{\alpha}=\left[K_{ts}\right]^{\alpha\beta}P_{ac;bd}^{\beta},
\end{equation}

\begin{equation}
P_{ab;cd}^{\alpha}=\left[K_{su}\right]^{\alpha\beta}P_{ad;bc}^{\beta},
\end{equation}

\begin{equation}
P_{ab;cd}^{\alpha}=\left[K_{us}\right]^{\alpha\beta}P_{ac;db}^{\beta}.
\end{equation}

From these equations we can see that $K_{ts}=K_{st}=K_{ts}^{-1}$
and that $K_{su}^{2}=K_{us}=K_{su}^{-1}$. The explicit form of $K_{ts}$,
$K_{su}$ and $K_{us}$ is given in \citep{Dokshitzer2005} eq $A.28$
and the paragraph thereafter:

\begin{equation}
K_{ts}=\left(\begin{array}{cccccc}
\frac{1}{2} & 0 & 1 & \frac{1}{2} & -\frac{1}{N} & \frac{1}{N}\\
0 & \frac{1}{2} & \frac{1}{2}\left(N^{2}-4\right) & -1 & \frac{2-N}{2N} & \frac{-N-2}{2N}\\
\frac{1}{N^{2}-1} & \frac{1}{N^{2}-1} & \frac{1}{N^{2}-1} & \frac{1}{N^{2}-1} & \frac{1}{N^{2}-1} & \frac{1}{N^{2}-1}\\
\frac{1}{2} & -\frac{2}{N^{2}-4} & 1 & \frac{N^{2}-12}{2N^{2}-8} & \frac{1}{N+2} & -\frac{1}{N-2}\\
-\frac{N(N+3)}{4(N+1)} & -\frac{N(N+3)}{4(N+1)(N+2)} & \frac{N^{2}(N+3)}{4(N+1)} & \frac{N^{2}(N+3)}{4(N+1)(N+2)} & \frac{N^{2}+N+2}{4(N+1)(N+2)} & \frac{N+3}{4(N+1)}\\
\frac{(N-3)N}{4(N-1)} & -\frac{(N-3)N}{4(N-2)(N-1)} & \frac{(N-3)N^{2}}{4(N-1)} & -\frac{(N-3)N^{2}}{4(N-2)(N-1)} & \frac{N-3}{4(N-1)} & \frac{N^{2}-N+2}{4(N-2)(N-1)}
\end{array}\right),
\end{equation}

\begin{equation}
K_{su}=\left(\begin{array}{cccccc}
-\frac{1}{2} & 0 & -1 & -\frac{1}{2} & \frac{1}{N} & -\frac{1}{N}\\
0 & -\frac{1}{2} & \frac{1}{2}\left(4-N^{2}\right) & 1 & \frac{N-2}{2N} & \frac{N+2}{2N}\\
\frac{1}{N^{2}-1} & \frac{1}{N^{2}-1} & \frac{1}{N^{2}-1} & \frac{1}{N^{2}-1} & \frac{1}{N^{2}-1} & \frac{1}{N^{2}-1}\\
\frac{1}{2} & -\frac{2}{N^{2}-4} & 1 & \frac{N^{2}-12}{2N^{2}-8} & \frac{1}{N+2} & -\frac{1}{N-2}\\
-\frac{N(N+3)}{4(N+1)} & -\frac{N(N+3)}{4(N+1)(N+2)} & \frac{N^{2}(N+3)}{4(N+1)} & \frac{N^{2}(N+3)}{4(N+1)(N+2)} & \frac{N^{2}+N+2}{4(N+1)(N+2)} & \frac{N+3}{4(N+1)}\\
\frac{(N-3)N}{4(N-1)} & -\frac{(N-3)N}{4(N-2)(N-1)} & \frac{(N-3)N^{2}}{4(N-1)} & -\frac{(N-3)N^{2}}{4(N-2)(N-1)} & \frac{N-3}{4(N-1)} & \frac{N^{2}-N+2}{4(N-2)(N-1)}
\end{array}\right),
\end{equation}

\begin{equation}
K_{us}=\left(\begin{array}{cccccc}
-\frac{1}{2} & 0 & 1 & \frac{1}{2} & -\frac{1}{N} & \frac{1}{N}\\
0 & -\frac{1}{2} & \frac{1}{2}\left(N^{2}-4\right) & -1 & \frac{2-N}{2N} & \frac{-N-2}{2N}\\
-\frac{1}{N^{2}-1} & -\frac{1}{N^{2}-1} & \frac{1}{N^{2}-1} & \frac{1}{N^{2}-1} & \frac{1}{N^{2}-1} & \frac{1}{N^{2}-1}\\
-\frac{1}{2} & \frac{2}{N^{2}-4} & 1 & \frac{N^{2}-12}{2N^{2}-8} & \frac{1}{N+2} & -\frac{1}{N-2}\\
\frac{N(N+3)}{4(N+1)} & \frac{N(N+3)}{4(N+1)(N+2)} & \frac{N^{2}(N+3)}{4(N+1)} & \frac{N^{2}(N+3)}{4(N+1)(N+2)} & \frac{N^{2}+N+2}{4(N+1)(N+2)} & \frac{N+3}{4(N+1)}\\
-\frac{(N-3)N}{4(N-1)} & \frac{(N-3)N}{4(N-2)(N-1)} & \frac{(N-3)N^{2}}{4(N-1)} & -\frac{(N-3)N^{2}}{4(N-2)(N-1)} & \frac{N-3}{4(N-1)} & \frac{N^{2}-N+2}{4(N-2)(N-1)}
\end{array}\right).
\end{equation}

Which indeed obey the rules above.

\subsubsection{Interaction Force}

Suppose we connect the two incoming gluons with another gluon. This
gluon has the form shown in figure \ref{fig:interaction force}, therefor:

\begin{figure}
\[
\xymatrix{ & {a}\ar@{~>}[dr] &  & {c}\\
 &  & *=0{}\ar@{~>}[dd]\ar@{~>}[ur]\\
{V_{ab;cd}=} &  &  &  & {=NP_{ac;bd}^{\left(t\right)8_{s}}=N\left[K_{ts}\right]^{8_{s}\alpha}P_{ab,cd}^{\alpha}}\\
 &  & *=0{}\ar@{~>}[dr]\\
 & {b}\ar@{~>}[ur] &  & {d}
}
\]

\caption{Interaction force \label{fig:interaction force}}
\end{figure}

\begin{align}
f^{aea^{\prime}}f^{beb^{\prime}}P_{ab;cd}^{\alpha} & =V_{a^{\prime}b^{\prime};ab}P_{ab;cd}^{\alpha}\nonumber \\
 & =N\left[K_{ts}\right]^{8_{s}\alpha}P_{a^{\prime}b^{\prime};ab}^{\alpha}P_{ab;cd}^{\alpha}\nonumber \\
 & =N\left[K_{ts}\right]^{8_{s}\alpha}P_{a^{\prime}b^{\prime};cd}^{\alpha}\nonumber \\
 & =c_{\alpha}P_{a^{\prime}b^{\prime};cd}^{\alpha},
\end{align}

for $c_{\alpha}=\left\{ \frac{N}{2},0,N,\frac{N}{2},-1,1\right\} $.

\subsubsection{Dimensions of the Representation}

By contracting two indices of the projector:

\begin{equation}
P_{ab;ab}^{\alpha}=K^{\alpha}.
\end{equation}

Here (eq. $3.3$): 
\[
K_{\alpha}=\left\{ N^{2}-1,\frac{1}{2}\left(N^{2}-4\right)\left(N^{2}-1\right),1,N^{2}-1,\frac{1}{4}(N-1)N^{2}(N+3),\frac{1}{4}(N-3)N^{2}(N+1)\right\} 
\]
is the dimension of the representation. This relation has the consequence:

\begin{equation}
P_{ab;cb}^{\alpha}=\delta_{ac}\frac{K^{\alpha}}{N^{2}-1}.
\end{equation}

\subsubsection{Completeness Relation }

We have that (eq. $3.5$):

\begin{equation}
\mathfrak{1}=\delta_{ac}\delta_{bd}=\sum_{\alpha}P_{ab;cd}^{\alpha}.
\end{equation}

\subsection{Proof of Incoming-Outgoing Symmetry \label{subsec:Proof-of-Incoming-Outgoing}}

We haven't found proof of this relation in the literature. Although
the proof is easy we'll write it here. We'll prove this relation for
every representation individually using the explicit forms in (\ref{eq: explicit forms}):

\paragraph{$1$}

It's trivial $\frac{1}{N^{2}-1}\delta_{ab}\delta_{cd}=\frac{1}{N^{2}-1}\delta_{cd}\delta_{ab}$.

\paragraph{$8_{s}$}

It's trivial $\frac{N}{N^{2}-4}d_{abe}d_{cde}=\frac{N}{N^{2}-4}d_{cde}d_{abe}$.

\paragraph{$8_{a}$}

It's somewhat less trivial but still easy to see that $-\frac{f^{aeb}f^{cde}}{N}=-\frac{f^{cebd}f^{abe}}{N}$.
Note that we change the order in both form factors so the total sign
is kept $\left(-1\right)^{2}=1$.

\paragraph{$10$}

We note that $\mathfrak{1}$ and $X$ have this symmetry. Then using
the explicit form the symmetry is easy to see since it's true to each
part. It should be noted that this symmetry is not true for either
$\overline{10},10$ alone, but is true for the sum $\overline{10}+10$.

\paragraph{$27/0$}

This symmetry gives:

\begin{equation}
W_{+}=Tr\left(t^{b}t^{d}t^{a}t^{c}\right)\rightarrow Tr\left(t^{b}t^{c}t^{a}t^{d}\right)=W_{-}
\end{equation}

and therefore $W_{+}+W_{-}$ is symmetric. We see that both $P^{0}$
and $P^{27}$ are symmetric using their explicit forms (as they can
be written as a combination of $P^{1},P^{8_{s}},\mathfrak{1},X$,
and $W_{+}+W_{-}$).

\section{Comments on Different Formalisms\label{sec:formalisms}}
\par In this Appendix we shall make  comments on the relation of the formalism
used in this paper for the calculation of $_2GPD$, and the formalism used in \cite{Manohar2012,Diehl2016}

\subsection{The Sudakov Suppression Factor.}
 \par First, consider the calculation of Sudakov formfactor using DDT procedure.
The anomalous dimensions
 in equation (\ref{eq:Modified-DGLAP}), that  include LO and NLO transverse logarithms,
exactly coincide with the logarithms corresponding to the anomalous dimensions $\gamma_\mu$ in \cite{Manohar2012,Diehl2016} 
(see e.g. Eq.  (11) in \cite{Manohar2012} for octet channel and extension to other channels in \cite{Diehl2016,buffing2021}). 
By LO we mean the double logarithmic  term first found in  \cite{Mekhfi1988}, and NLO means single transverse logarithms.
\par Let us consider the issue of  next to leading  transverse logarithms in Sudakov formfactor in more detail. 
We shall see that  the full answer coming from DGLAP equations is universal,
both for leading (LO) and next to leading (NLO) logarithms. Also the double logarithmic (LO) piece in Sudakov formfactor is universal.
On the other hand, the distribution between single logarithms in DGLAP like equation and Sudakov formfactor may depend on regularisation scheme. 
\par Recall that the full evolution equation  has the form (\ref{eq:Modified-DGLAP})  and  its solution  in general can be written in 
the factorized form (\ref{eq:Sudakov split}):
\begin{equation} \bar D^B_A(x,k^2,Q^2)=S^B_A(k^2,Q^2)\cdot \tilde D^B_A(x,k^2,Q^2)
\end{equation}
The Sudakov formfactor contains, for general  channel double and single transverse logs, but no dependence  on longitudinal
variable $x$, while the function $\tilde D^B_A(x,k^2,Q^2)$ contains only single logarithms and satisfies DGLAP like integrodifferential equation. In this paper we used the DDT regularisation scheme with the $z=1$ singularity regularised by
\begin{equation}
\frac{1}{1-z+\Delta},\  \Delta=k^2/Q^2,
\label{ewq}
\end{equation}
while it is conventional to use Altarelli-Parisi kernels where the regularisation in the $ z=1$ limit is given by + prescription
and delta functions, i.e. 
\begin{equation}
P_{qq}(z)=\frac{2z}{(1-z)_+}+1-z-\frac{3}{2}\delta(1-z)
\end{equation}
Here we shall call  this scheme AP scheme to distinguish from DDT regularisation. We shall show that both schemes lead to the same  full evolution equation
for full function $\bar D$, but the natural split between Sudakov formfactor  S and function $\tilde D$  depends on the scheme,
although the product of course remains the same.
\par To illustrate this point we shall consider here the equation for fermion ladder $D_F^F$. There are two possibilities: the singlet and the octet channel. Consider first the singlet channel.
\subsubsection{Proof of the Singlet Case}

First we'll prove the equivalence of DGLAP evolution equations in
the singlet case, between ``DDT formalism'' which we used in the
text and the perhaps more common Altarelli--Parisi formalism. The
evolution equation in \cite{Manohar2012} is:

\begin{equation}
\frac{\partial D_{F}^{F}\left(x,k^{2},Q^{2}\right)}{\partial ln\left(Q\right)}=\frac{\alpha_{s}}{\pi}\intop_{x}^{1}\frac{dz}{z}C_{F}^{F}P_{qq}\left(z\right)D_{F}^{F}\left(\frac{x}{z},Q^{2}\right)
\end{equation}
where $P_{qq}$ is the corresponding AP kernel, so that the equation has the form:
\begin{equation}
\frac{\partial D_{F}^{F}\left(x\right)}{\partial ln\left(Q^{2}\right)}=\frac{\alpha_{s}}{2\pi}\intop_{x}^{1}\frac{dz}{z}C_{F}^{F}\left[\frac{1+z^{2}}{\left(1-z\right)_{+}}+\frac{3}{2}\delta\left(1-z\right)\right]D_{F}^{F}\left(\frac{x}{z}\right)=
\end{equation}

\[
=\frac{\alpha_{s}}{2\pi}\intop_{0}^{1}\frac{dz}{z}\left\{ C_{F}^{F}\left[\frac{1+z^{2}}{1-z}\right]D_{F}^{F}\left(\frac{x}{z}\right)-zC_{F}^{F}\left(\frac{2}{1-z}-\frac{3}{2}\right)D_{F}^{F}\left(x\right)\right\} 
\]
where we explicitly carried the + substraction.
On the other hand the corresponding equation  in DDT scheme
 \cite{Dokshitzer1980}, reads:

\begin{equation}
\frac{\partial D_{F}^{F}\left(x\right)}{\partial ln\left(Q^{2}\right)}=\frac{\alpha_{s}}{4\pi}\intop_{0}^{1}\frac{dz}{z}\left\{ \Phi_{F}^{F}\left(z\right)D_{F}^{F}\left(\frac{x}{z}\right)-z^{2}D_{F}^{F}\left(x\right)\left(\Phi_{F}^{F}\left(z\right)+\Phi_{F}^{G}\left(z\right)\right)\right\} =
\end{equation}

\[
=\frac{\alpha_{s}}{4\pi}\intop_{0}^{1}\frac{dz}{z}\left\{ C_{F}^{F}\left[2\cdot\frac{1+z^{2}}{1-z+\Delta}\right]D_{F}^{F}\left(\frac{x}{z}\right)-2\cdot z^{2}C_{F}^{F}\left(\frac{1+z^{2}}{1-z+\Delta}+\frac{(1-z)^{2}+1}{z}\right)D_{F}^{F}\left(x\right)\right\} =
\]

\[
=\frac{\alpha_{s}}{2\pi}\intop_{0}^{1}\frac{dz}{z}\left\{ C_{F}^{F}\left[\frac{1+z^{2}}{1-z+\Delta}\right]D_{F}^{F}\left(\frac{x}{z}\right)-zC_{F}^{F}\left(\frac{2}{1-z+\Delta}-\frac{3}{2}-3z+\frac{3}{2}\right)D_{F}^{F}\left(x\right)\right\} =
\]

\[
=\frac{\alpha_{s}}{2\pi}\intop_{0}^{1}\frac{dz}{z}\left\{ C_{F}^{F}\left[\frac{1+z^{2}}{1-z+\Delta}\right]D_{F}^{F}\left(\frac{x}{z}\right)-zC_{F}^{F}\left(\frac{2}{1-z+\Delta}-\frac{3}{2}\right)D_{F}^{F}\left(x\right)\right\} -D_{F}^{F}\left(x\right)\frac{\alpha_{s}}{2\pi}\underset{=0}{\underbrace{\intop_{0}^{1}dzC_{F}^{F}\left(-3z+\frac{3}{2}\right)}}
\]

Since the integral converges we can set $\Delta=0$ and we see the
two equations coincide.

\subsubsection{The  non-Singlet Case.}
 Consider now the octet case.  For
brevity we suppress the representation notation $\alpha$ (\cite{Manohar2012}
considers only the octet representation). The comparison of the two of the evolution equations in this case would be more tricky. This is because \cite{Manohar2012} uses the normalization scale $\mu$ which is integrated from the lower cut-off (in \cite{Manohar2012} it is $\Lambda_{QCD}$ while in our case it's the splitting scale $k$) to the hard scale $Q$ while in this paper we write the evolution in term of the hard scale $Q$. The evolution equation in  \cite{Manohar2012} is (e.g eq. 14 in that paper):

\begin{equation}
\frac{\partial\overline{D}_{F}^{F}\left(x,\mu\right)}{\partial ln\left(\mu\right)}=\frac{\alpha_{s}}{\pi}\intop_{x}^{1}\frac{dz}{z}C_{F}^{F}P_{qq}\left(z\right)D_{F}^{F}\left(\frac{x}{z},\mu\right)+\frac{\alpha_{s}}{\pi}N\left[ln\left(\frac{\mu}{Q}\right)+\frac{3}{4}\right]\overline{D}_{F}^{F}\left(x\right).\label{eq:mu formalism}
\end{equation}

Actually the denominator in the logarithm of \cite{Manohar2012} is $p_{1}^{-}$
(defined in that paper) rather then $Q$ but they are of the same
order of magnitude so in NLO we can use either. In order to go from
$\mu$ evolution to $Q$ evolution we shall integrate this equation
and then differentiate it with respect for $Q$, i.e:

\begin{equation}
\overline{D}_{F}^{F}\left(x,k^{2},Q^{2}\right)=\intop_{ln\left(k\right)}^{ln\left(Q\right)}\frac{\partial\overline{D}_{F}^{F}\left(x,\mu\right)}{\partial ln\left(\mu\right)}dln\left(\mu\right),
\end{equation}

and then

\begin{equation}
\frac{\partial D_{F}^{F}\left(x,k^{2},Q^{2}\right)}{\partial ln\left(Q\right)}=\frac{\partial}{\partial ln\left(Q\right)}\intop_{ln\left(k\right)}^{ln\left(Q\right)}\frac{\partial\overline{D}_{F}^{F}\left(x,\mu\right)}{\partial ln\left(\mu\right)}dln\left(\mu\right).
\end{equation}

The first term of (\ref{eq:mu formalism}) as well as the $\frac{3}{4}$
part are $Q$ independent and are therefor trivial to compute. The
only non-trivial term is the one proportional to $ln\left(\frac{\mu}{Q}\right)$
for which we compute:

\begin{equation}
\frac{\partial}{\partial ln\left(Q\right)}\intop_{ln\left(k\right)}^{ln\left(Q\right)}ln\left(\frac{\mu}{Q}\right)dln\left(\mu\right)=\frac{\partial}{\partial ln\left(Q\right)}\left(-\frac{1}{2}ln^{2}\left(\frac{k}{Q}\right)\right)=ln\left(\frac{k}{Q}\right)
\end{equation}

So we see that the evolution equation of \cite{Manohar2012} can be
written in term of $Q$ rather then $\mu$ as:

\begin{equation}
\frac{\partial\overline{D}_{F}^{F}\left(x\right)}{\partial ln\left(Q^{2}\right)}=\frac{\alpha_{s}}{2\pi}N\left[ln\left(\frac{k^{2}}{Q^{2}}\right)+\frac{3}{4}\right]\overline{D}_{F}^{F}\left(x\right)+\frac{\alpha_{s}}{2\pi}\intop_{x}^{1}\frac{dz}{z}\overline{C}_{F}^{F}P_{qq}\left(z\right)\overline{D}_{F}^{F}\left(\frac{x}{z}\right)
\label{nsap}
\end{equation}

In DDT scheme  however:

\begin{equation}
\frac{\partial\overline{D}_{F}^{F}\left(x\right)}{\partial ln\left(Q^{2}\right)}=\frac{\alpha_{s}}{4\pi}\intop_{0}^{1}\frac{dz}{z}\overline{\Phi}_{F}^{F}\left(z\right)\overline{D}_{F}^{F}\left(\frac{x}{z}\right)-z^{2}\overline{D}_{F}^{F}\left(x\right)\left(\Phi_{F}^{F}\left(z\right)+\Phi_{F}^{G}\left(z\right)\right)=\label{eq:DDT non singlet}
\end{equation}

\[
=\frac{\alpha_{s}}{2\pi}\intop_{0}^{1}\frac{dz}{z}\overline{C}_{F}^{F}\left[\frac{1+z^{2}}{1-z+\Delta}\right]\overline{D}_{F}^{F}\left(\frac{x}{z}\right)-z^{2}C_{F}^{F}\left(\frac{1+z^{2}}{1-z+\Delta}+\frac{(1-z)^{2}+1}{z}\right)\overline{D}_{F}^{F}\left(x\right)=
\]

\begin{multline*}
=\frac{\alpha_{s}}{2\pi}\intop_{0}^{1}\frac{dz}{z}\overline{C}_{F}^{F}\left[\frac{1+z^{2}}{1-z+\Delta}\right]\overline{D}_{F}^{F}\left(\frac{x}{z}\right)-z^{2}\overline{C}_{F}^{F}\left(\frac{1+z^{2}}{1-z+\Delta}+\frac{(1-z)^{2}+1}{z}\right)\overline{D}_{F}^{F}\left(x\right)-\\
\overline{D}_{F}^{F}\left(x\right)\frac{\alpha_{s}}{2\pi}\intop_{0}^{1}dzz\left(C_{F}^{F}-\overline{C}_{F}^{F}\right)\left(\frac{1+z^{2}}{1-z+\Delta}+\frac{(1-z)^{2}+1}{z}\right)
\end{multline*}

The first integral is just the new kernel as in \cite{Manohar2012}
(after we set $\Delta=0$), The second integral can  be  calculated
analytically and  we have (to leading order in $\Delta$):
\begin{equation}
\frac{\partial\overline{D}_{F}^{F}\left(x\right)}{\partial ln\left(Q^{2}\right)}=\frac{\alpha_{s}}{2\pi}\intop_{x}^{1}\frac{dz}{z}\overline{C}_{F}^{F}P_{qq}\left(z\right)\overline{D}_{F}^{F}\left(\frac{x}{z}\right)+\overline{D}_{F}^{F}\left(x\right)\frac{\alpha_{s}}{2\pi}\left(C_{F}^{F}-\overline{C}_{F}^{F}\right)\left(\frac{3}{2}+2\cdot log\left(\frac{k^{2}}{Q^{2}}\right)\right)=\label{eq:NLO}
\end{equation}

\[
=\frac{\alpha_{s}}{2\pi}\intop_{x}^{1}\frac{dz}{z}\overline{C}_{F}^{F}P_{qq}\left(z\right)\overline{D}_{F}^{F}\left(\frac{x}{z}\right)+\overline{D}_{F}^{F}\left(x\right)\frac{\alpha_{s}}{2\pi}N\left(\frac{3}{4}+log\left(\frac{k^{2}}{Q^{2}}\right)\right)
\]
This is just Equation  \ref{nsap}.
 We see that  also  in the non-singlet case  both regularisation schemes lead to identical equations.

\subsubsection{ Different Sudakov Factors}

As we have seen in the previous subsection the evolution functions is the same, at least including  the NLO logarithms, and therefore
any observable arising from it must be the same. On the other hand
we define the Sudakov factor as in (\ref{eq:Sudakov}). When expanding
this expression to NLO for quark evolution (at $N=3$) we have: 

\begin{equation}
\overline{S}_{8_{A}}=e^{-\sum_{C}\left(C_{F}^{C}-\overline{C}_{F}^{C}\right)\intop_{k_{0}^{2}}^{Q^{2}}\frac{dk^{2}}{k^{2}}\frac{\alpha_{s}\left(k^{2}\right)}{4\pi}\intop_{0}^{1}dzzV_{F}^{C}\left(z\right)}=e^{-\int_{k_{0}^{2}}^{Q^{2}}\frac{dk^{2}}{k^{2}}\frac{\alpha_{s}}{4\pi}N\left(\frac{137}{54}+2\cdot log\left(\frac{k^{2}}{Q^{2}}\right)\right)}.\label{eq:Sudakov DDT}
\end{equation}

This is similar in LO but different in NLO then the ``extra'' logarithms
given in (\ref{eq:NLO}) which should give:

\[
\overline{S}_{8_{A}}=e^{-\left(C_{F}^{F}-\overline{C}_{F}^{F}\right)\intop_{k_{0}^{2}}^{Q^{2}}\frac{dk^{2}}{k^{2}}\frac{\alpha_{s}\left(k^{2}\right)}{4\pi}\intop_{0}^{1}dzz\sum_{C}V_{F}^{C}\left(z\right)}=e^{-\int_{k_{0}^{2}}^{Q^{2}}\frac{dk^{2}}{k^{2}}\frac{\alpha_{s}}{4\pi}N\left(\frac{3}{2}+2\cdot log\left(\frac{k^{2}}{Q^{2}}\right)\right)}.
\]

This is due to the different way we chose to define the tree-level
evolution equation (compare Eq. (\ref{eq:Reduced DGLAP}) and (\ref{eq:DDT non singlet})
 to the first part of Eq.  (\ref{eq:NLO})), which ``absorbs''
some of the NLO logs. We stress again that once multiplying
the solutions of (\ref{eq:Reduced DGLAP}) 
with the Sudakov factor (\ref{eq:Sudakov}) 
we'll get the same distributions to NLO as if we solved (\ref{eq:NLO}).

\subsection{Regularization of $z\rightarrow1$ divergence}
\par Let us now discuss the regularisation in section \ref{subsec:Regularization}.  We use  the fundamental
solutions of evolution equations/the Green function, that are singular in $x\rightarrow 1$
limit. On the other hand, the evolution equations themselves are not singular 
\cite{Manohar2012,Diehl2016}, and it is easy to check that solving these evolution 
equations does not involve singularities.
It's natural to ask then whether
these divergences, encountered in section \ref{subsec:Regularization}, are physical or are purely the mathematical artifact of our method
of computing $_{\left[1\right]}\overline{D}$, using Green functions. If so, does the
regularization method introduced gives the same results as  other
methods?

 In the following, we'll show that these divergences are of mathematical nature and that our method of regularizing them gives the same result as a different approach to this problem, that avoids singularities altogether.

First let's outline another method for the computation of $_{\left[1\right]}\overline{D}$.
Let $F_{k}\left(x_{1},x_{2},q_{1},q_{2}\right)$ be the distribution
that resulted from splitting at a scale $k$, i.e:

\begin{equation}
\phantom{}^{\alpha}F_{k}^{AB}\left(x_{1},x_{2},q_{1}=k,q_{2}=k\right)=\sum_{E}\frac{\alpha_{s}\left(k^{2}\right)}{2\pi}G_{h}^{E}\left(x_{1}+x_{2},k^{2}\right)V_{E}^{A}\left(x_{1}+x_{2}\right)\frac{n_{A}}{n_{E}}\phantom{}^{\alpha}\overline{C}_{A}^{B}
\end{equation}

Note that the parton type  notation is consistent with that appearing in  Eqs.  (\ref{eq: PH}-\ref{eq:D 1-2}). It's now possible to perform the evolution in $q_{1}$ and $q_{2}$
up to to $Q_{1}$ and $Q_{2}$. The evolution equation for $\phantom{}_{\left[1\right]}^{\alpha}F_{k}^{AB}$
is actually just (\ref{eq:Reduced DGLAP}) but in each of the variables $x_{i},Q_{i}$
alone \citep{Manohar2012,buffing2021,Diehl2021}:

\begin{equation}
\frac{\partial\phantom{}^{\alpha}F_{k}^{AB}\left(x_{i},q_{i}\right)}{\partial q_{i}}=\frac{\alpha_{s}\left(q_{i}^{2}\right)}{4\pi}\sum_{C}\intop_{0}^{1}\frac{dz}{z}\left[\phantom{}^{\alpha}\overline{\Phi}_{C}^{A}\left(z\right)\phantom{}^{\alpha}F_{k}^{CB}\left(\frac{x_{i}}{z},q_{i}\right)-z^{2}\phantom{}^{\alpha}\overline{\Phi}_{C}^{A}\left(z\right)\phantom{}^{\alpha}F_{k}^{CB}\left(x_{i},q_{i}\right)\right].\label{eq:alternative DGLAP}
\end{equation}

There is also a rapidity evolution, but as explained before, it only
contributes single logarithmic corrections to the Sudakov factor. We
also need to add the effect of the Sudakov factors and to integrate
over the initial splitting scale $k$, so we find that:

\begin{equation}
\phantom{}_{\left[1\right]}^{\alpha}\overline{D}_{h}^{AB}\left(x_{1},x_{2},Q_{1},Q_{2}\right)=\intop_{Q_{0}^{2}}^{min\left(Q_{1}^{2},Q_{2}^{2}\right)}\frac{dk^{2}}{k^{2}}\phantom{}^{\alpha}\overline{S}_{A}\left(k^{2},Q_{1}^{2}\right)\phantom{}^{\alpha}\overline{S}_{B}\left(k^{2},Q_{2}^{2}\right)\phantom{}^{\alpha}F_{k}^{AB}\left(x_{1},x_{2},Q_{1},Q_{2}\right).\label{eq: D alternative}
\end{equation}

Since (\ref{eq:alternative DGLAP}) has no divergences in it,
even for $\phantom{}^{\alpha}\overline{C}_{A}^{B}<0$, $F_{k}$ is
well defined without need for any regularization. $F_{k}$ can be
numerically evaluated at $Q_{1},Q_{2}$ for every $k$ using numerical
methods to solving PDE's. The integral in (\ref{eq: D alternative})
can then be computed numerically using the values of $F_{k}$ for
each $k$.

Does this method gives the same result as the regularization introduced
in the text? Because (\ref{eq: D alternative}) is hard to compute
numerically, we will demonstrate this in a toy model that leaves only
the essential ingredients. Let $\widetilde{F}\left(x,k\right)$ be
a function that satisfies

\begin{equation}
\phantom{}^{\alpha}\widetilde{F}\left(x,k=1.4\ GeV\right)=\widetilde{F}_{0}\left(x\right)
\end{equation}

for some known function $F_{0}$ and obeys the differential equation:

\begin{equation}
\frac{\partial\phantom{}^{\alpha}\widetilde{F}\left(x,k\right)}{\partial k}=\frac{\alpha_{s}\left(k^{2}\right)}{4\pi}\intop_{0}^{1}\frac{dz}{z}\left[\phantom{}^{\alpha}\overline{\Phi}_{G}^{G}\left(z\right)\phantom{}^{\alpha}\widetilde{F}\left(\frac{x}{z},k\right)-z^{2}\phantom{}^{\alpha}\overline{\Phi}_{G}^{G}\left(z\right)\phantom{}^{\alpha}\widetilde{F}\left(x,k\right)\right],\label{eq:toy model}
\end{equation}

i.e containing only gluon evolution. These assumptions simplify the numerical
computations greatly while still keeping the problem of negative
color factor for $\alpha=27$. We'll now evaluate $\widetilde{F}$
for several values of $k$ using the two methods. First using the
fundamental solutions (we had to modify the solutions given in Appendix
\ref{sec:-at-the-x-1} to include only gluon evolution, i.e by setting $\phantom{}^{\alpha}\overline{C}_{F}^{F}=\phantom{}^{\alpha}\overline{C}_{G}^{F}=\phantom{}^{\alpha}\overline{C}_{F}^{G}=0$)
we write:

\begin{equation}
\phantom{}^{\alpha}\widetilde{F}\left(x,k\right)=\intop_{0}^{1}\frac{dz}{z}\phantom{}^{\alpha}\widetilde{F}_{0}\left(z\right)\phantom{}^{\alpha}\widetilde{D}\left(\frac{x}{z},k\right).
\end{equation}

\begin{figure}[H]
\begin{centering}
\includegraphics[scale=0.75]{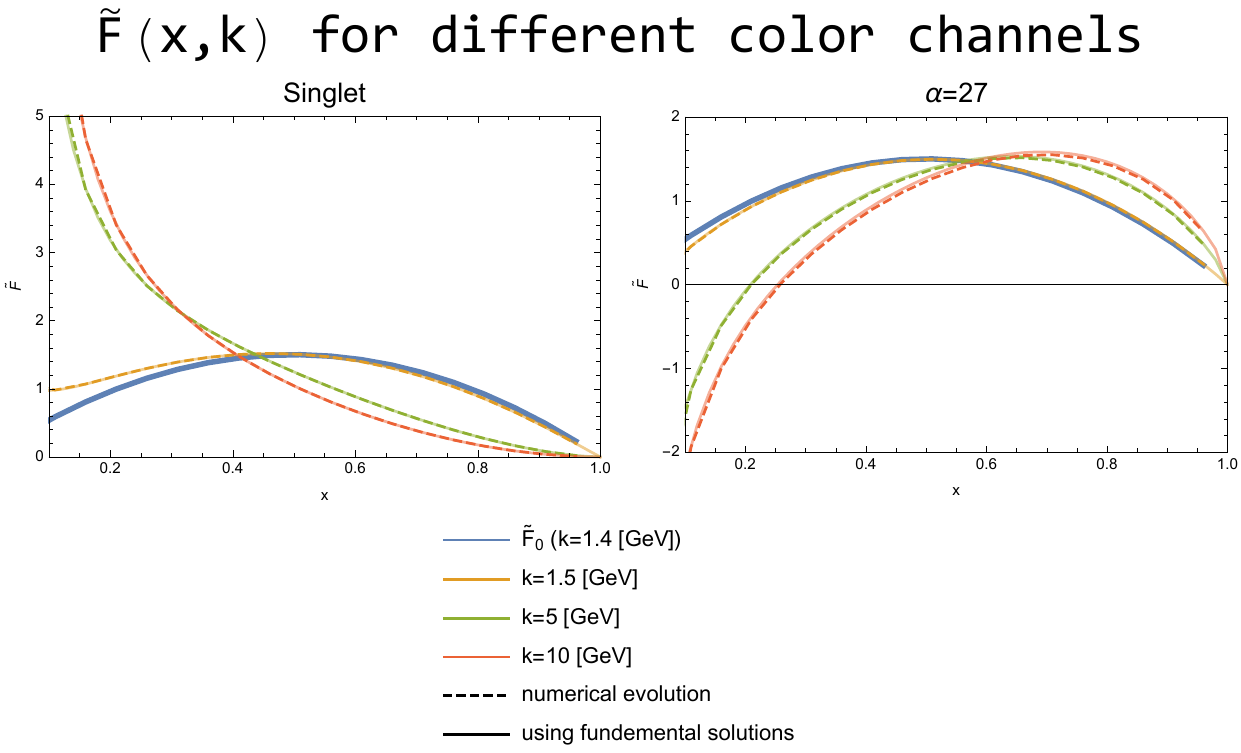}
\par\end{centering}
\caption{Comparison of the numerical evolution to the fundamental solution
method. It can be seen that the results of both methods match. \label{fig: Comparison of methods}}
\end{figure}

For $\alpha=27$ this integral diverges and we use the $1d$ version
of the regularization given in section \ref{subsec:Regularization}. On the other hand we
directly evaluated (\ref{eq:toy model}) using the Runge--Kutta method
with $k$ spacing of $0.025\ \left[GeV\right]$ (a similar numerical
methods as was used in \citep{Manohar2012,Diehl2021}). The results
for $F_{0}=6\cdot x\left(1-x\right)$ at $k=1.5,10,30$ $\left[GeV\right]$
for both $\alpha=27$ and the singlet channel (which is used as a
consistency test) are shown in Fig. \ref{fig: Comparison of methods}
. It should be stressed that the fact that $\widetilde{F}$ can be
negative for non-singlet channels is not a surprise \citep{Diehl2021}. 

We see that the two methods produce the same results. Because the
use of fundamental solutions is much easier numerically (once they
were computed) and also keeps the physical picture more transparent,
we adopt this method in this paper.

\bibliographystyle{aapmrev4-2}
%aapmrev4-2.bst 2019-01-14 (MD) hand-edited version of aapmrev4-1.bst
%Control: key (0)
%Control: author (8) initials jnrlst
%Control: editor formatted (1) identically to author
%Control: production of article title (-1) disabled
%Control: page (0) single
%Control: year (1) truncated
%Control: production of eprint (0) enabled
%

\end{document}